\documentclass[12pt]{article}
\usepackage{epsfig}
\usepackage{axodraw}
\usepackage{epsfig}                            
\usepackage{graphicx}
\usepackage{rotate}
\usepackage{latexsym}
\newcommand\pubnumber{}
\newcommand\pubdate{\today}
\newcommand\hepnumber{hep-ph/0108252}

\def\csumb{Dipartimento di Fisica Teorica, Universit\`a di Torino, Italy\\
INFN, Sezione di Torino, Italy}
\def\support{\footnote{Work supported by the
European Union under contract HPRN-CT-2000-00149.}} 

\def\Title#1{\begin{center} {\Large\bf #1 } \end{center}}
\def\Author#1{\begin{center}{ \sc #1} \end{center}}
\def\Address#1{\begin{center}{ \it #1} \end{center}}

\newcommand\pubblock{\rightline{\begin{tabular}{l} \pubnumber\\
         \pubdate\\ \hepnumber \end{tabular}}}
\newenvironment{Abstract}{\begin{quotation}  }{\end{quotation}}

\def\Acknowledgments{\bigskip  \bigskip \begin{center}
          \large\bf Acknowledgments\end{center}}

\makeatletter
\def\section{\@startsection{section}{0}{\z@}{5.5ex plus .5ex minus
 1.5ex}{2.3ex plus .2ex}{\large\bf}}
\def\subsection{\@startsection{subsection}{1}{\z@}{3.5ex plus .5ex minus
 1.5ex}{1.3ex plus .2ex}{\normalsize\bf}}
\def\subsubsection{\@startsection{subsubsection}{2}{\z@}{-3.5ex plus
-1ex minus  -.2ex}{2.3ex plus .2ex}{\normalsize\sl}}

\renewcommand{\@makecaption}[2]{%
   \vskip 10pt
   \setbox\@tempboxa\hbox{\small #1: #2}
   \ifdim \wd\@tempboxa >\hsize     
       \small #1: #2\par          
     \else                        
       \hbox to\hsize{\hfil\box\@tempboxa\hfil}
   \fi}

 \def\citenum#1{{\def\@cite##1##2{##1}\cite{#1}}}
\def\citea#1{\@cite{#1}{}}
 
\newcount\@tempcntc
\def\@citex[#1]#2{\if@filesw\immediate\write\@auxout{\string\citation{#2}}\fi
  \@tempcnta\z@\@tempcntb\m@ne\def\@citea{}\@cite{\@for\@citeb:=#2\do
    {\@ifundefined
       {b@\@citeb}{\@citeo\@tempcntb\m@ne\@citea\def\@citea{,}{\bf ?}\@warning
       {Citation `\@citeb' on page \thepage \space undefined}}%
    {\setbox\z@\hbox{\global\@tempcntc0\csname b@\@citeb\endcsname\relax}%
     \ifnum\@tempcntc=\z@ \@citeo\@tempcntb\m@ne
       \@citea\def\@citea{,}\hbox{\csname b@\@citeb\endcsname}%
     \else
      \advance\@tempcntb\@ne
      \ifnum\@tempcntb=\@tempcntc
      \else\advance\@tempcntb\m@ne\@citeo
      \@tempcnta\@tempcntc\@tempcntb\@tempcntc\fi\fi}}\@citeo}{#1}}
\def\@citeo{\ifnum\@tempcnta>\@tempcntb\else\@citea\def\@citea{,}%
  \ifnum\@tempcnta=\@tempcntb\the\@tempcnta\else
  {\advance\@tempcnta\@ne\ifnum\@tempcnta=\@tempcntb \else\def\@citea{--}\fi
    \advance\@tempcnta\m@ne\the\@tempcnta\@citea\the\@tempcntb}\fi\fi}
\makeatother

\newcommand{\nl}{\nonumber\\}
\newcommand{\nn}{\nonumber}
\newcommand{\ds}{\displaystyle}

\newcommand{\lpar}{\left(}                            
\newcommand{\rpar}{\right)} 
\newcommand{\lrbr}{\left[}
\newcommand{\rrbr}{\right]}

\newcommand{\bq}{\begin{equation}}                    
\newcommand{\eq}{\end{equation}}
\newcommand{\bqa}{\arraycolsep 0.14em\begin{eqnarray}}
\newcommand{\eqa}{\end{eqnarray}}
\newcommand{\ba}[1]{\begin{array}{#1}}
\newcommand{\ea}{\end{array}}
\newcommand{\ben}{\begin{enumerate}}
\newcommand{\een}{\end{enumerate}}
\newcommand{\bei}{\begin{itemize}}
\newcommand{\eei}{\end{itemize}}
\newcommand{\eqn}[1]{Eq.(\ref{#1})}
\newcommand{\eqns}[2]{Eqs.(\ref{#1})--(\ref{#2})}

\newcommand{\tabn}[1]{Tab.~\ref{#1}}

\newcommand{\fig}[1]{Fig.~\ref{#1}}
\newcommand{\figs}[2]{Figs.~\ref{#1}--\ref{#2}}
\newcommand{\sect}[1]{Section~\ref{#1}}
\newcommand{\sects}[2]{Section~\ref{#1} and \ref{#2}}

\newcommand{\hsp}{\hspace{.5mm}}

\def\Re{\mathop{\operator@font Re}\nolimits}
\def\Im{\mathop{\operator@font Im}\nolimits}
\newcommand{\ord}[1]{{\cal O}\lpar#1\rpar}

\newcommand{\ib}{i}
\newcommand{\asums}[1]{\sum_{#1}}

\newcommand{\asum}[3]{\sum_{#1=#2}^{#3}}
\newcommand{\ph}{\gamma}

\newcommand{\mw}{M_{_W}}

\newcommand{\mz}{M_{_Z}}

\newcommand{\mf}{m_f}

\newcommand{\me}{m_e}

\newcommand{\mzs}{M^2_{_Z}}

\newcommand{\mfs}{m^2_f}

\newcommand{\mes}{m^2_e}

\newcommand{\mone}{m_1}
\newcommand{\mtwo}{m_2}
\newcommand{\mtre}{m_3}

\newcommand{\mlone}{m}

\newcommand{\mones}{m^2_1}
\newcommand{\mtwos}{m^2_2}
\newcommand{\mtres}{m^2_3}
\newcommand{\mfors}{m^2_4}

\newcommand{\Lambdi}[1]{\Lambda_{#1}}

\newcommand{\ect}{e^3}

\newcommand{\seffsf}[1]{\sin^2\theta^{#1}_{\rm{eff}}}

\newcommand{\sla}[1]{/\!\!\!#1}

\newcommand{\spro}[2]{{#1}\cdot{#2}}
\newcommand{\gadu}[1]{\gamma_{#1}}

\newcommand{\li}[2]{\mathrm{Li}_{#1}\lpar\displaystyle{#2}\rpar} 

\newcommand{\egam}[1]{\Gamma\lpar#1\rpar}               
\newcommand{\intmomi}[2]{\int\,d^{#1}#2}

\newcommand{\intfx}[1]{\int_{\scriptstyle 0}^{\scriptstyle 1}\,d#1}
\newcommand{\intfxy}[2]{\int_{\scriptstyle 0}^{\scriptstyle 1}\,d#1\,
                        \int_{\scriptstyle 0}^{\scriptstyle #1}\,d#2}

\newcommand{\aff}[2]{A_{#1}\lpar #2\rpar}                   
\newcommand{\sbff}[1]{B_{#1}}                    

\newcommand{\bff}[4]{B_{#1}\lpar #2;#3,#4\rpar}

\newcommand{\scff}[1]{C_{#1}}                    
                
\newcommand{\cff}[7]{C_{#1}\lpar #2,#3,#4;#5,#6,#7\rpar}    
 
\newcommand{\sdff}[1]{D_{#1}}                    
\newcommand{\dffp}[7]{D_{#1}\lpar #2,#3,#4,#5,#6,#7;}       
\newcommand{\dffm}[4]{#1,#2,#3,#4\rpar}

\newcommand{\Jds}[5]{{\bar{J}}_{#1}\lpar #2,#3;#4,#5 \rpar}

\newcommand{\lpi}{\ln\pi}

\newcommand{\Ddr}{{\ds\frac{1}{{\bar{\varepsilon}}}}}

\newcommand{\Ddrh}{{\ds\frac{1}{\hat{\varepsilon}}}}

\newcommand{\dre}{\varepsilon}

\newcommand{\epp}{\varepsilon'}

\newcommand{\ep}{\epsilon}

\newcommand{\pinchc}[2]{C^{(#1)}_{#2}}

\newcommand{\tHs}{\mu}

\newcommand{\tHss}{\mu^2}
\newcommand{\Reb}{{\rm{Re}}}
\newcommand{\Imb}{{\rm{Im}}}
\newcommand{\tpfi}{\lpar 2\pi\rpar^4\ib}

\newcommand{\dpropi}[1]{d_{#1}}

\newcommand{\dpropii}[2]{d_{#1}\lpar #2\rpar}

\newcommand{\imom}{q}

\newcommand{\imoms}{q^2}

\newcommand{\Trmom}{Q}

\newcommand{\gmv}{Q^2}
\newcommand{\Trmoms}{Q^2}
\newcommand{\Prmoms}{P^2}

\newcommand{\pone}{p_1}
\newcommand{\ptwo}{p_2}
\newcommand{\ptre}{p_3}

\newcommand{\pones}{p_1^2}
\newcommand{\ptwos}{p_2^2}

\newcommand{\upar}[1]{u}

\newcommand{\ssG}{{\scriptscriptstyle{G}}}

\newcommand{\ssL}{{\scriptscriptstyle{L}}}

\newcommand{\ssN}{{\scriptscriptstyle{N}}}

\newcommand{\ssR}{{\scriptscriptstyle{R}}}
\newcommand{\ssS}{{\scriptscriptstyle{S}}}

\newcommand{\ssU}{{\scriptscriptstyle{U}}}
\newcommand{\ssV}{{\scriptscriptstyle{V}}}

\newcommand{\ssZ}{{\scriptscriptstyle{Z}}}

\newcommand{\bqas}{\begin{eqnarray*}}
\newcommand{\eqas}{\end{eqnarray*}}

\newcommand{\eilc}{\gamma}

\def\app#1#2 {{\it Acta. Phys. Pol.} {\bf#1},#2}
\def\cpc#1#2 {{\it Computer Phys. Comm.} {\bf#1},#2}
\def\np#1#2 {{\it Nucl. Phys.} {\bf#1},#2}
\def\pl#1#2 {{\it Phys. Lett.} {\bf#1},#2}
\def\prep#1#2 {{\it Phys. Rep.} {\bf#1},#2}
\def\prev#1#2 {{\it Phys. Rev.} {\bf#1},#2}
\def\prl#1#2 {{\it Phys. Rev. Lett.} {\bf#1},#2}
\def\zp#1#2 {{\it Zeit. Phys.} {\bf#1},#2}
\def\sptp#1#2 {{\it Suppl. Prog. Theor. Phys.} {\bf#1},#2}
\def\mpl#1#2 {{\it Modern Phys. Lett.} {\bf#1},#2}
\def\jetp#1#2 {{\it Sov. Phys. JETP} {\bf#1},#2}
\def\fpj#1#2 {{\it Fortschr. Phys.} {\bf#1},#2}
\def\afp#1#2 {{\it Acta.Phys. Polon.} {\bf#1},#2}
\def\err#1#2 {{\it Erratum} {\bf#1},#2}
\def\ijmp#1#2 {{\it Int. J. Mod. Phys} {\bf#1},#2}
\def\nc#1#2 {{\it Nuovo Cimento} {\bf#1},#2}
\def\ap#1#2 {{\it Ann. Phys.} {\bf#1},#2}
\def\cmp#1#2 {{\it Comm. Math. Phys.} {\bf#1},#2}
\def\el#1#2 {{\it Europhys. Lett.} {\bf#1},#2}
\def\hpa#1#2 {{\it Helv. Phys. Acta} {\bf#1},#2}
\def\yf#1#2 {{\it Yad. Fiz.} {\bf#1},#2}
\def\nim#1#2 {{\it Nucl. Instrum. Meth.} {\bf#1},#2}
\def\spz#1#2 {{\it Sov. Pisma Zhetf} {\bf#1},#2}
\def\jetpl#1#2 {{\it JETP Lett.} {\bf#1},#2}
\def\sjnp#1#2 {{\it Sov. J. Nucl. Phys.} {\bf#1},#2}
\def\ptp#1#2 {{\it Progr. Theor. Phys. (Kyoto)} {\bf#1},#2}
\def\rmp#1#2  {{\it Rev. Mod. Phys.} {\bf#1},#2}
\def\zhetf#1#2 {{\it ZhETF} {\bf#1},#2}
\def\prs#1#2 {{\it Proc. Roy. Soc.} {\bf#1},#2}
\def\phys#1#2 {{\it Physica} {\bf#1},#2}

\newcommand{\intfxx}[2]{\int_{\scriptstyle 0}^{\scriptstyle 1}\,d#1\,
                        \int_{\scriptstyle 0}^{\scriptstyle 1}\,d#2}
\newcommand{\lbpa}{\Bigl(}                            
\newcommand{\rbpa}{\Bigl)}                            
\def\bfi{\begin{figure}}
\def\efi{\end{figure}}
\newcommand{\intmomsii}[3]{\int\,d^{#1}#2\,d^{#1}#3}

\begin{document}
\begin{titlepage}
\pubblock

\vfill
\def\thefootnote{\fnsymbol{footnote}}
\Title{An Approach Toward the Numerical\\
Evaluation of Multi-Loop Feynman Diagrams}
\vfill
\Author{Giampiero Passarino\support}
\Address{\csumb}
\vfill
\begin{Abstract}
A scheme for systematically achieving accurate numerical evaluation of 
multi-loop Feynman diagrams is developed. This shows the feasibility of a 
project aimed to produce a complete calculation for two-loop predictions 
in the Standard Model.
As a first step an algorithm, proposed by F.~V.~Tkachov and based on the
so-called generalized Bernstein functional relation, is applied to one-loop 
multi-leg diagrams with particular emphasis to the presence of infrared 
singularities, to the problem of tensorial reduction and to the classification
of all singularities of a given diagram.
Successively, the extension of the algorithm to two-loop diagrams is examined.
The proposed solution consists in applying the functional relation to the 
one-loop sub-diagram which has the largest number of internal lines. 
In this way the integrand can be made smooth, a part from a factor which is 
a polynomial in $x_{\ssS}$, the vector of Feynman parameters needed for the 
complementary sub-diagram with the smallest number of internal lines. Since 
the procedure does not introduce new singularities one can distort the 
$x_{\ssS}$-integration hyper-contour into the complex hyper-plane, thus 
achieving numerical stability. The algorithm is then modified to deal
with numerical evaluation around normal thresholds. Concise and practical
formulas are assembled and presented, numerical results and comparisons with 
the available literature are shown and discussed for the so-called sunset 
topology.
\end{Abstract}
\vfill
\vfill
\begin{center}
\end{center}
\end{titlepage}
\def\thefootnote{\arabic{footnote}}
\setcounter{footnote}{0}
\section{Introduction}
The evaluation of multi-loop Feynman diagrams has a long history dating
back to the days of renormalization of QED (for an historical review see 
ref.~\cite{Pais:1986nu}). In this respect there are theories more
simple than others, noticeably QED and also to some extent QCD, where
we have few masses and the analytical approach can be pushed very far.
Typical examples are the calculations of $g-2$ in QED, see for instance 
ref.~\cite{Mastrolia:2000va}, or recent four-loop calculations in 
QCD~\cite{Chetyrkin:2000dq}. Conversely the full electroweak Lagrangian
shows several masses, ranging over a wide interval of values, therefore
making the analytical evaluation of Feynman diagrams a complicated task.

The modern era of loop-calculus begins with the work of 't Hooft and 
Veltman~\cite{'tHooft:1979xw} for arbitrary one-loop diagrams. As a result, an 
arbitrary scalar triangle diagram is expressed as the combination of $12$ 
di-logarithms (sometimes called Spence-functions) while $108$ di-logarithms 
are needed for the general box diagram. 

In recent years we have been witnessing a huge amount of work in the direction
of analytical evaluation of multi-loop diagrams. To be more precise we should
distinguish between reduction procedures to bring specific classes of
physics problems to some specific classes of integrals and calculational 
procedures for those specific classes of integrals.

It would be impossible to quote the hundreds of papers that have been dealing 
with the subject and we are forced to make some selection. In the following we 
will briefly review the relevant literature, confining the search to those 
papers that are not directly connected to the present work. Other important
references will appear in the body of the paper.

Following and improving the techniques of~\cite{'tHooft:1979xw}
dimensionally regulated pentagon integrals have been 
considered in~\cite{Bern:1994kr}. At the same time the methodology for 
one-loop $n$-point gauge theory amplitudes has been further developed 
in~\cite{Bern:1994zx} and in~\cite{Bern:1995zn}.

The algebraic reduction of one loop graphs, introduced 
in~\cite{Passarino:1979jh}, has been reconsidered in~\cite{Fleischer:2000hq}.
More generally, the problem of reducing Feynman graph amplitudes to a 
minimal set of basic integrals has been discussed in~\cite{Tarasov:1998nx}.

An example of systematic approach to multi-loop calculations with the use of
the program SCHOONSCHIP~\cite{Veltman:1991xb} is due to the Dubna group and 
a summary of the results can be found in ref.~\cite{Vladimirov:1979ib}. 
Meanwhile, algebraic programs like Mincer have been 
introduced~\cite{Gorishnii:1989gt}.

An important breakthrough in this line of research is also represented by the 
algebraic approach, the so-called method of integration by parts. The simplest
identity of this type was buried in a proof in~\cite{'tHooft:1972fi}. 
A complete summary of everything relevant for 3-loop
self-energies, as used in the next-to-next order (NNLO) calculations, is 
collected in~\cite{ftes}.
The concept and calculability of massless three-loop self-energies was
reported in~\cite{Tkachov:1981wb}.
Further details for three-loop self-energies can be found 
in~\cite{Chetyrkin:1981qh}.
Also important in this context are explicit solutions of the recurrence
relations, see for instance ref.~\cite{Tkachov:1983xk}.

Another line of development is related to the evaluation of massive integrals,
e.g.\ in the context of effective heavy-mass theory. The most recent reference
is~\cite{Broadhurst:1995ha}.
Finally a group of people were successfully using the so-called method
of uniqueness with various improvements, see ref.~\cite{Belokurov:1979qp}
and also~\cite{Kazakov:1988mu},\cite{Usyukina:1993jd} 
and~\cite{Smirnov:2001ie}.

Another significant development is represented by the method of differential 
equations, first discussed in~\cite{Kotikov:1991kg} and fully developed
in~\cite{Remiddi:1997ny} (see also ref.~\cite{Laporta:2000dd}).
Explicit solutions of the recurrence relations with respect to space-time
dimension are shown in~\cite{Tarasov:2000vg}.
Generalized recurrence relations for two-loop propagator integrals with 
arbitrary masses are discussed in~\cite{Tarasov:1997kx}.

Furthermore, there is the subject of asymptotic expansions of Feynman diagrams.
The first journal report of existence of simple formulas for OPE (those used 
in NNLO calculations) was in ref.~\cite{Tkachov:1983st}, while a summary of 
the rules was presented in ref.~\cite{Gorishnii:1983su}.
The euclidean variant of the theory appeared in ref.~\cite{Pivovarov:1986kt}
and \cite{Pivovarov:1993du}.
The theory was reviewed -- and the key trick to extend it to Minkowski
space indicated -- in ref.~\cite{Tkachov:1993zi}.
The final step was the extension of the theory to arbitrary
diagrams in Minkowski space (including real phase-space diagrams), see
for instance ref.~\cite{Tkachov:1997aq}. 
Successive contributions can be found in ref.~\cite{Fleischer:1998bq}.

The techniques of Gegenbauer polynomials has proven to be useful for massless 
self-energies, mostly in two loops but with non-canonical powers of 
propagators, see ref.~\cite{Tkachov:1979yc} and ref.~\cite{Chetyrkin:1980pr}.

Another meaningful technique, namely momentum expansion of multi-loop 
Feynman diagrams, was introduced in~\cite{Tarasov:1996bz}.
Furthermore, the connection between Feynman integrals having different values 
of the space-time dimension was shown in~\cite{Tarasov:1996br}, and the
conformal mapping and Pade approximants to the calculation
of various two-loop Feynman diagrams in~\cite{Fleischer:1994dc}.

Another significant attempt towards the analytical evaluation of multi-loop
diagrams is represented by the work of ref.~\cite{Broadhurst:1985vq}.

As far as applications to the standard model are concerned, two-loop 
self-energies have been discussed in~\cite{Fleischer:2000vb}.
From the point of view of calculations with a direct relevance to standard 
model phenomenology we quote the next-to-leading two-loop electroweak top 
corrections of ref.~\cite{Degrassi:1996mg}, the fermionic two-loop 
contributions to muon decay and $\mw-\mz$ interdependence of 
ref.~\cite{Freitas:2000nv} and the complete two-loop QED contributions to the 
muon lifetime~\cite{vanRitbergen:1999yd}.

All these approaches, with very few exceptions, have a common motivation:
to compute analytically as much as possible of multi-loop Feynman diagrams.
Soon or later this approach will collapse and one can easily foresee that
this failure will show up at the level of a complete two-loop calculation in 
the standard model which is required to produce quantities as $\seffsf{l}$ 
with a theoretical precision of $1\,\times\,10^{-6}$. For this reason one is 
lead to consider an alternative approach to the whole problem, namely to 
abandon the analytical way in favor of a fully automatic, numerical evaluation
of multi-loop diagrams. The Holy Graal (General Recursive Applicative and 
Algorithmic Language) requires fast and accurate procedures to deal with the 
singularities of an arbitrary diagram.

The outline of the paper will be as follows: in \sect{btt} we briefly review
the Bernstein-Tkachov (BT) theorem and in \sect{olc} we consider its 
application to one-loop diagrams. In particular, the infrared divergent 
triangle graph is discussed in \sect{tir}, while other one-loop functions are 
studied in \sects{oolf}{oolidf}. Reduction of tensor one-loop integrals is 
examined in \sect{redu}. Singularities of one-loop diagrams and the connection
between generalized Bernstein functional relation and Landau singularities is 
briefly addressed in \sect{singu}. Multi-loop diagrams are introduced in 
\sect{mld} and the main result of this paper, the minimal BT approach, is 
presented in \sect{mbta}.
In \sects{s3t}{evs3p} we discuss the evaluation of the so-called sunset 
two-loop diagram (or sunrise, depending on the mood of the author). 
Integration with complex Feynman parameters is introduced
in \sect{icp} and a solution for numerical evaluation around and at the normal
threshold is given in \sect{nt}. Special cases of the sunset diagram are 
discussed in \sect{sc} where numerical results are shown and comparison are 
performed with those results that are available in the literature. Tensor 
integrals are introduced and discussed in \sect{ti}. Technical details are 
presented in several appendices.
\section{Bernstein-Tkachov theorem\label{btt}}
For a fast and accurate numerical evaluation of multi-loop diagrams there
is some interesting proposal in the literature that has not yet received the 
due attention.

The Bernstein-Tkachov theorem tells us that for any finite set of polynomials
$V_i(x)$, where $x = \,\lpar x_1,\dots,x_{\ssN}\rpar$ is a vector of Feynman
parameters, there exists an identity of the following form:
\bq
{\cal P}\,\lpar x,\partial\rpar \prod_i\,V_i^{\mu_i+1}(x) = B\,
\prod_i\,V_i^{\mu_i}(x).
\label{functr}
\eq
where ${\cal P}$ is a polynomial of $x$ and $\partial_{i} = 
\partial/\partial x_i$; $B$ and all coefficients of ${\cal P}$ are polynomials 
of $\mu_i$ and of the coefficients of $V_i(x)$.
The proof of the theorem uses methods of abstract algebra and we refer to
the paper by F.~V.~Thachov~\cite{Tkachov:1997wh} for details.
Given any Feynman diagram the $\mu_i$ in \eqn{functr} will be of the form
$-n_i-\varepsilon/2$ where $n_i$ are positive integers and 
$n = 4 - \varepsilon$ with $n$ being the space-time dimension. One can apply 
recursively \eqn{functr} till the moment when all powers are of the form
$N_i - \varepsilon/2$ with $N_i \ge 0$. The Laurent expansion in
$\varepsilon$ yields the final form of the integrand.
As pointed out in ref.\cite{Bardin:2000cf}, with $N_i = 0$ the imaginary part 
of the integrand is discontinuous, with $N_i = 1$ the integrand is continuous, 
with $N_i = 2$ has continuous first derivative and so on.
In the next section we will show the solution for arbitrary one-loop diagrams.
\subsection{The one-loop case\label{olc}}
\begin{figure}[ht]
\vspace*{-8mm}
\[
  \vcenter{\hbox{
  \begin{picture}(140,130)(-15,-15)
    \Line(30,50)(73,75)        \Text(50,72)[cb]{$\mone$}
    \Line(73,75)(73,25)        \Text(50,25)[cb]{$\mtwo$}
    \Line(73,25)(30,50)        \Text(82,50)[cb]{$\mtre$}
    \Line(0,50)(30,50)    
    \Line(100,100)(73,75)   
    \Line(100,0)(73,25)   
    \LongArrow(10,57)(20,57)   \Text(13,65)[cb]{$\pone$}
    \LongArrow(89,98)(82,91)   \Text(79,103)[cb]{$\ptre$}
    \LongArrow(89,2)(82,9)     \Text(79,-3)[cb]{$\ptwo$}
  \end{picture}}}
\]
\vspace{-0.75cm}
\caption[]{The three-point Green function. All external momenta are flowing 
inwards.\label{fig:threepoint}}
\end{figure}
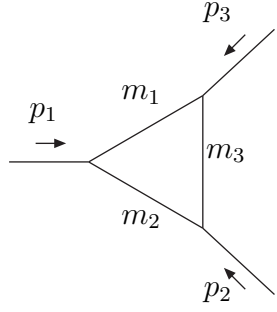

For arbitrary one-loop diagrams we have a universal master formula, again due 
to F.~V.~Tkachov~\cite{Tkachov:1997wh} (see also ref.~\cite{Bardin:2000cf}
and ref.~\cite{Passarino:2001sq}). Any one-loop Feynman diagram $G$, 
irrespectively from the number of external legs, is expressible as
\bq
G = \int_{\ssS}\,dx\,V^{-\mu}(x),
\eq
where the integration region is $x_i \ge 0, \,\asums{i}\,x_i \le 1$ and
where $V(x)$ is a quadratic form of $x$,
\bq
V(x) = x^t\,H\,x + 2\,K^t\,x + L.
\eq
The solution to the problem of determining the polynomial ${\cal P}$ is as 
follows:
\bq
{\cal P} = 1 - {{\,\lpar x+X\rpar\,\partial_x}\over {2\,\lpar\mu+1\rpar}},
\qquad
X = K^t\,H^{-1}, \qquad B = L - K^t\,H^{-1}\,K.
\label{rol}
\eq
Let us consider an example, the scalar $3$-point function $\scff{0}$ of
\fig{fig:threepoint}
The corresponding expression is written in terms of a $2$-dimensional
integral. We start from
\bqa
\scff{0} &\equiv& \cff{0}{p^2_1}{p^2_2}{\,\lpar p_1+p_2\rpar^2}{m_1}{m_2}{m_3} 
\nl
{}&=& \frac{\tHs^{\varepsilon}}{i\,\pi^2}\,\int\,d^nq\,{1\over{
\,\lpar q^2+m^2_1\rpar\,\lpar\,\lpar q+p_1\rpar^2+m^2_2\rpar\,
\,\lpar\,\lpar q+p_1+p_2\rpar^2 + m^2_3\rpar}},
\eqa
where $n = 4-\varepsilon$ is the space-time dimension and $\tHs$ is the 
arbitrary unit of mass.
After introducing Feynman parameters, we obtain
\bq
C_0 = \intfxy{x}{y}\, V^{-1-\varepsilon/2}(x,y),  \qquad
V(x,y) = a\,x^2 + b\,y^2 + c\,xy + d\,x + e\,y + f - i\,\delta
\eq
where $\delta \to 0_+$ and where the coefficients of the quadratic form $V$ are
\bqa 
a&=&-\ptwos, 
\qquad 
b = -\pones, 
\qquad 
c =  \pones+\ptwos-\gmv, 
\qquad
d =  \ptwos+ \mtwos - \mtres,
\nl
e&=&-\ptwos+\gmv+\mones-\mtwos, 
\qquad
f =  \mtres.
\eqa
Therefore, the quadratic form in Feynman parameters is specified by
\bqa
H = \frac{1}{2}\,
\,\lpar
\ba{lr}
2\,a & c \\
c & 2\,b 
\ea
\rpar,
\qquad
K^t = \frac{1}{2}\,\lpar d,e\rpar,
\qquad
L = f - i\,\delta.
\eqa
The main idea beyond the BT-theorem is that we can now integrate by parts
and {\em raise powers} in the integrand. The first step will be as follows: 
define $(X,Y) = (H^{-1})^t K$ and derive
\bqa
\scff{0} &=& \frac{1}{B}\,\lpar 1 - \frac{1}{\ep}\rpar\,\intfxy{x}{y}\,
V^{-\ep/2}\,\lpar x,y\rpar + \frac{1}{B\ep}\,\intfx{x}\,\,\Bigl[
\,\lpar 1 + Y\rpar\,V^{-\ep/2}\,\lpar x,1\rpar  \nl
{}&-& Y\,V^{-\ep/2}\,\lpar x,0\rpar + \,\lpar 1 + X\rpar\,V^{-\ep/2}\,
\lpar 1,x\rpar - \,\lpar x + X\rpar\,V^{-\ep/2}\,\lpar x,x\rpar\Bigr].
\label{ronce}
\eqa
Although this formulas could be used directly as it stands,
better numerical convergency is reached when we {\em raise} again the power
of $V$. 
This requires applying again \eqn{rol} to the first term in \eqn{ronce}
with $\mu= - \ep$. For the terms inside the second integral of \eqn{ronce}
we have now several quadratic forms in one variable to which the algorithm has
to be applied:
\bqa
\ba{ll}
1) &\; V(x,1) = ax^2+(c+d)x+b+e+f-i\delta \\
2) &\; V(x,0) = ax^2+dx+f-i\delta  \\
3) &\; V(1,x) = bx^2+(c+e)x+a+d+f-i\delta \\
4) &\; V(x,x) = (a+b+c)x^2+(d+e)x+f-i\delta.
\ea
\eqa
The corresponding coefficients are
\bqa
\ba{llll}
1) &\; H = a, &\;\;     K = \frac{1}{2}\,(c+d), &\;\; L = b+e+f-i\delta; \\
2) &\; H = a, &\;\;     K = \frac{1}{2}\,d,     &\;\; L = f-i\delta;  \\
3) &\; H = b, &\;\;     K = \frac{1}{2}\,(c+e), &\;\; L = a+d+f-i\delta; \\
4) &\; H = a+b+c, &\;\; K = \frac{1}{2}\,(d+e), &\;\; L = f-i\delta.
\ea
\eqa
The full result can be represented as the sum of a two-dimensional integral,
a one-dimensional integral and a remainder,
\bq
C_0 = \Delta^{-1}\,\,\Bigl[\Delta^{-1}\,C^d_0 + C^s_0 + C^0_0\Bigr].  
\label{toz}
\eq
Several combinations of the external parameters, not to be confused with
Gram's determinants, enter into the final expression. They are
\bq
G_2 = 4 a b-c^2, \qquad G_{11} = a+b+c,  \quad G_{12} = a,  \quad
G_{13} = b.  
\label{grams}
\eq
From them additional auxiliary quantities are constructed,
\bqa
\Delta &=& f G_2-(b d^2-c d e+a e^2),  \nl
\Delta_1 &=& f G_{11} - \frac{(d+e)^2}{4},  \quad
\Delta_2 = f G_{12} - \frac{d^2}{4},  \quad
\Delta_3 = (a+d+f),  
\label{deltas}
\eqa
and also
\bqa
a_x &=& 2 d b-c e,  \qquad a_y = 2 e a-d c,  \nl
a_{1x} &=& \frac{d+e}{2},  \qquad a_{2x} = \frac{d}{2},  \qquad
a_{3x} = \frac{c+e}{2}.
\eqa
The three contributions of \eqn{toz} will then be written in terms of the 
auxiliary function $V_{\ssL}(x,y) = V(x,y)\ln V(x,y)$. We obtain
\bqa
C^d_0 &=&  2\,G^2_2\,\intfxy{x}{y} V_{\ssL}(x,y), 
\eqa
\bqa
C^s_0 &=&  \frac{1}{2}\,\intfx{x} \Bigl\{ V_{\ssL}(x,x)   
   (a_x-a_y)\,\,\Bigl[ G_2 \Delta^{-1} 
    + \frac{3}{2}\, G_{11} \Delta_1^{-1}\Bigr] +
  V_{\ssL}(x,0)  a_y \,\Bigl[ G_2 \Delta^{-1} 
  + \frac{3}{2}\, G_{12} \Delta_2^{-1}\Bigr]   \nl
{}&-&
  V_{\ssL}(1,x)  \,\Bigl[ G_2 \lpar \Delta^{-1} a_x 
  + \frac{3}{2}\, _{13} \Delta_3^{-1}  
+ \frac{1}{2} G_2\rpar   
  + \frac{3}{2}\,   a_x G_{13} \Delta_3^{-1}\Bigr]\Bigr\}, \nl
{}&{}& \\
C^0_0 &=&  - \frac{1}{4}\, G_2 \Delta_3^{-1}\Bigl[ a_{3x}  (b + c + e )
       + \frac{1}{2}\, G_{13} (  \frac{4}{3} b + c + e )\Bigr] \nl
{}&+&
 \frac{1}{12}\,G_2^2 \Delta^{-1}   ( 3 a + b + \frac{3}{2} c + 4 d 
   + 2 e + 6 f )
+ \frac{1}{4}  a_x a_{1x} \Delta_1^{-1}  ( a + b + c + d + e )\nl
{}&-&
\frac{1}{4} a_x a_{3x} \Delta_3^{-1}   ( b + c + e )
       + \frac{1}{6}\, a_x G_{11} \Delta_1^{-1}   ( a + b + c + 
        \frac{3}{4} d + \frac{3}{4} e )\nl
{}&-&
 \frac{1}{8}\,  a_x G_{13} \Delta_3^{-1}  ( \frac{4}{3} b
      + c + e )
       - \frac{1}{4}\,  a_y a_{1x} \Delta_1^{-1}  ( a + b + c + d + e) \nl
{}&+&
 \frac{1}{4}\, a_y a_{2x} \Delta_2^{-1}   ( a +  d )
- \frac{1}{6}\,  a_y G_{11} \Delta_1^{-1}   (  a + b +
       c + \frac{3}{4} d + \frac{3}{4} e ) 
+ \frac{1}{6}  a_y G_{12} \Delta_2^{-1}   
         ( a + \frac{3}{4} d ) \nl
{}&+&
\frac{1}{4}\,V_{\ssL}(0,0) \,\Bigl[ a_x a_{1x} \Delta_1^{-1} 
-  a_y a_{1x} \Delta_1^{-1} +  a_y a_{2x} \Delta_2^{-1}\Bigr]  \nl   
{}&-&
\frac{1}{4}\, V_{\ssL}(1,0) \,\Bigl[ G_2 a_{3x} \Delta_3^{-1}
+ a_x a_{3x} \Delta_3^{-1} +4\,  a_y a_{2x} \Delta_2^{-1} 
+ a_y G_{12} \Delta_2^{-1}\Bigr] \nl
{}&+&
\frac{1}{4}\,V_{\ssL}(1,1)\,\Bigl[ G_2 a_{3x} \Delta_3^{-1}   
 + G_2 G_{13} \Delta_3^{-1} -  a_x a_{1x} \Delta_1^{-1} 
+  a_x a_{3x} \Delta_3^{-1} \nl
{}&-&
  a_x G_{11} \Delta_1^{-1} +  a_x G_{13} \Delta_3^{-1} 
  + a_y a_{1x} \Delta_1^{-1}
  + a_y G_{11} \Delta_1^{-1}\Bigr].
\eqa
It should be mentioned that in this approach the number of terms of the
integrand really does not matter. The only relevant request to be made is that
the algorithm is capable of producing a smooth integrand. Furthermore the
algorithm should be flexible enough to deal with all singularities, including
infrared ones.
\subsubsection{Infrared divergent $\scff{0}$-function\label{tir}}

As a first example that this method can handle infrared divergences (IR) in the
context of dimensional regularization we consider the case of the IR-divergent
$C_0$-function.
From \eqn{rol} one can prove that $B = 0$ when
\bq
a = b = f = m^2, \quad  \quad c = e = s - 2\,m^2, \quad d = -2\,m^2,
\label{c0ir}
\eq
i.e.\ for the case where an infrared divergency is developed. 
Here we introduced $s = -(p_1+p_2)^2$.
Since the zeros of $B$ correspond to leading Landau singularities
(see \sect{singu}) we recover the connection between infrared singularities
and the more general class of Landau ones: as shown in \eqn{c0ir}, an infrared
singularity for $C_0$ manifests itself independently of $s$.
As a consequence we have that the quadratic form $V$ satisfies the following 
equation:
\bq
\,\Bigl[1+P_x\,\frac{\partial}{\partial_x}+P_y\,\frac{\partial}{\partial_y}
\Bigr]\,V^{\mu+1}(x,y) = 0,  
\label{eqir}
\eq
where the solution for the polynomial $P$ is given by
\bq
P_{\{x;y\}} = \frac{\{1-x;-y\}}{2(\mu+1)}. 
\eq
We use \eqn{eqir} to write
\bq
\intfxy{x}{y}\,V^{-1-\varepsilon/e}(x,y) = - \frac{1}{\varepsilon}\,
\intfx{x}\,V^{-1-\varepsilon/e}(x,x).  
\eq
Since the IR pole at $n = 4$ will remain in the final answer we must keep
trace of all $n$-dependent factors,
\bqa
\cff{0}{-m^2}{-m^2}{-s}{m}{0}{m} &=&
-\,\tHs^{\varepsilon}\pi^{-\varepsilon/2}\,\egam{1+\frac{\varepsilon}{2}}
\frac{1}{\varepsilon}\,\intfx{x}\,V^{-1-\varepsilon/e}(x,x)
\nl
{}&=& -\,\frac{1}{2}\,\tHs^{\varepsilon}\,\Ddr\,
\intfx{x}\,V^{-1-\varepsilon/e}(x,x),
\label{ecir}
\eqa
where we have introduced the usual ultraviolet and infrared 
regulators~\cite{Bardin:1999ak}
\bq
\Ddr = \frac{2}{\dre} - \eilc - \lpi, \qquad
\Ddrh =\frac{2}{\epp} + \eilc + \lpi,
\eq 
and recall that these regulators satisfy the following relevant 
identity~\cite{Bardin:1999ak}:
\bq
\Ddr + \Ddrh = 0.
\eq
Therefore, from \eqn{ecir} we recover the familiar result for the 
infrared $\scff{0}$-function, see ref~\cite{Bardin:1999ak} and also
ref.~\cite{'tHooft:1979xw}. 
The power in $V(x,x)$ can be {\em raised} once more with the help of
the relation
\bq
\,\Bigl[ 1 + \frac{1}{2\,(\mu+1)}\,\lpar\frac{1}{2} - x\rpar\,\partial_x\Bigr]
V^{1+\mu}(x,x) = - \frac{1}{4}\,\lpar s - 4\,m^2\rpar\,V^{\mu}(x,x).
\eq
Collecting the various terms we obtain the final expression for
the $\scff{0}$-function,
\bqa
C^{\rm IR}_0 &=& \frac{1}{s - 4\,m^2}\,\Bigl\{ -\,\frac{1}{2}\,\Ddrh\,
\Bigl[ 1 + \frac{1}{2}\,\intfx{x}\,\ln\frac{V(x,x)}{m^2}\Bigr]  
- \frac{1}{2}  \nl
{}&+& \frac{1}{4}\,\ln\frac{m^2}{\tHss}\,\lpar 1 + 
\frac{1}{2}\,\ln\frac{m^2}{\tHss}\rpar -
\frac{1}{4}\,\intfx{x}\,\ln\frac{V(x,x)}{\tHss}\,\Bigl [3 + \frac{1}{2}\,
\ln\frac{V(x,x)}{\tHss}\Bigr]
\Bigr\}.
\eqa
In conclusion the BT-algorithm automatically extracts the poles in $n - 4$,
both ultraviolet and infrared.
\subsubsection{Other one-loop functions\label{oolf}}
The procedure outlined above can be applied to any one-loop scalar function.
Here we only show how it works for the massless pentagon
(the massive case is as easy as the massless one). One can prove that
\bqa
E_0 &=& 
\int_{\scriptstyle 0}^{\scriptstyle 1}\,dx_1\,
                        \int_{\scriptstyle 0}^{\scriptstyle 1-x_1}\,dx_2\,
                        \int_{\scriptstyle 0}^{\scriptstyle 1-x_1-x_2}\,dx_3\,
                        \int_{\scriptstyle 0}^{\scriptstyle 1-x_1-x_2-x_3}\,
                         dx_4  
\, \,\Bigl[x^t\,H^{-1}\,x + 2\,K^t\,x\Bigr]^{-3-\varepsilon/2},  \nl
\label{penta}
\eqa
where the $4\,\times\,4$ matrix $H$ is defined in terms of Mandelstam 
invariants, $s_{ij} = - \,\lpar p_i+p_j\rpar^2$,
by the following form:
\[ H = \frac{1}{2}\,\left(\begin{array}{cccc}
0 & -s_{51} & s_{12}-s_{34} & s_{45} \\
- & -2\,s_{51} & -s_{34}-s_{51} & s_{23}-s_{51} \\
- & - & -2\,s_{34} & - s_{34} \\
- & - & - & 0
\end{array}\right)\]
and the vector $K$ becomes
\bq
K_1 = 0, \quad K_2 = \frac{1}{2}\,s_{51}, 
\quad K_3 = \frac{1}{2}\,s_{34}, \quad K_4 = 0,  
\eq
Starting from \eqn{penta} we {\em raise} the power from $-3-\varepsilon/2$
to $1-\varepsilon/2$ and, successively, integrate by parts. Note that,
for the massless pentagon the coefficient $B$ becomes,
\bq
B = \frac{1}{16}\, s_{12} s_{23} s_{34} s_{45} s_{51}.
\eq
For the fully massive pentagon we derive, after four iterations of the
BT-algorithm (from power $-3-\varepsilon/2$ to power $1-\varepsilon/2$),
an expression which, remarkably enough, has no $4$-dimensional integrals.
In \tabn{ikount} we give the number of terms for the scalar box, pentagon
and hexagon functions, where in all cases we have {\em raised} powers up to
$1-\varepsilon/2$ an then we have expanded around $\varepsilon = 0$ up to
terms of $\ord{1}$.
\begin{table}[ht]\centering
\begin{tabular}{|l|r|r|r|r|r|r|}
\hline
 & & & & & & \\
diagram & $5$-dim & $4$-dim & $3$-dim & $2$-dim & $1$-dim & $0$-dim \\
 & & & & & & \\
\hline
 & & & & & & \\
$D_0\,\rm{Box}$ & - & -& $2$ & $54$ & $217$ & $98$ \\
 & & & & & & \\
$E_0\,\rm Pentagon$ & - & - & $9$ & $189$ & $945$ & $12826$ \\
 & & & & & & \\
$F_0\,\rm Hexagon$ & $1$ & $44$ & $693$ & $4581$ & $13860$ & $228846$ \\
 & & & & & & \\
\hline
\hline
\end{tabular}
\vspace*{3mm}
\caption[]{Number of terms in scalar box, pentagon and hexagon functions where
the BT-algorithm has been applied repeatedly up to powers $1 -
\varepsilon/2$.\label{ikount}}
\end{table}
\normalsize
Once more, the number of terms is not the relevant issue, only the smoothness 
of the integrand matters. No attempt has been performed, so far, towards
a systematic study of the singularities (zeros of $B$) of multi-leg,
$n \ge 5$, one-loop diagrams.
\subsubsection{Other one-loop infrared-divergent functions\label{oolidf}}
One of the nice features of the method is that the coefficient $B$ in
\eqn{functr} contains all divergences of the diagram, including the infrared
pole, as discussed for the triangle one-loop diagram in \sect{tir}.
For other one-loop diagrams there is no need to have separate derivations for
the infrared cases. All of them are reducible to $\scff{0}$-functions. Typical
example is the box diagram $\sdff{0}$, where we can use the following 
decomposition~\cite{Passarino:1979jh} (see also ref.~\cite{Bardin:1999ak}):
\bqa
\label{6.4.5a_7}
&&\dffp{0}{-\mes}{-\mes}{-\mfs}{-\mfs}{\Trmoms}{\Prmoms}\dffm{0}{\me}{0}{\mf}
=\frac{1}{\Trmoms}\Bigl[ -\Jds{\ph\ph}{\Trmoms}{\Prmoms}{\me}{\mf}
\nl &&\hspace{1cm}
  +\cff{0}{-\mes}{-\mfs}{\Prmoms}{\me}{0}{\mf} 
  +\cff{0}{-\mfs}{-\mes}{\Prmoms}{\mf}{0}{\me}\Bigr],
\nl
\label{6.4.5a_8}
&&\dffp{0}{-\mes}{-\mes}{-\mfs}{-\mfs}{\Trmoms}{\Prmoms}\dffm{0}{\me}{\mz}{\mf}
=\frac{1}{\Trmoms+\mzs}\Bigl[ -\Jds{\ph\ssZ}{\Trmoms}{\Prmoms}{\me}{\mf}
\nl &&\hspace{1cm}
  -\cff{0}{-\mes}{-\mfs}{\Prmoms}{\me}{\mz}{\mf}
  +\cff{0}{-\mfs}{-\mes}{\Prmoms}{\mf}{0}{\me}\Bigr],
\nl
\label{6.4.5a_9}
&&\dffp{0}{-\mes}{-\mes}{-\mfs}{-\mfs}{\Trmoms}{\Prmoms}\dffm{\mz}{\me}{0}{\mf}
=\frac{1}{\Trmoms+\mzs}\Bigl[\Jds{\ssZ\ph}{\Trmoms}{\Prmoms}{\me}{\mf}
\nl &&\hspace{1cm}
  +\cff{0}{-\mes}{-\mfs}{\Prmoms}{\me}{0}{\mf}
  -\cff{0}{-\mfs}{-\mes}{\Prmoms}{\mf}{\mz}{\me}\Bigr].
\nn
\eqa
All infrared divergent terms are contained in the $\scff{0}$-functions and
the additional integrals are as follows:
\bqa
\label{6.4.5a_6}
\ib\pi^2\Jds{\ph\ph}{\Trmoms}{\Prmoms}{\me}{\mf}&=&
\tHs^{4-n}\,\intmomi{n}{\imom}
\frac{2\imom\cdot\lpar\imom+\Trmom\rpar}
{\dpropii{0}{0}\dpropii{1}{\me}\dpropii{2}{0}\dpropii{3}{\mf}}\hsp,
\\
\ib\pi^2\Jds{\ph\ssZ}{\Trmoms}{\Prmoms}{\me}{\mf}&=&
\tHs^{4-n}\,\intmomi{n}{\imom}
\frac{2\imom\cdot\Trmom}
{\dpropii{0}{0}\dpropii{1}{\me}\dpropii{2}{\mz}\dpropii{3}{\mf}}\hsp,
\nl
\ib\pi^2\Jds{\ssZ\ph}{\Trmoms}{\Prmoms}{\me}{\mf}&=&
\tHs^{4-n}\,\intmomi{n}{\imom}
\frac{2\Trmom\cdot\lpar\imom+\Trmom\rpar}
{\dpropii{0}{\mz}\dpropii{1}{\me}\dpropii{2}{0}\dpropii{3}{\mf}}\hsp,
\nn
\eqa
The propagators are defined by
\bqa
\ba{ll}
\dpropi{0} = \imoms+\mones-\ib\delta, 
& \qquad
\dpropi{1} = \lpar\imom+\pone\rpar^2+\mtwos-\ib\delta,   
\\
\dpropi{2} = \lpar\imom+\pone+\ptwo\rpar^2+\mtres-\ib\delta,
& \qquad
\dpropi{3} = \lpar\imom+\pone+\ptwo+\ptre\rpar^2+\mfors-\ib\delta,\qquad
\ea
\eqa
with all four-momenta flowing inwards and we have explicitly written the 
internal masses.
\subsubsection{Reduction\label{redu}}
\begin{figure}[h]
\vspace*{-5mm}
\[
  \vcenter{\hbox{
  \begin{picture}(130,80)(0,0)
  \ArrowLine(0,50)(40,50)
  \PhotonArc(65,50)(25,0,180){2}{7}
  \ArrowArc(65,50)(25,-180,0)
  \ArrowLine(90,50)(130,50)
  \end{picture}}}
\]
\vspace*{-10mm}
\caption[]{Fermionic self-energy in QED.\label{fig:fermionse}}
\end{figure}
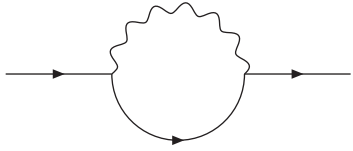

The standard procedure in dealing with one-loop diagrams is based on the 
introduction of scalar functions, $\sbff{0}, \scff{0}$ etc, and of tensor 
integrals that are reduced to linear combinations of scalar 
ones~\cite{Passarino:1979jh}.
The functional relation, \eqn{functr}, can be written in a more general
form starting from
\bq
\int_{\ssS}\,dx\,Q(x)\,\prod_{i}\,V^{\mu_i}_i(x),
\eq
where $Q(x)$ is an arbitrary polynomial of $x$. Hence we are led to consider
also the tensorial integrals in parametric space with no pre-reduction in
momentum space and no appearance of standard Gram's determinants. Consider 
some  simple example: we begin with the one-loop electron self-energy in QED,
see \fig{fig:fermionse},
\vskip 5pt
\bqa
\Sigma &=& -\,ie^2\,\pi^{2-\ep/2}\,\mu^{\ep}\,\egam{\frac{\ep}{2}}\,
\intfx{x}\,\,\Bigl[ i\,\lpar 2 - \ep\rpar\,x\sla{p} + \,\lpar 4 - \ep\rpar\,
\me\Bigr]\,\chi^{-\ep/2},  \nl
\chi(x) &=& \,\lpar 1 - x\rpar\,\lpar x\,p^2 + \mes\rpar.
\label{defchi}
\eqa
After integration by parts we obtain
\bqa
\Sigma &=& -\,i\pi^2e^2\,\,\Bigl[ \Sigma_p\,i\sla{p} + \Sigma_m\,\me\Bigr],
\nl
\Sigma_p &=& \Ddr + \frac{p^2}{\,\lpar p^2+\mes\rpar^2}\,\,\Bigl[ 
\frac{\gamma}{2}\,\frac{\,\lpar p^2+\mes\rpar^2}{p^2} - \frac{1}{3}\,\mes -
\frac{1}{6}\,p^2  \nl
{}&+& \intfx{x}\,\lpar \frac{1}{2} - \frac{1}{2}\,
\frac{\mes}{p^2}- 4\,x\rpar\,\chi\,\ln\frac{\chi}{\tHss}\Bigr],
\nl
\Sigma_m &=& {\ds\frac{4}{{\bar{\varepsilon}}}} + 
\frac{p^2}{\,\lpar p^2+\mes\rpar^2}\,\,\Bigl[ 
\,\lpar 2\,\gamma - \ln\frac{\mes}{\tHss} + \frac{1}{2}\rpar\,
\frac{\,\lpar p^2+\mes\rpar^2}{p^2} + 
 p^2\,\ln\frac{\mes}{\tHss} - \frac{2}{3}\,p^2  \nl
{}&+& \mes\,\lpar 3\,
\ln\frac{\mes}{\tHss} -2\rpar - 6\,\intfx{x}\,\chi\,\ln\frac{\chi}{\tHss}
\Bigr].
\eqa
Note that the general idea is now different from the one which is at the basis 
of the analytical approach. The number of terms in the final expression does 
not matter and the only relevant requisite is the smoothness of the integrand 
accompanied by the absence of high negative powers of true Gram's determinants 
in the reduction procedure. However, in this case a new and unpleasant
feature shows up, namely the evaluation of the self-energy is numerical 
unstable at threshold, $p^2 \approx - \mes$.

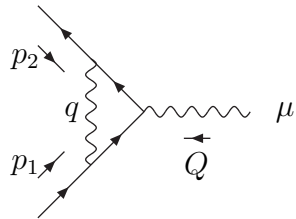
\begin{figure}[h]
\vspace*{-5mm}
\[
  \vcenter{\hbox{
  \begin{picture}(80,80)(0,0)
  \ArrowLine(40,40)(20,60)
  \ArrowLine(20,60)(0,80)
  \ArrowLine(0,0)(20,20)
  \ArrowLine(20,20)(40,40)
  \Photon(40,40)(80,40){2}{5}
  \Photon(20,60)(20,20){2}{5}
  \Text(90,40)[lc]{$\mu$}
  \ArrowLine(0,65)(10,55)
  \ArrowLine(0,15)(10,25)
  \ArrowLine(65,30)(55,30)
  \Text(-10,60)[cl]{$\ptwo$}
  \Text(-10,20)[cl]{$\pone$}
  \Text(10,40)[lc]{$\imom$}
  \Text(60,25)[tc]{$\Trmom$}
  \end{picture}}}
\]
\caption[]{QED vertex diagram.\label{fig:QEDvertex}}
\end{figure}

Another example is given by the one-loop vertex of \fig{fig:QEDvertex}
which is reducible to the following expression:
\bq
\Lambdi{\mu}= -\tpfi\,\frac{\ib\ect}{16\pi^2}
\,\Bigl[\gadu{\mu}V_{1}\,\lpar\Trmoms;\mlone,\mlone\rpar  
+\sigma_{\mu\nu}\,\lpar\pone + \ptwo \rpar_{\nu}
V_{2}\,\lpar\Trmoms;\mlone,\mlone\rpar\Bigr].
\label{QED_vertex}
\eq 
For $V_1$ we obtain
\bqa
V_1\,\lpar Q^2;m,m\rpar &=& \pi^2e^3\intfxy{x}{y}\,\Bigl\{\,\Bigl[ -2\,\mes\,
V^m_1 -2\,Q^2\,V^q_1\Bigr]\,\chi^{-1} - V^{\rm sing}_1\Bigr\},  \nl
V^m_1 &=& x^2 + 3\,x -3,  \quad
V^q_1 = y^2 -xy + x -1,  \nl
V^{\rm sing}_1 &=& \frac{1}{2}\,\lpar \ep -2\rpar^2\,\lpar \Ddr - 
\ln\frac{\chi}{\tHss}\rpar,  \nl
\chi(x,y) &=& x^2\mes + y\,\lpar x - y\rpar\,Q^2.
\eqa
Note that we have already expanded in $\ep$ because {\em raising} of powers
can actually be done even in four dimensions. This new result can be derived 
by the following example: for $\mu = -1+\ep$ we write
\bq
\,\Bigl[1 - {{\,\lpar x+X\rpar\,\partial_x}\over {2\,\lpar\mu+1\rpar}}\Bigr]\,
V^{\ep} = B\,V^{-1+\ep},
\eq
and expand in $\ep$ before integration by parts. By equating the coefficients
of $\ep^n$ we derive, for instance, that
\bq
B\,V^{-1} = 1 - \frac{1}{2}\,\lpar x + X\rpar\,\partial_x\,\ln\,V,
\qquad \mbox{etc},
\eq
showing that we can expand first and then integrate by parts. A similar result
can be derived with $\mu= \ep$, giving
\bq
B\,\ln\,V = V\,\ln\,V + \frac{1}{2}\,\lpar x + X\rpar\,\partial_x\,\,\Bigl[
V\,\lpar 1 - \ln\,V\rpar\Bigr].
\eq
\vskip 5pt
\subsubsection{Zeros of $B$\label{singu}}
The relation between the zeros of $B$ and the Landau singularities for an
arbitrary diagram $G$ remains to be examined, although $B = 0$ should 
contain all singularities of $G$. In this section we consider the one-loop 
three-point function and establish the result that the coefficient $B$
emerging after the first application of \eqn{functr} is connected with
the leading Landau singularity, the so-called anomalous threshold. Moreover, 
a second iteration produces several coefficients that are connected with 
sub-leading singularities, i.e. \ those of the contracted diagrams. This 
result holds also for higher one-loop functions, although the determination 
of the explicit form of the anomalous thresholds is, in general, an arduous 
task.

Note that $B$ is not a Gram determinant but its zeros are a vexation, as much 
as those of Gram determinants in the standard reduction 
procedure~\cite{Passarino:1979jh}.
As already observed in ref.~\cite{Tkachov:1997wh} the numerical advantage of
smoother integrands is lost near thresholds where numerical convergence may 
become poor. This fact is best illustrated with a simple example, the scalar
two point function,
\bq
\bff{0}{-s}{m}{m} = \Ddr - \ln\frac{m^2}{\tHss} + 2 - \beta\,
\ln\frac{\beta-1}{\beta+1},  \quad
\beta^2 = 1 - 4\,\frac{m^2}{s}.
\eq
The explicit result shows the well-known normal threshold at $s = 4\,m^2$.
If we apply \eqn{functr} the following result is obtained:
\bqa
\bff{0}{-s}{m}{m} &=& \pi^{-\ep/2}\,\tHs^{\ep}\,\egam{\frac{\ep}{2}}\,
\intfx{x}\,\chi^{-\ep/2}\,
= \frac{4}{\beta^2}\,\pi^{-\ep/2}\,\tHs^{\ep}\,
\frac{\egam{\frac{\ep}{2}}}{\ep-2}\,\Bigl[ (3-\ep)  \nl
{}&\times& \intfx{x}\,\chi^{1-\ep/2} - (m^2)^{1-\ep/2}\Bigr],
\eqa
with $\chi = s x^2 - s x + m^2 - i\,\delta$. Of course there is no pole
at $s = 4\,m^2$ but numerical convergence becomes questionable.
An interesting question is related to the singularity of $\sbff{0}$. Can we 
find it without having to perform the integration? The integrand is
$\ln(x^2 - x + \mu^2)$ with $\mu^2 = m^2/s$ and shows two branch points,
$x_{\pm}$, that are complex conjugated below threshold and pinch the contour 
at $x = 1/2$ when we are at threshold. We rewrite
\bq
\intfx{x}\,\ln\chi = - 2 + \ln\mu^2 + \frac{1}{2}\,\lpar 4\,\mu^2 - 1\rpar\,
\intfx{x}\,\chi^{-1}.
\eq
The integrand will the show poles at $x_{\pm}$ and we are more familiar with 
the fact that two poles, pinching the integration contour, produce
a singularity for the function~\cite{elop}.
In ref.~\cite{Tkachov:1997wh} an alternative method is suggested to deal with
regions near thresholds (see also ref.~\cite{Tkachov:1993ev}), while our 
solution to the problem will be discussed in \sect{icp}. 
Another example is represented by the triangle diagram. For the function 
\bq
\cff{0}{-m_a^2}{-m_b^2}{-s}{m_b}{M}{m_a}
\eq
we obtain the explicit form of the $\Delta$ factors of \eqn{deltas}:
\bq
\Delta_1 = -\frac{1}{4}\,\lambda\,\lpar s,m_a^2,m_b^2\rpar, \quad
\Delta_2 = -\frac{1}{4}\,\lambda\,\lpar M^2,m_a^2,m_b^2\rpar, \quad
\Delta_3 = -\frac{1}{4}\,\lambda\,\lpar M^2,m_a^2,m_b^2\rpar,
\eq
that correspond to the three possible cuts of the triangle diagram, 
$\lambda$ being the K\"allen function,
\bq
\lambda(x,y,z) = x^2 + y^2 +z^2 - 2\,\lpar xy + xz + yz\rpar.
\eq
The zeros of $\Delta_i$ correspond to non-leading Landau singularities,
a) normal thresholds at $s = (m_a+m_b)^2$ etc. and b) pseudo-thresholds at
$s = (m_a-m_b)^2$ etc. The latter do not show up in the physical Riemann sheet.
Furthermore, we also derive
\bq
\Delta =  \,\Bigl[ \,\lpar m_a^2-m_b^2\rpar^2 - M^2 s \Bigr] 
\, \,\Bigl[ s + M^2  - 2\,\lpar m_a^2 + m_b^2\rpar\Bigr].
\eq
For the special case $m_a = m_b = m$ we get
\bq
\Delta = - M^2 s\,\lpar s + M^2 - 4\,m^2\rpar,
\eq
and $s = 4\,m^2 - M^2$ corresponds to the leading Landau singularity, the 
anomalous threshold. Note that the general form of the leading Landau
singularity is
\bq
s = \frac{1}{2\,m^2_2}\,\Bigl\{ 2\,m^2_2\,(m^2_1+m^2_3) -
(m^2_1+m^2_2-M^2_1)\,(m^2_3+m^2_2-M^2_2) \pm
\Bigl[ \lambda(M^2_1,m^2_1,m^2_2)\lambda(M^2_2,m^2_3,m^2_2)\Bigr]^{1/2}
\Bigl\},
\eq
where we have introduced $p^2_i = - M^2_i$.

For all the zeros of $\Delta$ and $\Delta_i$ numerical convergence is at stake.
Here the method suffers the same disease of the usual appearance of 
Gram's determinants in the standard evaluation of $\scff{0}$ and in the 
reduction of vector and tensor triangles, although the origin of the
phenomenon is rather different. 
Finally, consider the simple case where $m_a = m_b = 0$. The exact expression 
is
\bq
\cff{0}{0}{0}{-s}{0}{M}{0} = -\,\frac{1}{s}\,\Bigl[ \zeta(2) -
\li{2}{1+\frac{s}{M^2}}\Bigr],
\label{c0ex}
\eq
where $\zeta$ is the Riemann zeta-function and $\li{2}{z}$ is the standard 
di-logarithm.
The zero of $G_{11}$ is at $s=0$, while $G_2$ has two zeros, respectively
at $s = 0$ and $s = - M^2$. Note that there is no pole at $s = 0$ in
\eqn{c0ex}, that the function is regular at $s = - M^2$ and that the only
singularity is the branch point of the di-logarithm.

Actually these problems are easily solved in the standard approach.
Consider the case of an arbitrary triangular graph for vanishing 
Gram's determinant:
\bq
\cff{0}{p_1^2}{p^2_2}{\gmv}{m_1}{m_2}{m_3} = \scff{0} = 
\asum{n}{0}{\infty}\,\cff{0n}{p_1^2}{p^2_2}{\gmv}{m_1}{m_2}{m_3}\,
\delta_3^n. 
\eq
The kinematics for vanishing Gram's determinant, $\Delta_3 \to 0$, is as 
follows:
\bqa
\Delta_3 &=& p_1^2p_2^2 - \lpar\spro{p_1}{p_2}\rpar^2 =
\delta_3\,P^4, \nl
p_1^2 &=& P^2, \qquad
p_2^2 = \lpar \alpha^2_0+\delta_3\rpar\,P^2, \qquad
\spro{p_1}{p_2} = - \alpha_0\,P^2, \nl
\gmv &=& -\lpar p_1+p_2\rpar^2 = - \lrbr\lpar 1-\alpha_0\rpar^2+\delta_3\rrbr
\, P^2.
\label{gramv}
\eqa
In lowest order the solution is to introduce three different
pinches of the basic three-point integral which correspond to the three 
ways of cutting the triangular graph:
\bq
\pinchc{1}{00} = B_0\lpar 23\rpar|_{\delta_3=0}, \quad
\pinchc{2}{00} = B_0\lpar 13\rpar|_{\delta_3=0}, \quad
\pinchc{3}{00} = B_0\lpar 12\rpar|_{\delta_3=0}
\eq
Then a solution, free of singularities, is
\bqa
\scff{00} &=& \frac{1}{N_0}\,\lrbr - \pinchc{3}{00} +
\lpar 1-\alpha_0 \rpar\,\pinchc{2}{00} + \alpha_0\,\pinchc{1}{00}\rrbr, \nl
N_0 &=& m_2^2-m_3^2+\lpar m_1^2-m_2^2\rpar \alpha_0 + P^2\alpha_0
\lpar 1-\alpha_0\rpar,
\eqa
where $\alpha_0$ is given in \eqn{gramv}. At next order we have $12$ terms:
\bqa
\scff{01} &=& {}  \frac{2}{3}\,N_0^{-2} 
+ \frac{1}{3}\,\Lambda N_0^{-3} \scff{00} + N_0^{-2} \scff{00} 
- \frac{2}{3} \frac{\mu_1^2}{1-\alpha_0} N_0^{-2} \aff{0}{1} \nl
{}&-& \frac{2}{3} \frac{\mu_2^2}{\alpha_0}   N_0^{-2} \aff{0}{2} 
+ \frac{2}{3} \beta_1\mu_3^2 N_0^{-2}  \aff{0}{3} 
- \frac{1}{3} b_{12} N_0^{-2} \pinchc{2}{00} \nl
{}&-& \frac{2}{3} \frac{1}{1-\alpha_0}  N_0^{-1} \pinchc{2}{00} 
+ \frac{1}{3} a_{12} N_0^{-2} \pinchc{1}{00} 
- \frac{2}{3} \frac{1}{\alpha_0}  N_0^{-1} \pinchc{2}{00}  \nl 
{}&+& \lpar 1-\alpha_0\rpar N_0^{-1} \pinchc{2}{01} 
+ \alpha_0 N_0^{-1} \pinchc{1}{01}.
\eqa
where we have introduced the following additional quantities:
\bqa
m_i^2 &=& - \mu_i^2\,P^2, \qquad
b_{12} = 1+\mu_1^2-\mu_2^2, \qquad
a^2_{12} = 4\,\mu_2^2 + \Lambda^2, \nl
\Lambda^2 &=& b_{12}^2 - 4\,\mu_1^2, \qquad
\beta_1 = \frac{1}{\alpha_0} + \frac{1}{1-\alpha_0}.
\eqa
At second order we have an expression with $48$ terms while $135$ terms
are needed at third order. For a general treatment of the reduction we
refer to the work in ref.~\cite{Devaraj:1998es}.
\subsection{Multi-loop diagrams\label{mld}}
\begin{figure}[th]
\vspace{0.5cm}
\[
  \vcenter{\hbox{
  \begin{picture}(150,0)(0,0)
  \Line(0,0)(50,0)
  \ArrowArc(75,0)(25,0,90)
  \ArrowArc(75,0)(25,-180,0)
  \ArrowArc(75,0)(25,90,180)
  \ArrowArc(50,25)(25,-90,0)
  \Line(100,0)(150,0)
  \Text(50,25)[cb]{$\scriptstyle r_1$}
  \Text(100,25)[cb]{$\scriptstyle r_2$}
  \Text(75,-35)[cb]{$\scriptstyle r_2+p$}
  \Text(80,-5)[cb]{$\scriptstyle r_1-r_2$}
  \end{picture}}}
\]
\vspace{0.5cm}
\caption[]{The two-loop diagram $S_4$. Arrows indicate the momentum flow.} 
\label{fig:s4}
\end{figure}
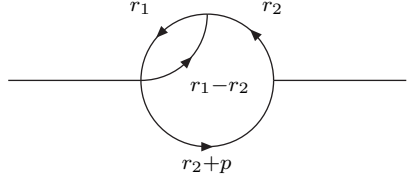

In moving to multi-loop diagrams we recall that any diagram $G$ with 
$N_L$ legs and $n_l$ loops is representable as~\cite{Cvitanovic:1974uf}
\bq
G = \lpar i\,\pi^{n/2}\rpar^{n_l}\,\Gamma\,\lpar N_L - \frac{n}{2}\,n_l\rpar\,
\int\,{{dx_{\ssG}\,\delta\,\lpar 1-x_{\ssG}\rpar}\over
{U^{n/2}\,\lpar V-i\,\delta\rpar^{N_L-nn_l/2}}},  
\label{defG}
\eq
where $\Gamma$ is the Euler gamma-function and where the integration measure 
can be written as
\bq
dx_{\ssG} = \prod_{i=1}^{N_l}\,dx_i, \qquad x_{\ssG} = \sum_{i=1}^{N_l}\,
x_i.
\eq
Furthermore, the polynomials $V$ and $U$ are defined by
\bqa
V &=& \asums{i} m^2_i\,x_i + \asums{i}\,q_i^2\,x_i - \frac{1}{U}\,
\asums{ij}\,B_{ij}\,\spro{q_i}{q_j}\,x_ix_j,  \nl
U &=& \asums{T}\prod_{x_i \in T}\,x_i = \mbox{det}\,\lpar U_{rs}\rpar, 
\quad U_{rs} = \asums{i} x_i\eta_{ir}\eta_{is},
\eqa
where $\eta_{is}$ is the projection of line $i$ along the loop $s$. 
Furthermore, $T$ is a co-tree and $B_{ij}$ are the parametric functions for
the given diagram. Although these functions can be determined completely by
the topological structure of the diagram $G$ we give a practical example
of how to construct $U$ and $V$ for the two-loop diagram $S_4$ of
\fig{fig:s4}
\vskip 5pt
After introducing Feynman parameters the integrand for $S_4$ contains a factor
$1/D^4$ with
\bq
D = \sum_{i=1}^{4}\,x_i\,\lpar q^2_i + m^2_i\rpar,  \qquad
q_1 = r_1, \quad q_2 = r_1 - r_2, \quad q_3 = r_2, \quad q_4 = r_2 + p,
\eq
where $r_s$ is the independent integration momentum around the loop $s$, $x_i$
are Feynman parameters with $\asums{i}\,x_i = 1$. The part of $D$ which is 
quadratic in $r_{1,2}$ will be written as
\bq
r^t\,U\,r, \qquad U_{11} = x_1+x_2, \quad U_{22} = x_2+x_3+x_4, \quad
U_{12} = U_{21} = - x_2.
\eq
Next we rewrite $U_{ij}$ as a sum,
\bq
U_{ij} = \sum_{l=1}^{4}\,\eta_{li}\eta_{lj}\,x_l,
\eq
and derive the coefficients $\eta$ as
\bq
\eta_{11} = \eta_{21} = \eta_{32} = \eta_{42} = +1,  \quad
\eta_{22} = -1, \eta_{31} = \eta_{41} = \eta_{12} = 0.
\eq
Furthermore, let $U$ be the determinant of the matrix $U_{ij}$, thus
\bq
U = {\rm det}\,\lpar U_{ij}\rpar = x_1\,x_{234} + x_2\,x_{34},
\eq
where $x_{ij\dots l} = x_i+x_j+\dots+x_l$. Momenta $p_i$ will then be defined
with $p_4 = p$ and $p_i = 0$ for $i < 4$. The following change of variables in 
the $r_1$-integral is then performed:
\bq
r^{\mu}_1 \to r^{\mu}_1 - \sum_{j=1}^{4}\,\sum_{t=1}^{2}\,x_j p^{\mu}_4
\eta_{it}\,\lpar U^{-1}\rpar_{1t}  
= r^{\mu}_1 - \sum_{t=1}^{2}\,x_4 p^{\mu} \eta_{4t}\,\lpar U^{-1}
\rpar_{1t}  = r^{\mu}_1 - x_4\,\frac{x_2}{U}\,p^{\mu}.
\eq
Similarly we change variable also in the $r_2$-integral,
\bq
r^{\mu}_2 \to \sum_{j=1}^{4}\,\sum_{r=1}^{2}\, x_j p^{\mu}_j \eta_{jr}
\,\lpar U^{-1}\rpar_{2r} 
= r^{\mu}_2 - x_4\,\frac{x_{12}}{U}\,p^{\mu}.
\eq
We derive the following result:
\bq
\sum_{i=1}^{4}\,x_i\,\lpar q^2_i + m^2_i\rpar \to r^t\,U\,r + V,
\eq
which defines the polynomial $V$ as
\bq
V = \sum_{i=1}^{4}\,x_i\,m^2_i + x_4\,p^2 - \frac{1}{U}\,x_{12}\,x^2_4\,p^2.
\eq
After a diagonalization of the symmetric matrix $U$,
\bq
\sum_{i'j'} (A^{-1})_{ii'}\,U_{i'j'}\,A_{j',j} = U_i\,\delta_{ij},
\eq
we perform a change of variables with unit Jacobian, $s_i = \sum_j A_{ij}r_j$,
and use
\bqa
{}&{}&
\int\,\prod_{i=1}^{n_l}\,ds_i\,\Bigl[ \sum_{i=1}^{n_l}\,U_i s^2_i + V 
\Bigr]^{-N_{\ssL}} = 
\int\,\prod_{i=2}^{n_l}\,ds_ids_1\,\Bigl[ U_1 s^2_1 + \sum_{i=2}^{n_l}\,
U_i s^2_i + V\Bigr]^{-N_{\ssL}}  \nl
{}&=& i\,\pi^{n/2}\,U^{-n/2}_1\,\frac{\egam{N_{\ssL}-n/2}}{\egam{N_{\ssL}}}\,
\int\,\prod_{i=2}^{n_l}\,ds_i\,\Bigl[ \sum_{i=2}^{n_l}\,U_i s^2_i + 
V\Bigr]^{n/2-N_{\ssL}} = \mbox{etc.},
\eqa
to obtain the result of \eqn{defG}.
Similarly for the diagram $S_3$ of \fig{fig:s3} we derive
\bq
U = x_1\,x_{23} + x_2\,x_3, \qquad V = \sum_{i=1}^{4}\,x_i\,m^2_i + 
\frac{1}{U}\,x_1x_2x_3\,p^2.
\eq
Note that UV-singularities come from $U$ so that, for finite diagrams, one 
should {\em raise} only the factor
\bq
{\tilde V} = U\,\,\Bigl[\asums{i} m^2_i\,x_i + \asums{i}\,q_i^2\,x_i\Bigr] - 
\asums{ij}\,B_{ij}\,\spro{q_i}{q_j}\,x_ix_j.
\eq
There are several comments to be made before we can start trying to apply the 
BT-theorem to any multi-loop diagram.
For two-loop diagrams $U$ is a quadratic form in the Feynman parameters $x$
and ${\tilde V}$ is a cubic, so that we have to construct a BT functional 
relation for a quintic or higher polynomial.
For $N_L = 3,4$, we have that the corresponding diagram is defined to consist of ${\tilde V}$ to a positive power and, in principle, it is enough to 
regularize $G$ with some $K_S$ operation ($S\in G$)~\cite{Cvitanovic:1974uf}.
For $N_L > 5$, we have that the diagrams contains ${\tilde V}$ to a negative 
power, $U$ to a positive power. Therefore, there is no UV-divergency and
we could {\em raise} the power for the cubic.

For self-energies at zero external momentum, $p^2 = 0$, the procedure is
quite straightforward. Consider, for example, the case $n_l= 2,N_L= 3$. 
First of all, we perform a projective transformation~\cite{'tHooft:1972fi}
\bq
x_i = A_i\,u_i/\asums{j}A_j\,u_j, \qquad A_i = \frac{1}{m^2_i},
\eq
to obtain ${\tilde V} = U$, where $U$ is a quadratic form. 
The important point being that we know how to {\em raise} powers for any 
quadratic form.

\begin{figure}[th]
\vskip 5pt
\[
  \vcenter{\hbox{
  \begin{picture}(150,90)(0,0)
  \CArc(75,50)(25,-180,180)
  \Line(50,50)(100,50)
  \Text(75,80)[cb]{$\scriptstyle m_1$}
  \Text(75,55)[cb]{$\scriptstyle m_2$}
  \Text(75,30)[cb]{$\scriptstyle m_3$}
  \end{picture}}}
\]
\vspace{-0.5cm}
\caption[]{The two-loop vacuum diagram with arbitrary masses.} 
\label{fig:vac}
\end{figure}
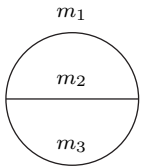

Finally, $N_L = 2\,(n_l+1)$ is a nice example of a single polynomial of degree
$n_l+1$. Of course, vacuum diagrams are also easy. At two-loop level
we have only the diagram of \fig{fig:vac} which becomes
\vskip 5pt
\bqa
G^{\rm vac}_2 &=& \,\lpar 4\,\pi\rpar^{\varepsilon-4}\,\mu^{\ep}\,\Gamma\,
\lpar \varepsilon-1\rpar\,\lpar K^tH^{-1}K\rpar^{-2}\,\intfxy{x}{y}  \nl
{}&\times& \,\Bigl[ 1 - {{\,\lpar x+K^tH^{-1}\rpar\partial}\over 
{\varepsilon-2}}\Bigr]\,
\,\Bigl[ 1 - {{\,\lpar x+K^tH^{-1}\rpar\partial}\over 
{\varepsilon}}\Bigr]\,U^{\varepsilon/2}(x,y),
\eqa
where $U$ can easily be derived according to the general rules presented above.
The same is obviously true for tadpoles. All of this, however, still does 
not represent a real multi-loop application of the BT-theorem. 

Additional problems are represented by the fact that ${\tilde V}$ is an highly 
incomplete polynomial with few non-zero coefficients. ${\cal P}$ can be found 
via a direct study of linear systems of $n$ equations (usually, $n \gg 100$) 
in $m$ variables and very often the matrix of coefficients has rank $< n$, so 
the system is generally impossible and we need to go to higher values of
$n,m$, i.e. \ beyond $n_{\rm min} = \min \{n<m\}$. 
A simple counting shows the following situation: given a compact 
representation for ${\cal P}$,
\bq
{\cal P} = P_n + P^i_{n+1}\partial_i + P_{n+2}^{ij}\partial_i\partial_j +
\dots
\eq
four variables and a cubic $V$ require $n_{\rm eq}= 126(330,715)$ and 
$n_{\rm var}= 155(415,871)$ for $n= 2(1,0)$ and first(second, third) 
derivatives. Simple examples already show that, for realistic 
polynomials $V$, one has to go beyond second order in derivatives.

\section{A minimal BT approach\label{mbta}}
The original proposal by F.~V.~Thachov requires knowledge of the 
polynomial ${\cal P}$ of \eqn{functr}, iterative application of the functional
relation \eqn{functr} followed by integration by parts. After a Laurent 
expansion in $\ep = 4-n$ one can achieve an arbitrary degree of smoothness of 
the integrand. In principle, the denominators $B$ will contain thousands of 
terms and may lead to large numerical instabilities around thresholds.

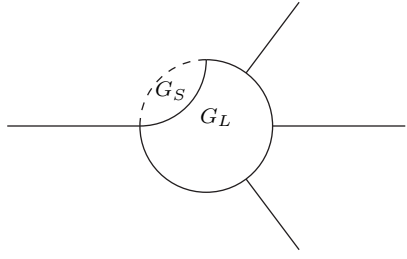
\begin{figure}[th]
\vspace{0.5cm}
\[
  \vcenter{\hbox{
  \begin{picture}(150,0)(0,0)
  \Line(0,0)(50,0)
  \CArc(75,0)(25,-180,90)
  \DashCArc(75,0)(25,90,180){3.}
  \CArc(50,25)(25,-90,0)
  \Line(100,0)(150,0)
  \Line(90,20)(110,46.67)
  \Line(90,-20)(110,-46.67)
  \Text(62,10)[cb]{$\scriptstyle G_{\ssS}$}
  \Text(79,0)[cb]{$\scriptstyle G_{\ssL}$}
  \end{picture}}}
\]
\vspace{0.5cm}
\caption[]{A generic two-loop diagram $G$; internal solid lines give the 
largest sub-diagram $G_{\ssL}$.} 
\label{fig:gtl}
\end{figure}

We have not been able to derive any compact form for the polynomial ${\cal P}$
of \eqn{functr}, even for the simplest two-loop topology. Instead, we have 
adopted a different strategy aimed to deal with arbitrary two-loop diagrams. 
It represents a compromise based on the simple observation that we know how to
apply the BT-iterative procedure for arbitrary one-loop diagrams. Therefore, 
given any two-loop diagram $G$ (for an illustration see~\fig{fig:gtl}) we 
apply the BT functional relation to $G_{\ssL}$, the one-loop sub-diagram of 
$G$ which has the largest number of internal lines. In this way the integrand 
for $G$ in $x$-space can be made smooth, a part from the factor $B$ of 
\eqn{functr} which is now a polynomial in $x_{\ssS}$, the Feynman parameters 
needed for the complementary sub-diagram of $G$ with the smallest number of 
internal lines, $G_{\ssS}$. 
The sub-diagram $G_{\ssS}$, after integration over its momentum, 
becomes an additional -- $x_{\ssS}$-dependent and with non-canonical power --
propagator for $G_{\ssL}$.

This procedure can be immediately generalized to any number of loops.
Furthermore, one should realize that the BT procedure does not introduce 
singularities through $B(x_{\ssS})$, a part from the singularities in the 
external parameters of the original diagram.
Therefore, before performing the $x_{\ssS}$-integration we move the
integration contour into the complex hyper-plane, thus avoiding the
crossing of apparent singularities. The idea of distorting the contour has 
been introduced in~\cite{Ghinculov:2001cz} and we have modified it to deal 
with the $x_{\ssS}$-integration.

A complete study of two-loop two-point functions (self-energies) has
shown~\cite{inp} that, for one iteration, a transformation of the
corresponding Feynman parameters can always be found that produces a 
coefficient $B$ which is $x_{\ssS}$-independent. In this way we are able to 
avoid distortion of the integration contour. However, this coefficient $B$ 
will vanish at some non-leading Landau singularity of the diagram where 
additional analytical work is needed before starting the numerical evaluation. A detailed derivation will be given in~\cite{inp}. 

In the following section we present a simple example that illustrates all
the features connected with integration in complex parametric space.
\vskip 5pt
\subsubsection{An example\label{examp}}
An instructive example is the following: suppose that we have to compute
\bq
I = \intfx{x}\,\bff{0}{-s}{x m}{m},
\label{iexa}
\eq
where $\sbff{0}$ is the scalar, one-loop, two-point function. 
Suppose also that we have no analytical expression available for $\sbff{0}$ 
which, of course, we have but, nevertheless, let us assume that we do not.
The integral of \eqn{iexa} becomes
\bqa
I &=& i\,\pi^{2-\ep/2}\,\egam{\frac{\ep}{2}}\,\intfx{x}\intfx{y}\,
\chi^{-\ep/2},  \nl
\chi(x,y) &=& s\,y^2 - \,\lpar s - m^2 + x^2m^2\rpar + x^2 m^2 - i\,\delta,
\label{ychi}
\eqa
with $\delta \to 0_+$. Then \eqn{rol} applies for
\bq
B = - \frac{1}{4\,s}\,\lambda\,\lpar s,x^2m^2,m^2\rpar, \qquad
Y = -\,\frac{1}{2\,s}\,\lpar s - m^2 + x^2 m^2\rpar,
\eq
with the following result:
\bq
\intfx{y}\,\chi^{-\ep/2} = - 4\,\frac{s}{\lambda}\,\intfx{y}\,
\,\Bigl[ 1 - \frac{1}{2-\ep}\,\lpar y + Y\rpar\,\frac{\partial}{\partial y}
\Bigr]\,\chi^{1-\ep/2}.
\eq
After integration by parts we obtain
\bq
\intfx{y}\,\chi^{-\ep/2} = - 4\,\frac{s}{\lambda}\,\frac{1}{2-\ep}\,\Bigl[
(3-\ep)\,\intfx{y}\,\chi^{1-\ep/2} 
- \,\lpar 1 + Y\rpar\,\lpar m^2\rpar^{1-\ep/2} + 
Y\,\lpar x^2 m^2\rpar^{1-\ep/2}
\Bigr].
\label{btexa}
\eq
This strategy is dubious if the roots of $\lambda = 0$ are inside
the interval $[0,1]$, on the real axis. These roots are easily evaluated
and are given by
\bq
x^+_{\pm} = \frac{1 \pm \mu}{\mu}, \qquad x^-_{\pm} = -\,
\frac{1 \pm \mu}{\mu}, \qquad \mu^2 s = m^2.
\eq
Therefore
\bq
x = \frac{\mu-1}{\mu}, \quad \mbox{for} \quad \mu > 1,  \qquad
x = \frac{1-\mu}{\mu}, \quad \mbox{for} \quad \mu < \frac{1}{2}, 
\eq
are the apparent singularities in the interval $[0,1]$. Crossing of
these points is avoided by moving the $x$-integration into the complex plane.
How to deform the integration path is suggested by simple considerations:
take the quadratic form $\chi$ of \eqn{ychi}; as a function of $x$ there are 
two roots,
\bq
\chi(x,y) = \mu^2\,\lpar 1 - y\rpar\,\lpar x + x_0 + i\,\delta\rpar\,
\,\lpar x - x_0 - i\,\delta\rpar,  \quad
x^2_0 = \frac{(1-\mu^2)y-y^2}{\mu^2(1-y)}.
\eq
Therefore, for $1-\mu^2 \le y \le 1$ we have two branch points at
$\pm\,i|x_0|$. Instead, for $0 \le y \le 1-\mu^2$ the branch points are
real and $\chi \le 0$ for $0 \le x \le x_0$. Furthermore, $x_0 \le 1$ for
$y \le y_-$ or $y \ge y_+$ with $y_{\pm} = (1 \pm \beta)/2$ and
$\beta^2 = 1-4\,\mu^2$. Thanks to the infinitesimal imaginary parts we have
\bq
\ln\chi = \ln\mu^2 + \ln\,\lpar 1 - y\rpar + \ln\,\lpar x + x_0 + i\,\delta
\rpar + \ln\,\lpar x - x_0 - i\,\delta\rpar,
\eq
and only the last logarithm contributes to the imaginary part. To avoid 
crossing the cut of the logarithm ($[-\infty,x_0]$) we choose any smooth
curve connecting $x = 0$ to $x = 1$ in the $x$ lower half-plane.
By comparing two expressions for $I$ (as defined in \eqn{iexa}), the first 
computed from the known analytical result for the $\sbff{0}$-function and a 
second one obtained through the double-integral originating from
\eqn{btexa} we find agreement in $8$-digits, as shown in \tabn{tab1}.
\begin{table}[hp]\centering
\begin{tabular}{|r|r|r|}
\hline
$\sqrt{s}\,$[GeV]  & \eqn{iexa} & \eqn{btexa}   \\
\hline
 & & \\
$1.1$ &  $1.12256370(1) + i\,0.042810523(5)$ &  $1.12256370(1) + 
i\,0.042810523(11)$ \\  
 & & \\
$10$  & $-2.52793882(1) + i\,3.09949189(0)$ & $-2.52793882(3) + i\,
3.09949189(3)$ \\
 & & \\
$100$ & $-7.20895671(4) + i\,3.14117375(0)$ & $-7.20895671(7) + i\,
3.14117375(7)$ \\
 & & \\
\hline
\hline
\end{tabular}
\vspace*{3mm}
\caption[]{$I$ from \eqn{iexa} or \eqn{btexa} with $m = 1$.\label{tab1}}
\end{table}
\normalsize
\subsection{The $S_3$ or sunset topology\label{s3t}}
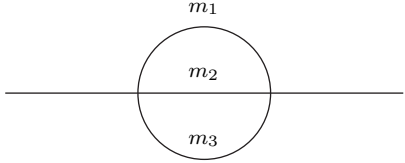
\begin{figure}[th]
\vskip 5pt
\[
  \vcenter{\hbox{
  \begin{picture}(150,90)(0,0)
  \Line(0,50)(50,50)
  \CArc(75,50)(25,-180,180)
  \Line(100,50)(150,50)
  \Line(50,50)(100,50)
  \Text(75,80)[cb]{$\scriptstyle m_1$}
  \Text(75,55)[cb]{$\scriptstyle m_2$}
  \Text(75,30)[cb]{$\scriptstyle m_3$}
  \end{picture}}}
\]
\vspace{-0.5cm}
\caption[]{The $S_3$ topology for two-loop self-energy diagrams.} 
\label{fig:s3}
\end{figure}

After some preliminar considerations we are ready to start with a realistic
example. Consider the simplest two-loop, two-point topology with three 
internal lines, the so-called sunset diagram illustrated in~\fig{fig:s3}. 
Literature relevant to the sunset (sunrise) topology is assembled, for
convenience, in ref.~\cite{sunsetrise}. The corresponding integral will be
\bq
\pi^4\,S_3 = \mu^{2\ep}\,\intmomsii{n}{q_1}{q_2}\,
\,\lpar q^2_1+m^2_1\rpar^{-1}\,
\,\lpar\,\lpar q_1-q_2+p\rpar^2+m^2_2\rpar^{-1}\,\lpar q^2_2 + m^2_3\rpar^{-1}.
\label{abgpi}
\eq
Here we introduce a special notation, $\lpar \alpha,m_1\mid 
\gamma,m_2,p\mid \beta,m_3\rpar,$  for
\bq
\mu^{2\ep}\,\intmomsii{n}{q_1}{q_2}\,
\,\lpar q^2_1+m^2_1\rpar^{-\alpha}\,
\,\lpar\,\lpar q_1-q_2+p\rpar^2+m^2_2\rpar^{-\gamma}\,
\lpar q^2_2 + m^2_3\rpar^{-\beta},
\eq
and $\lpar \mu_1\dots \mu_i\mid \nu_1\dots \nu_j\mid 
\alpha,m_1\mid \gamma,m_2,p\mid \beta,m_3\rpar$ for
\bq
\mu^{2\ep}\,\intmomsii{n}{q_1}{q_2}\,q^{\mu_1}_1\dots q^{\mu_i}_1
q^{\nu_1}_2\dots q^{\nu_j}_2 
\lpar q^2_1+m^2_1\rpar^{-\alpha}
\lpar\,\lpar q_1-q_2+p\rpar^2+m^2_2\rpar^{-\gamma} 
\lpar q^2_2 + m^2_3\rpar^{-\beta}.
\eq
\vskip 5pt
The best way of dealing with the sunset integral is to introduce the so-called
$\,\lpar\alpha\beta\gamma\rpar$ partial integration~\cite{'tHooft:1972fi} to 
obtain
\bq
S_3 = \sum_{i=1}^{3}\,m^2_i\,S_{3i} + S_{3p}.
\eq
With a suitable change of variables the four functions of \eqn{abgpi} can be 
cast into the following form:
\bqa
\pi^4\,S_{31} &=& \frac{\mu^{2\ep}}{\ep-1}\,\intmomsii{n}{q_1}{q_2}\,
\,\lpar q^2_1+m^2_3\rpar^{-1}\,
\,\lpar\,\lpar q_1-q_2-p\rpar^2+m^2_2\rpar^{-1}\,
\,\lpar q^2_2+m^2_1\rpar^{-2}  \nl
{}&=& (\ep-1)^{-1}\,\lpar 1,m_3\mid 1,m_2,-p\mid 2,m_1\rpar,  \nl
\pi^4\,S_{32} &=& \frac{\mu^{2\ep}}{\ep-1}\,\intmomsii{n}{q_1}{q_2}\,
\,\lpar q^2_1+m^2_1\rpar^{-1}\,
\,\lpar\,\lpar q_1-q_2+p\rpar^2+m^2_3\rpar^{-1}\,
\,\lpar q^2_2+m^2_2\rpar^{-2}  \nl
{}&=& (\ep-1)^{-1}\,\lpar 1,m_1\mid 1,m_3,p\mid 2,m_2\rpar,  \nl
\pi^4\,S_{33} &=& \frac{\mu^{2\ep}}{\ep-1}\,\intmomsii{n}{q_1}{q_2}\,
\,\lpar q^2_1+m^2_1\rpar^{-1}\,
\,\lpar\,\lpar q_1-q_2+p\rpar^2+m^2_2\rpar^{-1}\,
\,\lpar q^2_2+m^2_3\rpar^{-2}  \nl
{}&=& (\ep-1)^{-1}\,\lpar 1,m_1\mid 1,m_3,p\mid 2,m_2\rpar,  \nl
\pi^4\,S_{3p} &=& \frac{\mu^{2\ep}}{\ep-1}\,\intmomsii{n}{q_1}{q_2}\,\spro{q_2}{p}\,
\,\lpar q^2_1+m^2_1\rpar^{-1}\,
\,\lpar\,\lpar q_1-q_2+p\rpar^2+m^2_3\rpar^{-1}\,
\,\lpar q^2_2+m^2_2\rpar^{-2}  \nl
{}&=& (\ep-1)^{-1}\,\lpar 0\mid p\mid 1,m_1\mid 1,m_3,p\mid 2,m_2\rpar,
\nl
\label{s3cont}
\eqa
where, as usual, $n= 4 - \ep$. Their evaluation is discussed in \sect{es3i} 
and in \sect{evs3p}.
\subsubsection{Landau equations for $S_{33}$\label{landeq}}
Before starting the evaluation of the sunset diagram it is important to
know as much as possible about its singularities as a function of $p^2$ and 
of the internal masses. The corresponding Landau equations are
\bq
\alpha_1\,(q^2_1+m^2_1) = 0, \qquad
\alpha_2\,((q_1-q_2+p)^2+m^2_2) = 0, \qquad
\alpha_3\,(q^2_2+m^2_3) = 0, 
\label{land1}
\eq
and also
\bq
\alpha_1 q_{1\mu} + \alpha_2 (q_1-q_2+p)_{\mu} = 0, \qquad
- \alpha_2 (q_1-q_2+p)_{\mu} + \alpha_3 q_{2\mu} = 0.
\label{land2}
\eq
The leading Landau singularity occurs for $\alpha_i \ne 0, \forall i$.
We multiply the two equations \eqn{land2} by $q_{1\mu}, q_{2\mu}$ and
$p_{\mu}$ respectively. This gives an homogeneous system of six equations.
If all $\alpha_i$ are different from zero, the singularity will occur for
\bq
q^2_1= - m^2_1, \qquad q^2_2 = - m^2_3, \qquad
\spro{q_1}{q_2} = \frac{1}{2}\,\Bigl[ m^2_2 - m^2_1 - m^2_3 - s + 2\,
\spro{p}{(q_1-q_2)}\Bigr].
\eq
The Landau equations become as follows:
\bqa
{}&{}& - m^2_1 \alpha_1 + \Bigl[ \spro{q_2}{p} + \frac{1}{2} 
(s-m^2_1-m^2_2+m^2_3 \Bigr] \alpha_2 = 0,
\nl
{}&{}&\Bigl[ \spro{q_1}{p} - \spro{q_2}{p} - \frac{1}{2} 
(s+m^2_1-m^2_2+m^2_3)\Bigr]\,\alpha_1 +
\Bigl[ \spro{q_1}{p} - \frac{1}{2} (s+m^2_1-m^2_2-m^2_3)\Bigr]\,\alpha_2 = 0,
\nl
{}&{}&  \spro{q_1}{p}\, \alpha_1 + (  - s + \spro{q_1}{p} - \spro{q_2}{p} )\, 
\alpha_2 = 0,
\nl
{}&{}& \Bigl[ - \spro{q_2}{p} - \frac{1}{2} (s-m^2_1-m^2_2+m^2_3)\Bigr]\,
\alpha_2 + \Bigl[ \spro{q_1}{p} - \spro{q_2}{p} - \frac{1}{2} 
(s+m^2_1-m^2_2+m^2_3)\Bigr]\,\alpha_3 = 0,
\nl
{}&{}& \Bigl[ - \spro{q_1}{p} + \frac{1}{2} (s+m^2_1-m^2_2-m^2_3)\Bigr]\,
\alpha_2 - m^2_3 \alpha_3 = 0,
\nl
{}&{}& ( s - \spro{q_1}{p} + \spro{q_2}{p} )\,\alpha_2 + \spro{q_2}{p}\,
\alpha_3 = 0.
\label{land3}
\eqa
There are two compatibility conditions for the first three equations that can
be used to get $\spro{q_1}{p}$ and $\spro{q_2}{p}$,
\bq
\spro{q_1}{p} = \frac{m_1}{2}\,\frac{s - m^2_3 + (m_1+m_2)^2}{m_1+m_2},
\quad
\spro{q_2}{p} = - \frac{1}{2}\,\Bigl[ s + m^2_3 - (m_1+m_2)^2\Bigr].
\eq
Inserting these solutions into \eqn{land3} we obtain that
\bq
\alpha_1 = \frac{m_2}{m_1}\,\alpha_2,  \qquad
\alpha_3 = \frac{1}{2}\,\frac{m_2}{m^3_2}\,\lpar
             \frac{s-m^2_3}{m_1+m_2} - m_1 - m_2\rpar\,\alpha_2,
\eq
is a solution of the original system if and only if
\bq
s = (m_1 + m_2 \pm m_3)^2.
\eq
These are the leading singularities (plus permutations of masses),
referred to as normal and pseudo thresholds. Note that pseudo-threshold
does not correspond to a singularity on the physical Riemann sheet.
\subsubsection{Evaluation of $S_{3i}$\label{es3i}}
We will now compute $S_{33}$, as given in \eqn{s3cont}, and the 
remaining $S_{3i}$-terms will follow from $S_{33}$ with the corresponding 
permutations:
\bqa
31 &\;\mbox{with}\;& m_1 \leftrightarrow m_3, \quad p \leftrightarrow -\,p,
\nl
32 &\;\mbox{with}\;& m_2 \leftrightarrow m_3.
\label{permut}
\eqa
First a Feynman parameter $x$ is introduced, and the $q_1$-integration is 
performed giving
\bqa
S_{33} &=& i\,\pi^{-2-\ep/2}\,\frac{\mu^{2\ep}}{\ep-1}\,
\egam{\frac{\ep}{2}}\,\intfx{x}\intmomi{n}{q}\,
\,\lpar q^2+m^2_3\rpar^{-2}  \nl
{}&\times& \,\Bigl[ x(1-x)\,(q-p)^2 + xm^2_2 + (1-x)m^2_1\Bigr]^{-\ep/2},
\eqa
where we have introduced the Euler $\Gamma$-function. Next we change 
integration variable, $q' = q-p$, and introduce a $x$-dependent mass,
\bq
m^2_x = {{(1-x)m^2_1 + xm^2_2}\over {x(1-x)}}.
\eq
$S_{33}$ can be cast into the following form:
\bq
S_{33} = i\,\pi^{-2-\ep/2}\,\frac{\mu^{2\ep}}{\ep-1}\,\egam{\frac{\ep}{2}}\,
\intfx{x}\,\,\Bigl[ x(1-x)\Bigr]^{-\ep/2}\,\intmomi{n}{q}  
{{q^2+m^2_x}\over 
{\,\lpar q^2+m^2_x\rpar^{1+\ep/2}\,\lpar\,\lpar q+p\rpar^2+m^2_3\rpar^2}}.
\eq
According to some standard procedure we introduce a new Feynman parameter $y$
and write
\bqa
S_{33} &=& i\,\pi^{-2-\ep/2}\,\frac{\mu^{2\ep}}{\ep-1}\,
\egam{\frac{\ep}{2}}\,\intfx{x}\,\intfx{y}
\,\Bigl[ x(1-x)\Bigr]^{-\ep/2}\,\lpar 1-y\rpar^{\ep/2}\,y  \nl
{}&\times& \intmomi{n}{q}\,\lpar q^2+m^2_x\rpar\,\,\Bigl[
q^2 + 2\,y\,\spro{p}{q} + y\,\lpar p^2+m^2_3\rpar + \,\lpar 1-y\rpar\,
m^2_x\Bigr]^{-3-\ep/2}.
\eqa
Let $U$ be defined by
\bq
U = y\,(1-y)\,p^2 + y\,m^2_3 + (1-y)\,m^2_x - i\,\delta,
\label{defU}
\eq
with $\delta \to 0_+$, then $S_{33}$ becomes
\bqa
S_{33} &=& -\,\pi^{-\ep}\,\frac{\mu^{2\ep}}{\ep-1}\,
\frac{\egam{\ep/2}}{\egam{1+\ep/2}}\,
\intfx{x}\intfx{y}\,\,\Bigl[ x(1-x)\Bigr]^{-\ep/2}\,\lpar 1-y\rpar^{\ep/2}  \nl
{}&\times& y\,\Bigl[\egam{1+\ep}\,\lpar y^2p^2 + m^2_x\rpar\,U^{-1-\ep} +
\,\lpar 2 - \frac{\ep}{2}\rpar\,\egam{\ep}\,U^{-\ep}\Bigr].
\eqa
Next we introduce the Mandelstam variable $s$,
\bq
p^2 = -\,s, \qquad \mu^2_i = \frac{m^2_i}{s},
\eq
and scale $U$ accordingly:
\bq
U = s\,\chi, \qquad \chi = - y\,\lpar 1-y\rpar + \mu^2_3\,y + \mu^2_x\,
\,\lpar 1-y\rpar -i\,\delta.
\eq
For the rest of this discussion we assume that $s > 0$ and will postpone to
Appendix B the discussion of the case $s < 0$.

There is no need to evaluate numerically the poles at $\ep= 0(n = 4)$. 
We perform a Laurent expansion of the integrand around $\ep = 0$, after which
the double pole is easy to compute and gives
\bq
\frac{2}{\ep^2} + \Delta^2_{\ssU\ssV}, \qquad 
\Delta_{\ssU\ssV} = \gamma + \ln\pi + \ln\frac{s}{\tHss}.
\label{DUV}
\eq
In this result $\gamma= 0.577216$ is the Euler's constant and we have
included in the definition of {\em double-pole} a bunch of constants.
Also the single pole at $\ep = 0(n = 4)$ can be computed analytically. 
The $x-y$ integration is greatly simplified by the use of the following 
relation:
\bq
\frac{\mu^2_x - y^2}{\chi} = 1 - \frac{y}{\chi}\,
\frac{\partial}{\partial y}\,\chi.
\label{trel}
\eq
With the help of \eqn{trel} we find the following result for the single-pole:
\bq
-\,\frac{2}{\ep}\,\Delta_{\ssU\ssV} +
\,\lpar 2\,\ln\mu^2_3 - 3\rpar\,\lpar \Delta_{\ssU\ssV} - \frac{1}{\ep}\rpar,
\label{sppart}
\eq
where, again, a bunch of constants have been included into the definition of 
{\em single-pole}. A detailed derivation is presented in Appendix A.
Continuing our calculation we write 
\bqa
S_{33} &=& -\,\frac{\pi^{-\ep}}{\ep-1}\,\frac{\egam{\ep/2}}{\egam{1+\ep/2}}\,
\,\lpar\frac{s}{\mu^2}\rpar^{-\ep}\,
\intfx{x}\intfx{y}\,\,\Bigl[ x(1-x)\Bigr]^{-\ep/2}\,\lpar 1-y\rpar^{\ep/2}\,y
\nl
{}&\times& \,\Bigl[\egam{1+\ep}\,\lpar -y^2 + \mu^2_x\rpar\,\chi^{-1-\ep} +
\,\lpar 2 - \frac{\ep}{2}\rpar\,\egam{\ep}\,\chi^{-\ep}\Bigr].
\label{ints}
\eqa
The quadratic form $\chi$ is defined by
\bq
\chi = H\,y^2 + 2\,K\,y + L,
\eq
with coefficients
\bq
H = 1, \qquad K = -\,\frac{1}{2}\,\lpar 1 - \mu^2_3 + \mu^2_x\rpar,
\qquad L = \mu^2_x.
\eq
Our strategy will be to single out the $y$-integral and to {\em raise} the
power of $\chi$. The coefficients relevant for this case are:
\bqa
B &=& \mu^2_x - \frac{1}{4}\,\lpar 1 - \mu^2_3 + \mu^2_x\rpar^2 = 
-\,\frac{1}{4}\,\lambda\,\lpar 1,\mu^2_3,\mu^2_x\rpar,  \nl
Y &=& -\,\frac{1}{2}\,\lpar 1 - \mu^2_3 + \mu^2_x\rpar, \qquad \mu = - 1 - \ep,
\eqa
where $\lambda$ is the usual K\"allen function. Therefore we obtain
\bq
\chi^{-1-\ep} = -\frac{4}{\lambda\,\lpar 1,\mu^2_3,\mu^2_x\rpar}\,\,\Bigl[
1 + \frac{1}{2\,\ep}\,\lpar y + Y\rpar\,\frac{\partial}{\partial y}
\Bigr]\,\chi^{-\ep}.
\label{spur}
\eq
No spurious singularity at $\ep = 0$ is introduced in the r.h.s. of 
\eqn{spur} and to avoid apparent poles we use a new variant of the
basic functional relation, \eqn{functr}:
\bq
\chi^{-1-\ep} = -\,\frac{4}{\lambda\,\lpar 1,\mu^2_3,\mu^2_x\rpar}\,
\Bigl\{ 1 - \frac{1}{2}\,\lpar y+Y\rpar\,\frac{\partial}{\partial y}\,
\ln\chi - \ep\,\,\Bigl[ \ln\chi - \frac{\ep}{4}\,\lpar y+Y\rpar  
\frac{\partial}{\partial y}\,\ln^2\chi + \ord{\ep^2}\Bigr]\Bigr\}.
\label{trick}
\eq
Note that no new singularity in the variable $x$ is introduced by the 
procedure.
However, numerical instabilities could be introduced in the $x$-integration,
especially if the process of {\em raising} powers is continued, resulting
in a final expression with higher and higher negative powers of $\lambda$.
For this reason we are led to study the zeros of the K\"allen function, i.e.\
$\lambda\,\lpar 1,\mu^2_3,\mu^2_x\rpar = 0$.
Clearly, if $s < 0$ there are no real solutions for $x$. Otherwise solutions
are given by
\bq
\,\lpar \mu^2_x\rpar_{\pm} = \,\lpar 1 \pm \mu_3\rpar^2.
\eq
Consider now $\mu^2_x$ as a function of the variable $x$,
\bq
\mu^2_x = {{(1-x)\,\mu^2_1 + x\,\mu^2_2}\over {x\,(1-x)}}.
\eq
The minimum, for $\mu^2_x$, occurs at $x_{\pm} = \mu_1/(\mu_2 \pm \mu_1)$.
Only $x_+$ lies between $0$ and $1$, corresponding to  
\bq
\,\lpar \mu^2_x\rpar_{\rm min} = \,\lpar \mu_1 + \mu_2\rpar^2.
\eq
There are three distinct possibilities:
\begin{enumerate}

\item the root $(\mu^2_x)_+$ is below the minimum, i.e.\
$\mu_1 + \mu_2 - \mu_3 \ge 1$, therefore $\lambda$ can never be zero.
\item Only one root is above the minimum, i.e.\
$\lpar 1 - \mu_3\rpar^2 \le \,\lpar \mu_1+\mu_2\rpar^2 \le \,
\lpar 1 + \mu_3\rpar^2$, when there are two values of $x$ where $\lambda = 0$,
\bq
x^+_{\pm} = \frac{1}{2\,\lpar 1 + \mu_3\rpar^2}\,\,\Bigl[
\,\lpar 1 + \mu_3\rpar^2 + \mu^2_1 - \mu^2_2 \pm \lambda^{1/2}\,\lpar
\,\lpar 1 + \mu_3\rpar^2,\mu^2_1,\mu^2_2\rpar\Bigr].
\label{defxp}
\eq
\item Both roots are above the minimum, i.e.\
$\lpar \mu_1 + \mu_2\rpar^2 \le \,\lpar 1 - \mu_3\rpar^2$,
when we have four values of $x$ where $\lambda = 0$. The new pair of points is
given by
\bq
x^-_{\pm} = \frac{1}{2\,\lpar 1 - \mu_3\rpar^2}\,\,\Bigl[
\,\lpar 1 - \mu_3\rpar^2 + \mu^2_1 - \mu^2_2 \pm \lambda^{1/2}\,\lpar
\,\lpar 1 - \mu_3\rpar^2,\mu^2_1,\mu^2_2\rpar\Bigr].
\label{defxm}
\eq
\end{enumerate}
For the finite part of this Feynman diagram we have now
to understand when and where the imaginary part develops. Let us assume that
$s > 0$. Imaginary parts are present as soon as $\chi < 0$. There are two 
solutions for $y$ to the equation $\chi = 0$,
\bq
y_{\pm} = \frac{1}{2}\,\,\Bigl[ 1 - \mu^2_3 + \mu^2_x \pm \lambda^{1/2}
\,\lpar 1,\mu^2_3,\mu^2_x\rpar\Bigr].
\label{tsol}
\eq
Therefore, $\ln\chi$ develops an imaginary part for $y_- \le y \le y_+$,
if and only if $\lambda(1,\mu^2_3,\mu^2_x) > 0$. As before, there are only
three possibilities,
\begin{enumerate}
\item $\,\lpar 1 + \mu^2_3\rpar^2 \le \,\lpar \mu_1 + \mu_2 \rpar^2$, where
$\lambda$ is always positive,
\item $\,\lpar 1 - \mu_3\rpar^2 \le \,\lpar \mu_1 + \mu_2 \rpar^2 \le \,\lpar 
1 + \mu_3\rpar^2$, where we have two alternatives,
\bqa
\,\lpar \mu_1+\mu_2\rpar^2 \le \mu^2_x &\le& \,\lpar 1 + \mu_3\rpar^2, \qquad
\lambda \,\mbox{is negative},  \nl
\mu^2_x &\ge& \,\lpar 1 + \mu_3\rpar^2, \qquad
\lambda \,\mbox{is positive},  
\eqa
\item $\,\lpar \mu_1+\mu_2\rpar^2 \le \,\lpar 1 - \mu^2_3\rpar^2$ where we have
three alternatives,
\bqa
\,\lpar \mu_1+\mu_2\rpar^2 \le \mu^2_x &\le& \,\lpar 1 - \mu_3\rpar^2, \qquad
\lambda \,\mbox{is positive},  \nl
\,\lpar 1- \mu^2_3\rpar^2 \le \mu^2_x &\le& \,\lpar 1 + \mu_3\rpar^2, \qquad
\lambda \,\mbox{is negative},  \nl
\mu^2_x &\ge& \,\lpar 1 + \mu_3\rpar^2, \qquad
\lambda \,\mbox{is positive},  
\eqa
\end{enumerate}
Consider the original integral, \eqn{ints}; the integration hyper-contour can 
be distorted away from its original real location and when the possibility of 
this distortion ceases, we encounter a singularity of the 
function~\cite{elop}. The difficulty in locating the singularities lies in 
imagining what happens in the multi-dimensional (complex) space of integration.
In the case of \eqn{ints} the integrand shows poles at the zeros of
$\chi(x,y)$, i.e.\ for for $y = y_{\pm}(x)$ with $y_{\pm}$ given in \eqn{tsol}. 
We may distort the interval $y \in [0,1]$ but a pinch may appear for 
$y_+ = y_-$, which requires $\lambda(1,\mu^2_3,\mu^2_x) = 0$. Again, we can 
move the $x$-integration contour to avoid the points where this happens, 
i.e.\ for $\mu^2_x = ( 1 \pm \mu_3)^2$,
\bq
\mu^2_x = ( 1 \pm \mu^3 )^2, \qquad \mbox{for} \quad x = x^{\pm}_{\pm}.
\eq
However, if $x^+_+ = x^+_-$ or $x^-_+ = x^-_-$ a singularity may appear.
Note that for $\mu^2_x = (1+\mu_3)^2$ we will have
\bq
y_{\pm} = 1 + \mu_3, \qquad \chi = \lpar y - 1 - \mu_3\rpar^2,
\eq
while for $\mu^2_x = (1-\mu_3)^2$ we have
\bq
y_{\pm} = 1 - \mu_3, \qquad \chi = \lpar y - 1 + \mu_3\rpar^2,
\eq
so that, in the first case, the pinch should correspond to $y > 1$,
therefore there is no singularity.
In this case we will speak of the so-called pseudo-threshold, corresponding
to $s = ( m_1+m_2-m_3)^2$. It is a well known fact that the $S_3$ topology
is regular at any of the pseudo-thresholds. On the contrary, when
$\mu^2_x \to (1-\mu_3)^2$ (or $x \to x^-_{\pm}$) we have $y_+ \to y_-$
from opposite sides of the integration contour but the relevant part of
the integrand is $(\mu^2_x - y^2)/\chi$ which, for $y = 1 - \mu_3 + \Delta$
($\Delta \to 0$) and $\mu^2_x = (1-\mu_3)^2$ behaves as
\bq
{{\mu^2_x - y^2}\over {\chi}} \sim 
{{2\,\lpar 1 - \mu_3\rpar\,\Delta}\over {\Delta^2 - i\,\delta}}.
\eq
Therefore there is no pinch, even in this situation.
Note that the condition $x^-_+ = x^-_-$ corresponds to the so-called
normal threshold, $s = (m_1+m_2+m_3)^2$ which is, nevertheless, a 
singular point of the sunset diagram. The singularity clearly arises from the 
remaining $\chi^{-\ep}$ terms of the integrand, in the same way as the normal
threshold singularity arises in the one-loop self-energy. This rather
long discussion has been introduced for one specific purpose: if we
rewrite $S_{33}$ as
\bq
S_{33} = -\,\frac{\pi^{-\ep}}{\ep-1}\,
\frac{\egam{\ep/2}\egam{\ep}}{\egam{1+\ep/2}}\,
\,\lpar\frac{s}{\mu^2}\rpar^{-\ep}\,\intfx{x}\,{\cal S}_{33}(\ep,x),
\eq
then ${\cal S}_{33}(\ep,x)$ has no poles at $x = x^-_{\pm}$. In other words
$x = x^-_{\pm}$ are points of analyticity for
\bq
\intfx{y}\,\,\lpar 1-y\rpar^{\ep/2}\,y\,
\Bigl[\egam{1+\ep}\,\lpar -y^2 + \mu^2_x\rpar\,\chi^{-1} +
\,\lpar 2 - \frac{\ep}{2}\rpar\,\egam{\ep}\Bigr].
\eq
This fact remains obviously true after {\em raising} powers of $\chi$, 
operation that brings negative powers of $\lambda$ in the result and, due to 
analyticity of the integrand in those points, we can distort the integration 
contour.
One can reach this conclusion also by following an alternative derivation of 
the result. Starting from \eqn{ints} we apply \eqn{trick} only once. The 
integral will be of the following form:
\bq
S_{33} = \intfxx{x}{y}\,\lpar \frac{I_1}{\Lambda} + I_0\rpar, \qquad
\Lambda = \lambda(1,\mu^2_3,\mu^2_x).
\eq
A simple exercise shows that 
\bq
\intfx{y} I_1 = 0, \qquad \mbox{ for}\, \qquad \mu^2_x = (1-\mu_3)^2, 
\eq
so that we can replace
\bq
I_1(x,y) \to I^{\rm sub}(x,y) = I_1(x,y) - I_1\lpar \mu_x = 1-\mu_3\rpar.
\eq
Just one iteration of the {\em raising} procedure gives $I_{0,1}$ 
continuous, as compared with two iterations that give also continuous first
derivatives but, in this way, the whole integrand is well behaved around
the zeros of $\lambda(1,\mu^2_3,\mu^2_x)$.
As for the one-loop, two-point function, once a suitable
integral representation is found, the branch point corresponding to normal
threshold is not due to poles of the integrand.

We repeat that distortion of the contour is only dictated by the need of
achieving numerical convergence and must satisfy one criterion: the new 
contour cannot cross the cuts of $\chi^{-\ep}$. Since at the normal
threshold the two branch points approach each other on the real axis, there
we can no longer distort the contour and face numerical instabilities.
Therefore, the method cannot be applied, as it stands, when we are
exactly at threshold and an alternative algorithm will be presented in 
\sect{nt}.

Summarizing, me may say that BT method has the advantage of moving poles
into branch points and minimal BT method requires a knowledge of the
analytical properties of a one-loop sub-diagram of the whole diagram $G$,
a task much simpler than the one of knowing the whole analytical structure of
$G$.

After a discussion on the singularities of $\lambda$ and of the imaginary parts
we continue by performing an integration by parts after the first operation
of {\em raising} the powers. This will be achieved, after inserting
\eqn{trick} into \eqn{ints}, with the help of the following relations:
\bqa
\intfx{y}\,y^n\,\lpar 1 - y\rpar^{\ep/2}\,\frac{\partial}{\partial y}\,
\ln^m\chi(x,y) &=&
-\,\delta_{n0}\,\ln\chi(x,0) - \intfx{y}\,
\,\Bigl[ n\,y^{n-1}\,\lpar 1 - y\rpar^{\ep/2}  \nl
{}&-& \frac{\ep}{2}\,y^n\,
\,\lpar 1 - y\rpar^{\ep/2-1}\Bigl]\,\ln^m\chi(x,y),
\eqa
followed by a subtraction procedure that takes care of the $1-y$ terms,
\bqa
\intfx{y}\,y^n\,\lpar 1 - y\rpar^{\ep/2-1}\,\ln^m\chi(x,y) &=&
\frac{2}{\ep}\,\ln^m\chi(x,1)\,
+ \intfx{y}\,\lpar 1 - y\rpar^{\ep/2-1}  \nl
{}&\times& \,\Bigl[ y^n\,\ln^m\chi(x,y) - \ln^m\chi(x,1)\Bigr].
\eqa
The procedure of {\em raising} the power of $\chi$ can be repeated. Since
\bq
\chi(x,y) = y^2 + 2\,K\,y + L,  
\eq
we could use 
\bq
B\,\chi^{-\ep} = \,\Bigl[ 1 - \frac{1}{2\,\lpar 1 - \ep\rpar}\,
\,\lpar y + Y\rpar\,\frac{\partial}{\partial y}\Bigr]\,\chi^{1-\ep},
\eq
but it is better to expand this relation before integrating by parts, thus
avoiding the introduction of fictitious poles at $\ep = 0$. The result that 
we obtain by expanding in powers of $\ep$ and by subsequently equating 
coefficients of the same powers in $\ep$ is as follows:
\bqa
B\,\ln\chi &=& \chi\,\ln\chi + \frac{1}{2}\,\lpar y + Y\rpar\,
\frac{\partial}{\partial y}\,\Bigl[ \chi\,\lpar 1 - \ln\chi\rpar\Bigr],  \nl
B\,\ln^2\chi &=& \chi\,\ln^2\chi -\,\lpar y + Y\rpar\,
\frac{\partial}{\partial y}\,\Bigl[ \chi\,\lpar 1 - \ln\chi + \frac{1}{2}\,
\ln^2\chi\rpar\Bigr].
\eqa
In this way terms as $\ln\chi$ or $\ln^2\chi$ are transformed into
$\chi\,\ln\chi$ and $\chi\,\ln^2\chi$.
However, some care is needed in the presence of terms containing $\ln (1-y)$. 
In this case we proceed as follows:
\bqa
\intfx{y}\,y^n\,\ln \,\lpar 1 - y\rpar\,\ln\chi &=& \intfx{y}\,
y^n\,\ln\,\lpar 1 - y\rpar\,\lpar \ln\frac{\chi}{\mu^2_3} + \ln\mu^2_3\rpar \nl
{}&=& -\,\frac{1}{n+1}\,\ln\mu^2_3\sum_{j=1}^{n+1}\,\frac{1}{j} +
\intfx{y}\,y^n\,\ln\,\lpar 1 - y\rpar\,\ln\frac{\chi}{\mu^2_3},
\nl
\eqa
where use has been made of the relation $\chi(x,1) = \mu^2_3$. In this case the
quadratic form in $y$ to which we apply the algorithm is
\bq
\frac{y^2}{\mu^2_3} + 2\,\frac{K}{\mu^2_3}\,y + \frac{L}{\mu^2_3},
\eq
giving $B$ and $Y$ coefficients,
\bq
B = \frac{\lambda}{\mu^2_3}\, \qquad Y = K.
\eq
Therefore we will have terms of the following form:
\bq
\intfx{y}\,y^n\,\ln\,\lpar 1 - y\rpar\,\frac{\partial}{\partial y}\,\lpar
\chi\,\ln\chi\rpar,
\eq
which, after integration by parts, become
\bq
-\,\intfx{y}\,\chi\ln\frac{\chi}{\mu^2_3}\,\,\Bigl[ n\,y^{n-1}\,\ln\,\lpar 1 - y
\rpar + \frac{y^n}{y-1}\Bigr].
\eq
In this way we arrive at our final formulas for $S_{33}$ which is shown 
explicitly in Appendix C. The other $S_{3i}$ terms follow through the 
permutations of \eqn{permut}. 
\subsubsection{Evaluation of $S_{3p}$\label{evs3p}}
The remaining contribution is $S_{3p}$, defined in \eqn{s3cont}. Everything
proceeds as in \sect{es3i}, in particular the double-pole turns out to be zero
and the single-pole is
\bq
\frac{1}{2}\,s\,\Bigl[ \Delta_{\ssU\ssV} - \frac{1}{\ep}\Bigr],
\label{sp3p}
\eq
with $\Delta_{\ssU\ssV}$ defined in \eqn{DUV}.
The complete expression for $S_{3p}$ is also given in Appendix C.
Collecting the various terms we obtain the following expression for $S_3$:
\bqa
S_{3\,\rm double-pole} &=& \sum_{i=1}^{3}\,m^2_i\,\lpar \frac{2}{\ep^2} +
\Delta^2_{\ssU\ssV}\rpar,  \nl
S_{3\,\rm single-pole} &=& - 2\,\sum_{i=1}^{3}\,m^2_i\,
\frac{\Delta_{\ssU\ssV}}{\ep} +
\lpar \Delta_{\ssU\ssV} - \frac{1}{\ep}\rpar\,
\Bigl[ \sum_{i=1}^{3}\,m^2_i\,\lpar 2\,\ln\frac{m^2_i}{s} - 3\rpar + 
\frac{s}{2}\Bigr].
\eqa
The finite part will also be given in Appendix C.
\subsubsection{Integration in the complex plane\label{icp}}
The integrand for $S_3$ is now a smooth function but for inverse powers
of $\lambda$ whose zeros are known and their crossing should be avoided
by moving the $x$-integration from the interval $[0,1]$ to the complex plane.
Technically speaking, when we are exactly at the normal threshold, our method 
is no better than any other numerical method. 
Away from it, the actual shape of the integration path follows (as discussed 
in \sect{examp}) from a careful examination of the imaginary part of the 
logarithms appearing in the integrand.
The following recipe will produce the correct imaginary part for $S_3$.
If $(1-\mu_3)^2 \le (\mu_1+\mu_2)^2 \le (1+\mu_3)^2$ we compute $x^+_{\pm}$,
while for $(1-\mu_3)^2 \ge (\mu_1+\mu_2)^2$ we compute both $x^+_{\pm}$ and
$x^-_{\pm}$.
Next we will have to investigate the branch points of the logarithms. The 
argument of the logarithms is a quadratic form in $x$,
\bq
\xi(x,y) = - y\,\lpar y - 1 + \mu^2_3\rpar\,x^2 + \Bigl[
y\,\lpar y - 1 + \mu^2_3\rpar + \lpar 1 - y\rpar\,\lpar \mu^2_2 - \mu^2_1\rpar
\Bigr]\,x + \lpar 1 - y \rpar\,\mu^2_1.
\label{quad}
\eq
If the corresponding discriminant is positive we will have two real roots,
$x_{\ssL,\ssR}$, otherwise we have a pair of complex conjugated roots. 
In the latter case we select a contour which starts at $x = 0$ and bypasses
$x = x^+_{\pm}$ or $x = x^+_{\pm},\,x^-_{\pm}$ in the upper half-plane to 
return to $x = 1$. The distorted contour must avoid any crossing of the cuts
of the logarithm which will be, in this case, parallel to the imaginary 
$x$-axis.

If the roots are real me must distinguish between two cases. Let us denote 
the quadratic in $x$ of \eqn{quad} as $a x^2 + b x +c$, with $b^2 \ge 4\,ac$. 
If $a$ is positive then the cut is between $x_{\ssL}$ and $x_{\ssR}$, while 
for negative $a$ the cut is $[-\infty,x_{\ssL}]\,\cup\,[x_{\ssR},+\infty]$.
Furthermore, if $x = \alpha + i\,\beta$, the imaginary part of the logarithm is
$\beta ( 2\,\alpha a + b)$. This consideration immediately tells us the sign of
the imaginary part when $x$ approaches the real axis on the cut. This sign
is crucial when we distort the contour in the complex plane since, for
$x$ real and on the cut, the $i\,\delta$ prescription,
\bq
\ln ( a x^2 + b x + c ) \to \ln ( a x^2 + b x + c - i\,\delta),
\eq
gives $-\,i\pi$ for the imaginary part. To give an example, when the branch 
points are real and $a$ is positive, any contour should start at $x = 0$,
bypass $x^-_{\pm}$, return to the real axis from above and for 
$\Reb\,x \le -\,b/(2 a)$, leave the real axis in the lower half-plane for 
$\Reb\,x \ge -\,b/(2 a)$ and bypass $x^+_{\pm}$ to return to $x = 1$.
Typical examples are shown in \figs{cutplane1}{cutplane2}.

We now examine the behavior of the integrand close to the normal threshold.
Consider \fig{irxlr} which corresponds to $s = (m_1+m_2+m_3)^2$ for some
choice of masses: for $y$
below some value $y_0$ we have that $x_{\ssL,\ssR}$ are complex conjugated
and their real part is internal to the interval $[0,1]$. Just below
$y_0$ the two imaginary parts go to infinity and above $y_0$ the two branch
points become real but are external to $[0,1]$. Furthermore, for
some value $y_{\rm th}$ the two complex roots $x_{\ssL,\ssR}$ pinch the
integration contour at $x = x^-_- \equiv x^-_+$. Therefore, any distortion 
of it will inevitably cross the cut, giving the wrong answer. In a word, 
we cannot distort the contour any longer. Note that there is no crossing of 
poles implied here, only a serious problem of numerical convergence of the 
minimal BT approach due to negative powers of $\lambda(1,\mu^2_3,\mu^2_x)$ 
which exhibits a double zero at $x = x^-_- \equiv x^-_+$.
For different choices of the internal masses, at threshold, we may have
a different behavior of $x_{\ssL,\ssR}$ but there will always be a
$y_{\rm th}$ where they are complex conjugated, with their real part
approaching $x = x^-_- \equiv x^-_+$ and with imaginary parts pinching the 
real axis.

The situation at pseudo-threshold, $s= (m_1+m_2-m_3)^2$, is different
as one can see from \fig{irxlrp}. Below some value $y_0$ the roots
$x_{\ssL,\ssR}$ are complex conjugated but the absolute value of the
imaginary part has a minimum and, therefore no pinch will occur. Above
$y_0$ the roots are real but external to the interval $[0,1]$. 
Therefore a distortion of the integration contour, to avoid the point
$x = x^+_- \equiv x^+_+$, is always possible.
\subsubsection{The normal threshold for the sunset\label{nt}}
The arguments developed in \sect{icp} show that we have to use some other 
algorithm at the normal threshold $s = (m_1+m_2+m_3)^2$. Actually, for
$S_{3i}$ and $S_{3p}$ we have a much simpler derivation of the result which,
although different from the BT-approach, allows us to {\em raise} powers.
We consider again \eqn{ints} and apply directly \eqn{trel}. After the
expansion around $n = 4$ we have the following integrations by parts:
\bqa
\intfx{y}\,y^n\,\ln(1-y)\,\partial_y\,\ln\chi(x,y) &=&
- \intfx{y}\,\ln\frac{\chi(x,y)}{\chi(x,1)}\,\Bigl[ n y^{n-1}\,\ln(1-y) -
\frac{y^n}{1-y}\Bigr],  \nl
\intfx{y}\,y^n\ln\chi(x,y)\,\partial_y\,\ln\chi(x,y) &=&
\frac{1}{2}\,\Bigl[ \ln^2\chi(x,1) - \delta_{n,0}\,\ln^2\chi(x,0) - n\,
\intfx{y}\,y^{n-1}\,\ln^2\chi(x,y)\Bigr],  \nl
\intfx{y}\, y^n\,\partial_y\,\ln\chi(x,y) &=& 
\ln\chi(x,1) - \delta_{n,0}\,\ln\chi(x,0) - n\,
\intfx{y}\,y^{n-1}\,\ln\chi(x,y)\Bigr].
\eqa
Using these relations we obtain a very simple expression for the finite part
of $S_{33}$,
\bqa
S^{\rm fin}_{33} &=& \intfxx{x}{y}\,\Bigl[\ln\xi(x,y) +
\lpar \frac{\ln\xi(x,y)}{y-1}\rpar_+\Bigr] + \ln\mu^2_3\,\lpar
\ln\mu^2_3 - 4\rpar  + \frac{13}{2} + \frac{1}{2}\,\zeta(2),  \nl
S^{\rm fin}_{3p} &=& \intfxx{x}{y},y\,\ln\xi(x,y) + \frac{3}{8},
\label{simple}
\eqa
where $\xi = x(1-x)\chi$ and the `+'-distribution is defined by \eqn{plusd}
of Appendix C. Therefore, we have reached our goal of expressing
the result in terms of a smooth integrand, without obnoxious overall factors
that can vanish in certain regions with a consequent numerical instability.
The reason why we have not indicated \eqn{simple} as the main result of
our work is based on the fact that the same result cannot be generalized for
other topologies. Indeed, it is essential to have the factor $\mu^2_x-y^2$
and, therefore, \eqn{trel} can be applied only once.
However, we have now a result that can be used for the sunset diagram, without 
numerical instabilities, also at threshold. Numerical results will be shown
in the next sections. 

For other topologies we will use the minimal BT-approach everywhere but at the
corresponding normal thresholds and will adapt the present algorithm around
them. Actually there are cases where the result is again very simple.
\begin{itemize}
\item[--] {\em The $S_4$ topology}:
\end{itemize}
The $S_4$ topology is given by
\bqa
\pi^4\,S_{4} &=& \mu^{2\ep}\,\intmomsii{n}{q_1}{q_2}\,
{1\over {
\lpar q^2_1+m^2_1\rpar\,
\lpar\,\lpar q_1-q_2\rpar^2+m^2_2\rpar\,
\lpar q^2_2 + m^2_3\rpar\,
\lpar\lpar q_2 + p\rpar^2 + m^2_4\rpar}}.
\eqa
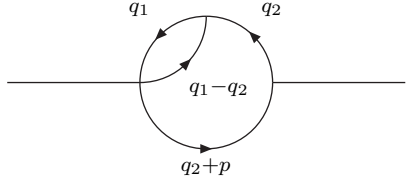
\begin{figure}[th]
\vspace{0.5cm}
\[
  \vcenter{\hbox{
  \begin{picture}(150,0)(0,0)
  \Line(0,0)(50,0)
  \ArrowArc(75,0)(25,0,90)
  \ArrowArc(75,0)(25,-180,0)
  \ArrowArc(75,0)(25,90,180)
  \ArrowArc(50,25)(25,-90,0)
  \Line(100,0)(150,0)
  \Text(50,25)[cb]{$\scriptstyle q_1$}
  \Text(100,25)[cb]{$\scriptstyle q_2$}
  \Text(75,-35)[cb]{$\scriptstyle q_2+p$}
  \Text(80,-5)[cb]{$\scriptstyle q_1-q_2$}
  \end{picture}}}
\]
\vspace{0.5cm}
\caption[]{The two-loop diagram $S_4$. Arrows indicate the momentum flow.} 
\label{tops4}
\end{figure}
We easily derive the following relation:
\bqa
S_4 &=& (\ep - 1)\,\intfx{x}\,S_{33}\lpar x^2 p^2;m_1,m_2,M_x\rpar,  \nl
M^2_x &=& - p^2 x^2 + (p^2 + m^2_4 - m^2_3) x + m^2_3,
\eqa
where
\bqa
(\ep - 1)\,S^{\rm fin}_{33}\lpar x^2 p^2;m_1,m_2,M_x\rpar &=& 
-\,\intfxx{y}{z}\,\Bigl[\ln\xi(y,z) +
\lpar \frac{\ln\xi(y,z)}{z-1}\rpar_+\Bigr] + \ln M^2_x\,\lpar
2 - \ln M^2_x\rpar  \nl
{}&-& \frac{7}{2} - \frac{1}{2}\,\zeta(2),
\eqa
and $\xi$ is given in \eqn{quad}, with $p^2 \to x^2p^2$. For other cases the 
result is more involved.
\begin{itemize}
\item[--] {\em Other topologies}:
\end{itemize}
The key observation is that any two-loop diagram can be written as
a multi-dimensional integral, over {\em external} Feynman parameters 
$x_{\rm ext}$, of a kernel which is nothing but a generalized sunset topology 
with masses that depend on $x_{\rm ext}$,
\bq
D = \prod_{i=1,N}\,\intfx{x^i_{\rm ext}}\,
\lpar \alpha,m_1\mid \gamma,m_2,p\mid \beta,m_3\rpar.
\eq
Around thresholds, where nothing better will work, we make use of recurrence 
relations, for instance
\bq
\lpar \alpha+1,m_1\mid \gamma,m_2,p\mid \beta,m_3\rpar =
- \frac{1}{\alpha}\,\frac{\partial}{\partial\,m^2_1}\,
\lpar \alpha,m_1\mid \gamma,m_2,p\mid \beta,m_3\rpar,
\eq
until we reach $\lpar 1,m_1\mid 1,m_2,p\mid 2,m_3\rpar$ where a stable 
numerical result can be produced. At this point, despite the poor reputation 
enjoyed by numerical differentiation, we will perform all differentiations 
with respect to masses numerically. Only at the end the final integration over
$x_{\rm ext}$ will be executed\footnote{However, a detailed study of two-loop
self-energies~\cite{inp} shows that, in these cases, numerical 
differentiation can be avoided in favor of alternative realizations of the
{\em raising} procedure}.
\subsubsection{Special cases and numerical results\label{sc}}
The whole algorithm presented in the previous sections has been translated 
into a FORM code~\cite{Vermaseren:2000nd} and the output has been used to 
create a FORTRAN code\footnote{FORTRAN may be archaic but never leaves you in 
solitude.} that computes $S_3$. Some of the results are shown in
the following sub-section where they are compared with the analytical results 
that are available in the literature for some special cases.
When some of the masses are zero the result for $S_3$ simplifies. For $m_1 = 0$
we obtain
\bq
S_3\lpar p^2;0,m_2,m_3\rpar = 
m^2_2\,S_{32}\lpar p^2;0,m_3,m_2\rpar + m^2_3\,S_{33}\lpar p^2;0,m_2,m_3\rpar 
+ S_{3p}\lpar p^2;0,m_3,m_2\rpar.
\label{zmass}
\eq
For $m_2 = 0$ or $m_3 = 0$ we use \eqn{zmass} and the following properties:
\bq
S_3\lpar p^2;m_1,m_2,m_3\rpar = S_3\lpar p^2;m_2,m_1,m_3\rpar  
= S_3\lpar p^2;m_3,m_2,m_1\rpar.
\eq
Finally, if two masses are zero we use the appropriate permutation and
\bq
S_3\lpar p^2;0,m_2,0\rpar =  m^2_2\,
S_{32}\lpar p^2;0,0,m_2\rpar + S_{3p}\lpar p^2;0,0,m_2\rpar.
\eq
For $m_1 = m_3 = 0$ we have an analytical result~\cite{Berends:1996ev}
(see also ref.~\cite{Caffo:1998du}).
With
\bq
z = \frac{s}{m^2} + i\,\delta, \qquad \Delta_{\ssU\ssV}(z)= 
\Delta_{\ssU\ssV} - \ln z, 
\eq
we derive the real and imaginary parts of $S_3$:
\bqa
\Reb S_3(p^2;0,m,0) &=&  m^2\,\Bigl[ \frac{2}{\ep^2} + \frac{1}{\ep}\,
   \lpar 3 - \frac{1}{2} z - 2 \Delta_{\ssU\ssV} +  2\,\ln z\rpar   
+ \frac{1}{2}\,\lpar z - \frac{1}{z}\rpar\,\ln\mid z-1\mid  \nl
{}&+& \Reb\,\li{2}{z} + 3 
+ \frac{1}{2}\,z\,\Delta_{\ssU\ssV}(z)  - \frac{13}{8}\,z  
+ \Delta^2_{\ssU\ssV}(z) - 3\, \Delta_{\ssU\ssV}(z) + 
\frac{3}{2}\,\zeta(2)  \Bigr],  \nl
{}&{}&  \nl
\Imb S_3(p^2;0,m,0) &=& m^2\,\Imb\,\Bigl[ \frac{1}{2}\,\lpar z - 
\frac{1}{z}\rpar\,\ln\lpar 1 - z \rpar + \li{2}{z}\Bigr].
\label{banal}
\eqa
For the numerical evaluation we have used the NAG~\cite{naglib} Fortran 
library subroutine D01EAF that integrates a vector of similar functions 
(e.g.\ real and imaginary parts of the scalar diagram) over the same 
multi-dimensional hyper-rectangular region. In certain regions, e.g.\
around thresholds, one obtains moderately accurate results and for these cases
we switch to the routine D01GDF, a multi-dimensional quadrature for general 
product regions, using number-theoretic methods.   

A detailed comparison is shown between the numerical approach based on the
minimal BT-algorithm and the analytical result of \eqn{banal} in 
\tabn{bacomp}. There we use $m_1 = m_3 = 0$, $m_2 = 1.765\,$GeV and
moreover we select $\ep = 1$ and $\tHs= 1\,$GeV. Note that there is no loss
of numerical precision near, but not too near, the normal thresholds.
\begin{table}[ht]\centering
\begin{tabular}{|l|l|l|l|l|l|}
\hline
 & & & & &\\
$\sqrt{s}\,$[GeV]  &  $\Reb\,S_3$  & $\Imb\,S_3$ & 
$\sqrt{s}\,$[GeV]  &  $\Reb\,S_3$  & $\Imb\,S_3$  \\
 & & & & &\\
\hline
 & & & & &\\
0.5 &    15.055283(146) &      0  & 4   &    18.534988(176) &     
-8.249166(53)\\
    &    15.055293      &      0  &     &    18.534996      &     
-8.249165\\
 & & & & &\\
1   &    15.440106(120) &      0  & 6   &    31.015475(323) &    
-32.328029(149)\\
    &    15.440114      &      0  &     &    31.015477      &    
-32.328028\\
 & & & & &\\
1.5 &    15.996235(151) &      0  & 8   &    59.575707(705)&    
-70.918158(379)\\
    &    15.996239      &      0  &     &    59.575722      & 
   -70.918157\\
 & & & & &\\
1.7 &    16.228067(165) &      0  & 10  &   106.89072(123)  &   
-123.22756(69)\\
    &    16.228068      &      0  &     &   106.89073       &   
-123.22756\\
 & & & & &\\
1.8 &    16.333331(172) &     -0.000196(4) & 50  &  6572.7884(387)   &  
-3862.1006(247)\\
    &    16.333332      &     -0.000196    &     &  6572.7846        &  
-3862.1002\\
 & & & & &\\
2   &    16.491314(184) &     -0.028545(7) & 100 & 33332.656(159)    & 
-15629.648(102)\\
    &    16.491316      &     -0.028545    &     & 33332.642         & 
-15629.648\\
 & & & & &\\
\hline
\hline
\end{tabular}
\vspace*{3mm}
\caption[]{The sunset topology $S_3$ for $m_1 = m_3 = 0$ and $m_2 = 
1.756\,$GeV as a function of $\sqrt{s}$. First entry is the minimal 
BT-approach, second entry is the analytical result of \eqn{banal}.
The UV-pole is $\ep = 1$ and the unit of mass is $\tHs = 1\,$GeV.}
\label{bacomp}
\end{table}
\normalsize
The requested precision for the integration routine is of a relative error of
$10^{-5}$. Comparison with the analytical result shows that the returned
numerical error is overestimated with a very good agreement over a wide
range of values for $\sqrt{s}$.

To investigate in more detail the situation around $s = m^2_2$ we have 
produced \tabn{bacompt}.
\begin{table}[ht]\centering
\begin{tabular}{|l|l|l|l|l|l|}
\hline
 & & & & & \\
$\sqrt{s}\,$[GeV]  &  $\Reb\,S_3$  & $\Imb\,S_3$  &
$\sqrt{s}\,$[GeV]  &  $\Reb\,S_3$  & $\Imb\,S_3$  \\
 & & & & & \\
\hline
 & & & & & \\
1.732 & 16.26311(2)  & 0  & 1.762 & 16.29475(1238)  & *  \\
      & 16.26311     & 0  &       & 16.29495        & -0.51$\times 10^{-6}$ \\
 & & & & & \\
\hline
 & & & & & \\
1.740 & 16.27171(10) & 0  & 1.772 & 16.30528(22)    & -0.9(5)$\times 10^{-5}$ \\
      & 16.27171     & 0  &       & 16.30529        & -0.96$\times 10^{-5}$ \\
 & & & & & \\
\hline
 & & & & & \\
1.750 & 16.28075(13643) & 0  & 1.780 & 16.31344(4)  & -0.324(8)$\times 10^{-4}$  \\
      & 16.28235        & 0  &       & 16.31344     & -0.323$\times 10^{-4}$  \\
 & & & & & \\
\hline
\hline
\end{tabular}
\vspace*{3mm}
\caption[]{A scan of the sunset, $S_3$, topology for $m_1 = m_3 = 0$ and 
$m_2 = 1.756\,$GeV around the normal threshold $s = (m_1+m_2+m_3)^2$. First 
entry is the minimal BT-approach, second entry is the analytical result of 
\eqn{banal}. A * indicates poor numerical convergence in the numerical 
evaluation. Note, however, that this table is only for illustration and,
in this region, one should always use \eqn{simple} (see \tabn{thresh}). The 
UV-pole is $\ep = 1$ and the unit of mass is $\tHs = 1\,$GeV.}
\label{bacompt}
\end{table}
\normalsize
From it we see that by approaching $\sqrt{s} = m_2$ up to
$1-\sqrt{s}/m_2 = 3.4 \times 10^{-3}$ we register an increasing numerical
error on the real part, although the relative deviation from the analytical 
result is better than $10^{-4}$ signalling, once more, the conservative
estimate of the numerical error. For the imaginary part, immediately after
threshold, the numerical convergence is poor in a situation where the 
result is very small. At $1.4 \times 10^{-2}$ away form threshold, however,
we register again an excellent agreement.

As explained in \sect{nt}, however, around the normal threshold we can produce
an accurate result by using \eqn{simple}. Results for a scan around and at
threshold are shown in \tabn{thresh}.
\begin{table}[ht]\centering
\begin{tabular}{|l|l|l|l|l|l|}
\hline
 & & & & & \\
$\sqrt{s}\,$[GeV]  &  $\Reb\,S_3$  & $\Imb\,S_3$  &
$\sqrt{s}\,$[GeV]  &  $\Reb\,S_3$  & $\Imb\,S_3$  \\
 & & & & & \\
\hline
 & & & & & \\
1.732 & 16.2631072(1) & 0  & 1.762 & 16.2949492(1)  & 
-0.512620701(1)$\times 10^{-6}$  \\
      & 16.2631072    & 0  &       & 16.2949492     & 
-0.512620700$\times 10^{-6}$ \\
 & & & & & \\
\hline
 & & & & & \\
1.740 & 16.2717059(1) & 0  & 1.772 &  16.3052845(1)  & 
-0.963868786(1)$\times 10^{-5}$ \\
      & 16.2717059    & 0  &       &  16.3052845     & 
-0.963868786$\times 10^{-5}$ \\
 & & & & & \\
\hline
 & & & & & \\
1.750 & 16.2823493(1) & 0  & 1.780 & 16.3134359(1) & 
-0.323117581(1)$\times 10^{-4}$  \\
      & 16.2823493    & 0  &       & 16.3134359    & 
-0.323117581$\times 10^{-4}$  \\
 & & & & & \\
\hline
 & & & & & \\
\mbox{threshold} &  16.2886745(1) & 0 &&& \\  
                 &  16.2886745    & 0 &&& \\
 & & & & & \\
\hline
\hline
\end{tabular}
\vspace*{3mm}
\caption[]{The same as in \tabn{bacompt} but using \eqn{simple}. Integration 
is performed with the NAG subroutine D01GDF.}
\label{thresh}
\end{table}
\normalsize
Another comparison concerns the imaginary part of $S_3$ which is finite in four
dimensions and can be derived analytically by reduction to Legendre form
of elliptic integrals. The complete expression is given in 
ref.~\cite{Berends:1996ev} and, here, we only give a detailed comparison.
\begin{table}[ht]\centering
\begin{tabular}{|c|c|c|c|c|c|}
\hline
 & & & & & \\
$\sqrt{s}\,$[GeV]  &  $\Imb\,S_3$  & $\Imb\,S_3$ (analytical) &
$\sqrt{s}\,$[GeV]  &  $\Imb\,S_3$  & $\Imb\,S_3$ (analytical) \\
 & & & & & \\
\hline
 & & & & & \\
255 &     -32.038(253)    &     -32.044 & 295 &   -6464.878(259)    &   -6464.877\\
    &     -32.046(237)    &             &     &   -6464.404(517)    & \\
 & & & & & \\
\hline
 & & & & & \\
265 &    -616.707(256)    &    -616.717 & 300 &   -7995.983(258)    &   -7995.056\\
    &    -617.210(201)    &             &     &   -7995.226(164)    &  \\
 & & & & & \\
\hline
 & & & & & \\
275 &   -1906.548(257)    &   -1906.560 & 400 &  -66930.101(220)    &  -66930.105\\
    &   -1906.891(327)    &             &     &  -66930.175(418)    &  \\
 & & & & & \\
\hline
 & & & & & \\
285 &   -3866.374(258)    &   -3866.289 & 500 & -171686.26(17)      & -171686.23\\
    &   -3866.388(421)    &             &     & -171684.32(67)      & \\
 & & & & & \\
\hline
\hline
\end{tabular}
\vspace*{3mm}
\caption[]{$\Imb\,S_3$, the imaginary part of the sunset topology, for 
$m_1 = m_3 = 80.448\,$GeV and $m_2 = 91.1875\,$GeV as a function of $\sqrt{s}$.
First entry is the minimal BT-approach, first row integrated with D01EAF, 
second row with D01GDF and \eqn{simple}; second entry is the analytical 
result of ref.~\cite{Berends:1996ev}.}
\label{icomp}
\end{table}
\normalsize
The result is shown in \tabn{icomp} where we compare two different methods
of numerical integration. We find better and better agreement
away from thresholds with a numerical error that, in general, overestimate 
the difference with the analytical result. Indeed, near threshold, we have
a relative difference of $0.017\%(0.005\%)$ with a returned numerical error of 
$0.79\%(0.74\%)$.

Finally, following the suggestion of ref.~\cite{Berends:1996ev}, we introduce
a special combination:
\bqa
S_{\rm c} &=& S_3\lpar p^2;m_1,m_2,m_3\rpar -
S_3\lpar p^2;m_1,0,m_3\rpar  \nl
{}&-& S_3\lpar p^2;0,m_2,m_3\rpar -
S_3\lpar p^2;0,0,m_3\rpar,
\label{defc}
\eqa
where infinite parts cancel. For small $\mid p^2\mid$, essentially for
$\mid p^2\mid < m^2_3$, the combination $S_{\rm c}$ is known in the
form of a multiple series in the variables
\bq
z_1 = \frac{m^2_1}{m^2_3}, \qquad
z_2 = \frac{m^2_2}{m^2_3}, \qquad
z_3 = \frac{s}{m^2_3}, 
\eq
with $s = - p^2$. The comparison is shown in \tabn{mscomp} for $p^2$
space-like or time-like, showing excellent agreement with the analytical
result derived from the expansion of a Lauricella function.
\begin{table}[ht]\centering
\begin{tabular}{|r|c|c|c|}
\hline
 & & & \\
$s\,$[GeV$^2$]  &  $\Reb\,S_{\rm c}$  & $\Reb\,S_{\rm c}$ (multiple series) & 
$10^5\,\times\,$ Dev. \\
 & & & \\
\hline
 & & & \\
$-$ 1  &   -70.686011  & -70.685698  & 0.44  \\
$+$ 1  &   -70.680106  & -70.679831  & 0.39  \\
 & & & \\
\hline
 & & & \\
$-$ 25 &   -70.756299  & -70.756203  & 0.14  \\
$+$ 25 &   -70.609231  & -70.609519  & 0.41  \\
 & & & \\
\hline
 & & & \\
$-$ 81 &   -70.921481  & -70.921413  & 0.10  \\
$+$ 81 &   -70.446044  & -70.446147  & 0.15  \\
 & & & \\
\hline
\hline
\end{tabular}
\vspace*{3mm}
\caption[]{$\Reb\,S_{\rm c}$ (see \eqn{defc} for the definition) for 
$m_1 = 10\,$GeV, $m_2 = 20\,$GeV and $m_3 = 100\,$GeV as a function of $p^2$. 
First entry is the minimal BT-approach, second entry is the analytical 
result of ref.~\cite{Berends:1996ev}.
The UV-pole is $\ep = 1$ and the unit of mass is $\tHs = 1\,$GeV.}
\label{mscomp}
\end{table}
\normalsize
\subsection{Tensor integrals\label{ti}}
Tensor integrals of the type
\bq
\lpar \mu_1\dots \mu_i\mid \nu_1\dots \nu_j\mid 1,m_1\mid 1,m_2,p
\mid 1,m_3\rpar,
\eq
are usually decomposed into scalar integrals. For a general discussion we 
refer to the work of ref.~\cite{Ghinculov:2001cz} where the following functions
are introduced:
\bq
(\ep-1){\cal P}^{ij}_{112} = \Bigl( 
{\underbrace{p\dots p}_i} 
\mid 
{\underbrace{p\dots p}_j}
\mid 1,m_1\mid 1,m_2,p\mid 2,m_3\Bigr) = \Bigl(i,p\mid j,p\mid
1,m_1\mid 1,m_2,p\mid 2,m_3\Bigr),
\eq
where $i+j \le 3$. It is easily shown that these integrals admit the following
representation:
\bq
{\cal P}^{ij}_{112} = \lpar\frac{\tHss}{\pi}\rpar^{\ep}
\,\frac{\egam{\ep}}{\ep-1}\,\sum_{n=-1}^{+1}\,
\intfxx{x}{y}\,\Bigl[x(1-x)\Bigr]^{-\ep/2}\,y\,(1-y)^{\ep/2}\,
Q^{ij}_n(x,y,\ep)\,U^{n-\ep}(x,y),
\eq
where $U$ is given in \eqn{defU}. Therefore, to each of these functions 
we apply the minimal BT-algorithm and derive expressions similar to those
obtained for $S_{33} = {\cal P}^{00}_{112}$. The polynomials $Q$ are
as follows:
\bqa
Q^{00}_{-1} &=& 2\,( m^2_x + y^2 p^2),  \quad
Q^{00}_0     = \frac{4}{\ep} - 1,  \quad
Q^{00}_1     = 0,  \nl
\nl
Q^{10}_{-1} &=& - 2\,x y p^2 ( m^2_x + y^2 p^2),  \quad
Q^{10}_0     = x y p^2 (1 - \frac{6}{\ep}),  \quad
Q^{10}_1     = 0,  \nl
\nl
Q^{20}_{-1} &=& 2\,x^2 y^2 p^4 (m^2_x + y^2 p^2) +
                2\,x (1-x) \frac{p^2}{\ep-2} ( m^2_x + y^2 p^2 )^2,  \nl
Q^{20}_0    &=& x^2 p^2 ( \frac{m^2_x + 9\,y^2 p^2}{\ep} - y^2 p^2) +
                2\,x (1-x) p^2 ( \frac{-2\,m^2_x-3\,y^2 p^2}{\ep} +
                \frac{m^2_x+2\,y^2 p^2}{\ep-2}),  \nl
Q^{20}_1    &=& x^2 p^2 (-\frac{3}{\ep} + \frac{5}{2}\,\frac{1}{\ep-1}) +
                x (1-x) p^2 ( \frac{6}{\ep} + \frac{2}{\ep-2} - \frac{15}{2}\,
                \frac{1}{\ep-1}),  \nl
\nl
Q^{30}_{-1} &=& -2\,x^3 y^3 p^6 (  m^2_x + y^2 p^2 )
       - 6\,x^2 (1-x) y \frac{p^4}{\ep-2} ( m^2_x + y^2 p^2 )^2,  \nl
Q^{30}_0    &=& x^3 y p^4 ( \frac{3\,m^2_x+13\,y^2 p^2}{\ep} + y^2 p^2 )
       + x^2 (1-x) y p^4 ( 6\,\frac{3\,m^2_x+4\,y^2 p^2}{\ep}
         - 6\,\frac{2\,m^2_x+3\,y^2 p^2}{\ep-2} ),  \nl
Q^{30}_1    &=& x^3 y p^4 \Bigl[ \frac{9}{\ep} - \frac{15}{2}\,\frac{1}{\ep-1}
          + \frac{3}{\ep+4} ( \frac{4}{\ep} \frac{5}{\ep-1} )\Bigr]
       + x^2 (1-x) y p^4 (  - 3\frac{36}{\ep} - \frac{18}{\ep-2} + 
          \frac{105}{2}\,\frac{1}{\ep-1} ),  \nl
\nl
Q^{01}_{-1} &=& 2 (1-y) p^2  (  m^2_x +  y^2 p^2 ),  \quad
Q^{01}_0     = p^2  ( 2\,\frac{2-3y}{\ep} - 1 + y  ),  \quad
Q^{01}_1     = 0,  \nl
\nl
Q^{02}_{-1} &=& 2 p^4 (1-y)^2 ( m^2_x + y^2 p^2),  \nl
Q^{02}_0    &=& p^2 \Bigl[ \frac{m^2_x+p^2(4-12\,y+9\,y^2)}{\ep} -
                p^2 (1-y)^2\Bigr],  \nl
Q^{02}_1    &=& p^2 ( -\frac{3}{\ep} + \frac{5}{2}\,\frac{1}{\ep-1} ),  \nl
\nl
Q^{03}_{-1} &=& 2\,p^6 ( 1 - y )^3 ( m^2_x + y^2 p^2)  \nl
Q^{03}_0    &=& p^4 \Bigl[ \frac{3\,(1-y) m^2_x + ( 4 - 18\,y + 27\,y^2 -
                          13\,y^3) p^2}{\ep} - (1-y)^2 p^2\Bigr],  \nl
Q^{03}_1    &=& p^4 \Bigl[ -9\,\frac{1-y}{\ep} + \frac{15}{2}\,
                \frac{1-y}{\ep-1} + \frac{3}{\ep+4}\,(\frac{4}{\ep} -
                \frac{5}{\ep-1}\Bigr],  \nl            
\nl
Q^{11}_{-1} &=& - 2\,x y (1-y) p^4 ( m^2_x + y^2 p^2 ),  \nl
Q^{11}_0    &=&  x p^2 \Bigl[ 3\,\frac{2\,m^2_x - 2\,y p^2 + 3\,y^2p^2}{\ep} +
                 y (1-y) p^2\Bigr],  \nl
Q^{11}_1    &=& x p^2 ( -\,\frac{3}{\ep} + \frac{5}{2}\,\frac{1}{\ep-1}),  \nl
\nl
Q^{21}_{-1} &=& 2 x^2 y^2 (1-y) p^6 ( m^2_x + y^2 p^2 ) + 
                2 x (1-x) (1-y) \frac{(m^2_x + y^2 p^2)^2}{\ep-2},  \nl
Q^{21}_0    &=& x^2 p^4 \Bigl[ \frac{(1-3\,y) m^2_x + y^2 p^2 
                (9 - 13\,y)}{\ep} - y^2 (1-y) p^2 \Bigr] \nl
{}&+&           x ( 1-x ) \Bigl[ \frac{2\,(3\,y-2) m^2_x + 2\,y^2 p^2
                4\,y-3)}{\ep} + \frac{2\,(1-2\,y) m^2_x + 2\,y^2 p^2
                (2-3\,y)}{\ep-2}\Bigr],  \nl
Q^{21}_1    &=& x^2 p^4 \Bigl[ \frac{3\,(3\,y-1)}{\ep} + \frac{5}{2}\,
                \frac{1-3\,y}{\ep-1} + 3\,\frac{1}{\ep+4} ( \frac{4\,y}{\ep} -
                \frac{5\,y}{\ep-1}\Bigr] \nl
{}&+&           x (1-x) ( 1-3\,y) \Bigl[ 6\,\frac{1}{\ep} + \frac{2}{\ep-2} -
                \frac{15}{2}\,\frac{1}{\ep-1}\Bigr],  \nl
\nl
Q^{12}_{-1} &=& -\, 2 x y (1-y)^2 p^6 ( m^2_x + y^2 p^2 ),  \nl
Q^{12}_0    &=& x p^4 \Bigl[ \frac{(2-3\,y) m^2_x - y (6-18\,y+13\,y^2) p^2}
                {\ep} + y (1-y)^2 p^2\Bigr],  \nl
Q^{12}_1    &=& x p^4 \Bigl[ \frac{-3\,(3\,y-2)}{\ep} + 5\,(1-\frac{3}{2})
                \frac{1}{\ep-1} + \frac{3}{\ep+4} (\frac{4}{\ep} -
                \frac{5}{\ep-1} )\Bigr].
\eqa
In our approach we avoid using derivatives with respect to the external
parameters since the poor reputation enjoyed by numerical differentiation leads
to search for techniques which avoid the explicit use of it.
Conversely our numerical integration is usually performed with the help of the
NAG~\cite{naglib} Fortran library subroutine D01EAF that computes 
approximations to the integrals of a vector of similar functions over the 
same multi-dimensional hyper-rectangular region. Therefore all relevant 
integrals, connected to the $S_3$ topology (any topology), are computed in 
one stroke without any noticeable loss of CPU time with respect to computing 
the scalar integral alone.
Therefore we introduce additional tensor integrals of the following form:
\bqa
(\ep-1)\,{\cal P}^{1|j}_{112} &=& \Bigl( 
\mu\mu
\mid 
{\underbrace{p\dots p}_j}
\mid 1,m_1\mid 1,m_2,p\mid 2,m_3\Bigr),  \nl
(\ep-1)\,{\cal P}^{2|i}_{112} &=& \Bigl( 
{\underbrace{p\dots p}_i}
\mid 
\mu\mu
\mid 1,m_1\mid 1,m_2,p\mid 2,m_3\Bigr),
\eqa
where $i,j = 0,1$. They also can be cast into the form
\bq
{\cal P}^{1,2|i}_{112} = \lpar\frac{\tHss}{\pi}\rpar^{\ep}
\,\frac{\egam{\ep}}{\ep-1}\,\sum_{n=-1}^{+1}\,
\intfxx{x}{y}\,\Bigl[x(1-x)\Bigr]^{-\ep/2}\,y\,(1-y)^{\ep/2}\,
Q^{1,2|i}_n(x,y,\ep)\,U^{n-\ep}(x,y),
\eq
with corresponding polynomials that will not presented in this paper.
Other integrals can be easily derived by applying 
$\,\lpar\alpha\beta\gamma\rpar$ partial integration, for instance
\bqa
{}&{}&
\Bigl(i,p\mid j,p\mid 1,m_1\mid 1,m_2,p\mid 1,m_3\Bigr) =  
\frac{1}{\ep - 1 - (i+j)/2}\,\Bigl[  \nl
{}&{}& m^2_1\,\Bigl(i,p\mid j,p\mid 2,m_1\mid 1,m_2,p\mid 1,m_3\Bigr) +  
(p^2+m^2_2)\,\Bigl(i,p\mid j,p\mid 1,m_1\mid 2,m_2,p\mid 1,m_3\Bigr) +  \nl
{}&{}&
m^2_3\,\Bigl(i,p\mid j,p\mid 1,m_1\mid 1,m_2,p\mid 2,m_3\Bigr) +  
\Bigl(i+1,p\mid j,p\mid 1,m_1\mid 2,m_2,p\mid 1,m_3\Bigr) +  \nl
{}&{}&
\Bigl(i,p\mid j+1,p\mid 1,m_1\mid 2,m_2,p\mid 1,m_3\Bigr)\Bigr].
\eqa
The tensor integrals are easily computed and, once the method will be
extended to cover all two-loop topologies, one may start organizing
a realistic calculation.
However, in any realistic calculation, the general strategy will be
substantially different form the one usually adopted in one-loop
exercises. It will be more convenient to write a Form code for any
two-loop diagram $G$, or set of diagrams $\sum_jG_j$, that produces one 
expression of the form
\bq
\int_{\ssS}\,dx\,\sum_{ij}\,Q_{ij}(x)\,V_{\ssG_j}^{-\mu_j}(x),
\eq
and to apply the method directly to this integral instead of applying first 
the tensor decomposition and then to compute, separately, the various $ij$
components of the total result. In a way one looses some attractive
feature of the analytical one-loop approach, where few universal building
blocks are needed and everything is algebraically reduced to those blocks.
But, again, this is a typical example of a procedure motivated by our ability 
to derive analytical results. The perspective, here, is radically different,
see ref.~\cite{Tkachov:1997wh} for an epistemological discussion.
\section{Conclusions}
In this paper we have considered the Bernstein-Tkachov algorithm, proposed 
in~\cite{Tkachov:1997wh}, which proves of some use in dealing with
fast and accurate numerical evaluation of arbitrary multi-loop Feynman 
diagrams.

As a prelude, we have applied the BT algorithm to one-loop diagrams with an
arbitrary number of external legs, where reliable analytical results have been
known for a long time, to show the feasibility of a project based on this new
strategy and to discuss, with simple examples, the infrared divergent cases 
and the new approach that one will have to adopt in dealing with reduction of 
tensor integrals. 

After this brief excursus in the one-loop world we have considered the 
application of the algorithm to genuine multi-loop diagrams. Vacuum and
tadpole diagrams and two-loop diagrams at zero external momentum allow
for a full implementation of the method but, in general, we have not
succeeded in deriving the basic ingredient, i.e. \ the polynomial {\cal P}
of \eqn{functr}\footnote{To the best of our knowledge no example
is known in the literature beyond the one-loop case of~\cite{Tkachov:1997wh}.
Working with some unrealistic case S.~Uccirati has shown how to deal with
cubic polynomials in two variables where, however, the coefficient $B$
of \eqn{functr} contains approximately $3\,\times\,10^4$ terms.}.
Therefore, we have introduced a variant of the original proposal, a minimal 
BT-approach, which can be described as the convolution of the Ghinculov and 
Yao techniques, especially the use of complex Feynman parameters, with the 
basic BT-algorithm of {\em raising} powers in Feynman integrands. The main 
idea is related to the simple observation that we know how to apply the 
BT-iterative procedure of {\em raising} powers for an arbitrary one-loop 
diagram. Therefore, given any two-loop diagram $G$ we minimally apply BT 
functional relation to the one-loop sub-diagram $L\,\in\,G$ which has the 
largest number of internal lines. 
In this way the integrand can be made smooth, a part from the factor $B$ of 
\eqn{functr} which is now a polynomial in $x_{\ssS}$, the set of Feynman 
parameters needed for the complementary one-loop sub-diagram $S\,\in\,G$
with the smallest number of internal lines. Since the BT procedure does not 
introduce singularities through $B$, a part from the singularities of $G$ 
itself in the external parameter space, before performing the 
$x_{\ssS}$-integration we move the contour into the complex hyper-plane, 
thus avoiding the crossing of apparent singularities. This is the procedure 
introduced in~\cite{Ghinculov:2001cz}.

The minimal BT-approach has been applied to the sunset two-loop diagram
and numerical results are shown, comparing them with the relevant literature.
Whenever available, published results for the sunset show excellent agreement 
with our method which is fast and accurate in all regions, time-like or
space-like external momentum, around normal thresholds and pseudo-thresholds.

However, when we are close enough to the normal threshold the integration
contour cannot be distorted and numerical instabilities appear. 

A preliminar analysis of other two-loop diagrams leads us to formulate the 
conjecture that these numerical instabilities are always connected with the 
leading Landau singularity of the original diagram.

In these regions we have found more convenient to apply another algorithm 
for {\em rasing} powers which gives a very simple result, although with 
discontinuous imaginary part of the integrand. Therefore, very accurate 
predictions can be made also at the normal threshold. The extension of this 
variant of the {\em rasing} procedure to other two-loop topologies may require 
the introduction of numerical differentiation with respect to masses which 
is, in general, an unsatisfactory branch of numerical analysis. However, 
close to normal threshold, this alternative procedure represents the safest 
way for a numerical evaluation of diagrams, unless one want to adopt specific 
expansions that can be found in the literature.
A preliminar analysis~\cite{inp} has shown that, for two-point functions and
one iteration, a transformation of the Feynman parameters can always be found 
that produces a coefficient $B$ $x_{\ssS}$-independent. This $B$ will vanish 
at some non-leading Landau singularity of the diagram where additional 
analytical work is needed before starting the numerical evaluation.

Work is in progress to extend the method to all two-loop, two-(three-)point
topologies and we plan to extend it to all two-loop diagrams with the evident
goal of performing a complete two-loop analysis of the standard model 
predictions, in particular of quantities like $\seffsf{l}$ at the level
of $1\,\times\,10^{-6}$ theoretical precision. 

\Acknowledgments

I would like to express my gratitude to Fyodor Tkachov for several discussions
on the general idea of evaluating multi-loop Feynman diagrams numerically.
I sincerely thank Dmitri Bardin for being stronger than my stubbornness
and for finally succeeding in drawing my full attention to the work of Fyodor
Tkachov during several walks in Dubna, in Summer 2000. The contribution of
Sandro Uccirati for evaluating pentagon and hexagon one-loop integrals
is also acknowledged.

\clearpage

\section{Appendix A}
In this Appendix we give details about the analytical derivation of double and
single UV poles for $S_{33}$, defined in \eqn{s3cont}. First we use the 
Laurent expansion of $\Gamma(z)$ around $z = 0$,
\bq
\egam{z}= \frac{1}{z}-\gamma + \frac{1}{2}\,\,\Bigl[ \gamma^2 + \zeta(2)\Bigr] x -
\frac{1}{6}\,\,\Bigl[ \gamma^3 + 3\,\zeta(2) + 2\,\zeta(3)\Bigr]\,z^2 +
\ord{z^3}.
\eq
Next we introduce
\bq
\chi(x,y) = {{\xi(x,y)}\over {x\,(1-x)}},
\eq
and expand in $\ep$ the various terms as follows:
\bqa
\,\lpar 1 - y\rpar^{\ep/2} &=& 1 + \frac{\ep}{2}\,\ln(1-y) + \frac{\ep^2}{8}\,
\ln^2(1-y) + \ord{\ep^3},  \nl
\,\Bigl[ x\,(1-x)\Bigr]^{l+\ep/2} &=& \,\Bigl[ x\,(1-x)\Bigr]^l\,\,\Bigl[
1 + \frac{\ep}{2}\,\ln\,\lpar x - x^2\rpar + \frac{\ep^2}{8}\,\ln^2
\,\lpar x - x^2\rpar + \ord{\ep^3}\Bigr],  \nl
\xi^{-1-\ep} &=& \xi^{-1}\,\,\Bigl[ 1 - \ep\,\ln\xi +
\frac{\ep^2}{2}\,\ln^2\xi + \ord{\ep^3}\Bigr].
\eqa
Some of the integrals can be performed directly by using the following 
relations:
\bqa
\intfx{y}\,y^n\,\ln^2(1-y) &=& \intfx{y}\,(1-y)^2\,\ln^2 y,  \nl
\intfx{y}\,y^n\,\ln^2 y &=& \frac{2}{(n+1)^3},  \nl
\intfx{y}\,y^n\,\ln (1-y) &=& -\,\frac{1}{n+1}\,\sum_{j=1}^{n+1}\,\frac{1}{j},
\nl
\intfx{y}\,y^n &=& \frac{1}{n+1},  \nl
\intfx{x}\,x^n\,\ln x\,\ln (1-x) &=& \frac{1}{n+1}\,\sum_{j=1}^{n+1}\,
\frac{1}{j}\,\lpar \frac{1}{n+1} + \frac{1}{j}\rpar - \frac{\zeta(2)}{n+1}.
\eqa
In this way we arrive at the following expression for $S_{33}$:
\bqa
S_{33} &=& \frac{2}{\ep^2} - \frac{2}{\ep}\,
\,\lpar 1 + \gamma + \ln\pi + \ln\frac{s}{\tHss}\rpar  
+ \,\lpar \gamma + \ln\pi\rpar^2 \nl
{}&+& \ln\frac{s}{\tHss}\,\lpar 2 + 2\,\gamma +
\ln\frac{s}{\tHss}\rpar + 2\,\ln\pi\,\lpar 1 + \ln\frac{s}{\tHss}\rpar + 2\,\gamma  \nl
{}&+& \intfxy{x}{y}\,\xi^{-1}\,
\,\lpar \frac{2}{\ep} - \gamma - \ln\pi - \ln\frac{s}{\tHss}\rpar
\,\,\Bigl[ -\,x y^3 (1-x) + y M^2_x\Bigr]  \nl
{}&+& \intfxy{x}{y}\,\xi^{-1}\,\,\Bigl[2 + \ln\,\lpar 1 - y\rpar\, +
\ln\,\lpar x - x^2\rpar\, - 2\,\ln\xi\Bigr]\,
\,\,\Bigl[ -\,x y^3 (1-x) + y M^2_x\Bigr]  \nl
{}&-& 4\,\intfxy{x}{y}\,y\,\ln\xi\,\lpar
\frac{1}{\ep} - \gamma - \lpi - \ln\frac{s}{\tHss}\rpar  \nl
{}&+& 2\,\intfxy{x}{y}\,y\,\,\Bigl[ \ln^2\xi - \ln\,\lpar 1 - y\rpar -
\ln\,\lpar x - x^2\rpar - \frac{3}{2}\,\ln\xi\Bigr]  \nl
{}&+& \frac{13}{4} + \frac{\zeta(2)}{2}.
\eqa
here we have introduced
\bq
M^2_x = (1-x)\,\mu^2_1 + x\,\mu^2_2 = x\,\lpar 1 - x\rpar\,\mu^2_x.
\eq
next we restore $\xi = x\,\lpar 1 - x\rpar\,\chi$ and make use of the relation
\eqn{trel}. After integration by parts, i.e.\
\bq
\intfx{y}\,\frac{y^n}{\chi}\,\frac{\partial}{\partial y}\,\chi =
\ln \mu^2_3 - n\,\intfx{y}\,y^{n-1}\,\ln\chi,
\eq
the final result, \eqns{DUV}{sppart}, follows.
\section{Appendix B}
In this appendix we consider the case $p^2 = -t$, with $t < 0$.
Since in this case
\bq
U(x,y) = - y\,\lpar 1 - y\rpar\,t + y\,m^2_3 + \,\lpar 1 - y\rpar\,m^2_x,
\eq
the quadratic form $\chi$, for $t$-channel diagrams is redefined as
\bq
U = -\,t\chi, \qquad
\chi = y\,\lpar 1 - y\rpar + y\,\mu^2_3 + \,\lpar 1 - y\rpar\,\mu^2_x,
\eq
with $\mu^2 = -m^2/t > 0$. Furthermore, when {\em raising} the power $\mu$
in $\chi^{\mu}$ for the $t$-channel case, we have to change coefficients
accordingly:
\bq
H = - 1, \qquad K = \frac{1}{2}\,\lpar 1 + \mu^2_3 - \mu^2_x\rpar, \qquad
L = \mu^2_x,
\eq
which gives
\bq
B = \frac{1}{4}\,\lambda\,\lpar 1,-\mu^2_3,-\mu^2_x\rpar, \qquad 
Y = - \frac{1}{2}\,\lpar 1 + \mu^2_3 - \mu^2_x\rpar,
\eq
and one should replace
\bq
\ln\frac{s}{\tHss} \to \ln\frac{-t}{\tHss}
\eq
in the pole terms. 
\section{Appendix C}
In this Appendix we give the explicit integrand for $S_{33}$, the other 
$S_{3i}$ terms following through the permutations of \eqn{permut}. 
First we introduce auxiliary variables
\bqa
X &=& x\,\lpar 1- x\rpar,  \qquad
A = X\,\lpar 1 - \mu^2_3\rpar + B,  \nl
B &=& X\,\mu^2_x, \qquad
\xi(x,y) = X\,\chi(x,y).
\eqa
The `+'-distribution is defined, as usual, by its action on a generic test
function $g(x)$:
\bq
\intfx{x}\,g(x)f_+(x) = \intfx{x}\,\,\Bigl[ g(x) - g(1)\Bigr]\,f(x).
\label{plusd}
\eq
We also use the notations
\bq
\ln_+\xi(x,y)= \ln\xi(x,y) - \ln\xi(x,1), \quad L_x = \ln X, \quad
L_y = \ln(1-y).
\eq
Then we derive the following expressions:
\bq
S_{33} = S^{\rm dp}_{33} + S^{\rm sp}_{33} + S^{\rm fin}_{33},  \quad
S^{\rm fin}_{33} = \frac{1}{\Lambda}\,\Bigl[ \frac{I_2}{\Lambda} +
I_1\Bigr] + I_0,  \quad
\Lambda = X^2\,\lambda\lpar 1,\mu^2_3,\mu^2_x\rpar.
\eq
Double and single pole parts are defined in \eqn{DUV} and in \eqn{sppart}
respectively. For the UV finite part we split the final result according to 
the dimension of the integrals over Feynman parameters,
\bqa
I_2 &=&  \intfxx{x}{y} I_{22} + \intfx{x} I_{21}  \nl
I_1 &=&  \intfxx{x}{y} I_{12} + \intfx{x} I_{11}  \nl
I_0 &=& \intfx{x} I_{01} + I_{00}  \nl
\eqa
The result is as follows:
\small
\bqa
I_{22} &=&\xi(x,y)\,X\, \Bigl[ 32 A B - 48 A^2 y + 128 B X y 
       - \xi(x,y)\,\lbpa 42 A + 96 X y \rbpa\Bigr]\,
         L_y  \ln_+\xi(x,y)
\nl
{}&{}&
       +  \xi(x,y)\,X\,\Bigl\{  
         - 32 A B - 8 A X (1-y) + A^2 (1+90 y)
\nl
{}&{}&
       +  X\,\Bigl[ - 296 y B - 32 B 
       + X (12 + 16 y )\Bigr]
       +  \xi(x,y) \,\Bigl[ 64 A + X ( 20 + 248 y )\Bigr]\Bigr\}\,
        L_x
\nl
{}&{}&
       +  \xi(x,y) \,X\,\Bigl[ 32 A B - 48 A^2 y +128 B X y 
       - \xi(x,y) \,\lbpa 42 A + 96 X y \rbpa\Bigr]\,
       \ln\xi(x,y) L_x
\nl
{}&{}&
       +  \xi(x,y)\,X\,\Bigl\{ 32 A B + 8 A X (1-y) - A^2 (1+90y)
       +  X
         \,\Bigl[ 296 y B + 32 B - X (12+16y)\Bigr]
\nl
{}&{}&
       - \xi(x,y) 
        \,\Bigl[ 64 A + X (20+248)y\Bigr]\Bigr\}\,
       \ln\xi(x,y) 
\nl
{}&{}&
       +  \xi(x,y) \,X\,\Bigl[
       - 32 A B + 48 A^2 y - 128 B X y 
       +  \xi(x,y) \,\lbpa
         42 A + 96 X y\rbpa\Bigr]\,\ln^2\xi(x,y)
\nl
{}&{}&
       +  \xi(x,y)  
     \,\lbpa - 6 A B X + 14 A X^2 - 3 A^2 X + A^2 B +
           8 B X^2 - 16 X^2\rbpa\,\,\lbpa\frac{\ln\xi(x,y)}{y-1}\rbpa_+
\nl
{}&{}&
       +   \xi(x,y) \,X\,
      \Bigl\{ - 4 A B + A X (14-2y) - A^2 (3+5y) +
        24 B X (y+1) - 16 (y+1) X^2
\nl
{}&{}&
       - \xi(x,y) 
       \,\Bigl[ 2 A + 16 X (1+y)\Bigr]\Bigr\}\,
        \ln_+\xi(x,y)
\nl
{}&{}&
\nl
{}&{}&
\nl
I_{21} &=& 
        - \frac{106}{9} A B X^2   
        + \frac{1658}{75} A X^3
        + 11 A^2 B X
        - \frac{193}{12} A^2 X^2
\nl
{}&{}&
        + 4 A^3_x X
        - 4 A^3_x B
        + \frac{13}{3} B X^3
        - \frac{158}{15} X^4 \
\nl
{}&{}&
       +  X\,\Bigl[ A 
         \,\lbpa  - 20 B^2 +\frac{62}{3} B X + \frac{586}{5} X^2\rbpa
       +  A^2 
         \,\lbpa 9 B - 72 X \rbpa
\nl
{}&{}&
       +  A^3 
         \,\lbpa 14 X - 4 B \rbpa
       +  X   \,\lbpa 24 B^2 - 44 B X - \frac{176}{3} X^2 \rbpa\Bigr]\,
          L_x
\nl
{}&{}&
       +   A X\,\lbpa  - 2 B^2 + 2 B X - A B + 5 A B^2 \rbpa\,
           \ln\xi(x,0)
       +  A^2  B^2\,\ln\xi(x,0) L_x 
\nl
{}&{}&
       +  \Bigl[ A\,\lbpa 22 X B^2 - 16 X^2 B 
        - 126 X^3\rbpa + A^2\,\lbpa
        - 12 X B
        + 77 X^2
        - 5  B^2\rbpa
\nl
{}&{}&
        + A^3_x\,\lbpa
        - 15 X
        + 5 B\rbpa
        - 24 B^2 X^2
        + 40 B X^3
        + 64 X^4 \Bigr]\,
        \ln\xi(x,1)
\nl
{}&{}&
       +  \Bigl[ 6 A B^2 X
         - 30 A X^3
         - 4 A^2 B X
         + 17 A^2 X^2
         - A^2 B^2
\nl
{}&{}&
         - 3 A^3_x X
         + A^3_x B
         - 8 B^2 X^2
         + 8 B X^3
         + 16 X^4 \Bigr]\,
        \ln\xi(x,1) L_x 
       -  A^2 B^2\,\ln^2\xi(x,0) 
\nl
{}&{}&
       +   A X
         \,\Bigl[  - 6 A B^2 X
        + 30 A X^3
        + 4 A^2 B X
        - 17 A^2 X^2
        + A^2 B^2
\nl
{}&{}&
        + 3 A^3_x X
        - A^3_x B
        + 8 B^2 X^2
        - 8 B X^3
        - 16 X^4 \Bigr]\,\ln^2\xi(x,1)
\nl
{}&{}&
\nl
{}&{}&
\nl
I_{12} &=&  \xi(x,y) \,\lbpa - 2 A + 16 X y\rbpa\,L_y  \ln_+\xi(x,y)
       +  \xi(x,y) \,\lbpa 7 A - 40 X y\rbpa\,L_x
\nl
{}&{}&
       +  \xi(x,y) \,\lbpa - 2 A + 16 X y \rbpa\,\ln\xi(x,y) L_x
       +   \xi(x,y) \,\lbpa - 7 A + 40 X y \rbpa\,\ln\xi(x,y)
\nl
{}&{}&
       +   \xi(x,y) \,\lbpa 2 A - 16 X y \rbpa\,\ln^2\xi(x,y)
       +  \lbpa - A X + A B - 2 B X + 2 X^2 \rbpa\,
        \lbpa\frac{\ln\xi(x,y)}{y-1}\rbpa_+
\nl
{}&{}&
       +  \xi(x,y) \,\lbpa - 2 A + 4 X\rbpa\,
       \lbpa\frac{\ln\xi(x,y)}{y-1}\rbpa_+
       +  \xi(x,y) \,\Bigl[ - 2 A + 4 X \,\lbpa 1 + y \rbpa \Bigr]\,
        \ln_+\xi(x,y)
\nl
{}&{}&
\nl
{}&{}&
\nl
I_{11} &=&  A X   \,\lbpa
         - \frac{40}{9} A X
         + 2 A^2
         - B X
         + \frac{11}{4} X^2 \rbpa
\nl
{}&{}&
       +  A X
         \,\lbpa   - \frac{133}{6} A X
         - 4 A B
         + 6 A^2
         + 9 B X
         + \frac{115}{6} X^2 \rbpa\,L_x
\nl
{}&{}&
       +   A X
         \,\lbpa \frac{149}{6} A X
        + 4 A B
        - 7 A^2
        - 7 B X
        - \frac{133}{6} X^2 \rbpa\,\ln\xi(x,1)
\nl
{}&{}&
       +  A
        \,\lbpa 7 A X  
        + A B
        - 2 A^2
        - 2 B X
        - 6 X^2 \rbpa\,\ln\xi(x,1) L_x
\nl
{}&{}&
       +  A X
         \,\lbpa  - 7 A X
        - A B
        + 2 A^2
        + 2 B X
        + 6 X^2 \rbpa\,\ln^2\xi(x,1)
\nl
{}&{}&
\nl
{}&{}&
\nl
I_{01} &=& - \frac{3}{2} \,\Bigl[ L_x\, + \ln\xi(x,1) \Bigr]  \qquad
I_{00} = \frac{13}{4} + \frac{1}{2} \zeta(2)
\eqa
\normalsize
Similarly we derive the following expression for $S_{3p}$ of \eqn{s3cont}:
\bq
S_{3p} = S^{\rm sp}_{3p} + S^{\rm fin}_{3p},  \qquad
S^{\rm fin}_{3p} = \frac{1}{\Lambda}\,\Bigl[ \frac{J_2}{\Lambda} +
J_1\Bigr] + J_0,  \quad
\Lambda = X^2\,\lambda\lbpa 1,\mu^2_3,\mu^2_x\rbpa,
\eq
where the single-pole part is given in \eqn{sp3p}. For the finite part we 
split again the various contributions according to
\bqa
J_2 &=&  \intfxx{x}{y} J_{22} + \intfx{x} J_{21}  \nl
J_1 &=&  \intfxx{x}{y} J_{12} + \intfx{x} J_{11}  \nl
J_0 &=& \intfx{x} J_{01} + J_{00}  \nl
\eqa
The result for the $J_i$ functions is as follows:
\small
\bqa
J_{22} &=&  \xi(x,y)   
        \Bigl[ A X B  ( 224 y - 32 )
         + 48  A^2 X   
         + 66  A^2 B 
\nl
{}&{}&
         - 64 A^3 y 
         - 200 X B^2 
         - 128 X^2 y B \Bigr]\,L_y  \ln_+\xi(x,y)
\nl
{}&{}&
       +   \xi^2(x,y)  
        \Bigl[ A X   ( 42 - 192 y )
         - 64 A^2 
         + 340 X B 
         + 96  X^2 y \Bigr]\,L_y  \ln_+\xi(x,y)
\nl
{}&{}&
       - 140  \xi^3(x,y)  X\,L_y  \ln_+\xi(x,y)
\nl
{}&{}&
       +  \Bigl[ A X   (  - 524 y B + 32 B )
          - 80 A^2 X y 
          - 148 A^2 B 
          + 139 A^3 y 
\nl
{}&{}&
          + 528 X B^2
          + 256 X^2 y B \Bigr]\,L_x \xi(x,y)   
\nl
{}&{}&
       +   \xi^2(x,y)   
         \Bigl[ A X   (  - 56 + 484 y )
          + 139 A^2 
          - 924 X B
          - 224 X^2 y \Bigr]\,L_x
       +  396 \xi^3(x,y)   X\,L_x
\nl
{}&{}&
       +   \xi(x,y)   
        \Bigl[ A X   ( 224 y B - 32 B )
         + 48  A^2 X y 
         + 66  A^2 B 
         - 64  A^3 y 
\nl
{}&{}&
         - 200 X B^2 
         - 128 X^2 y B \Bigr]\,\ln \xi(x,y) L_x
\nl
{}&{}&
       +   \xi^2(x,y)   
         \Bigr[ A X   ( 42 - 192 y )
          - 64  A^2 
          + 340 X B
          + 96  X^2 y \Bigr]\,\ln \xi(x,y) L_x
\nl
{}&{}&
       - 140  \xi^3(x,y)  X\,\ln \xi(x,y) L_x
\nl
{}&{}&
       +   \xi(x,y)   
        \Bigl[ A X   ( 524 y B - 32 B )
         + 80  A^2 X y 
         + 148 A^2 B 
\nl
{}&{}&
         - 139 A^3 y 
         - 528 X B^2
         - 256 X^2 y B \Bigr]\,\ln \xi(x,y)
\nl
{}&{}&
       +   \xi^2(x,y)   
         \Bigl[ A X   ( 56 - 484 y )
          - 139 A^2
          + 924 X B
          + 224 X^2 y \Bigr]\,\ln \xi(x,y)
\nl
{}&{}&
       - 396  \xi^3(x,y)   X\,\ln \xi(x,y)
\nl
{}&{}&
       +   \xi(x,y)   
        \Bigl[ A X   (  - 224 y B + 32 B )
         - 48  A^2 X y 
         - 66  A^2 B 
\nl
{}&{}&
         + 64  A^3 y 
         + 200 X B^2
         + 128 X^2 y B \Bigr]\,\ln^2 \xi(x,y)
\nl
{}&{}&
       +   \xi^2(x,y)   
        \Bigl[ A X   (  - 42 + 192 y )
         + 64  A^2 
         - 340 X B
         - 96 X^2 y \Bigr]\,\ln^2 \xi(x,y)
\nl
{}&{}&
       +  140  \xi^3(x,y)   X\,\ln^2 \xi(x,y)
\nl
{}&{}&
       +   \xi(x,y)    
       \Bigl[ - 4 A X B 
        + 4 A X^2
        -   A^2 X 
        +   A^2 B 
        + 4 X^2 B
        - 4 X^3 \Bigr]\,\lbpa\frac{\ln\xi(x,y)}{y-1}\rbpa_+
\nl
{}&{}&
       +   \xi(x,y)   
       \Bigl[ A X   ( 24 y B - 4 B )
        + 4 A X^2
        - A^2 X (1+y )
        + 8 A^2 B 
\nl
{}&{}&
        - 6 A^3 y 
        - 32 X B^2
        + 8 X^2 B (1+y)
        - 4 X^3 (1+y) \Bigr]\,\ln_+\xi(x,y)
\nl
{}&{}&
       +   \xi^2(x,y)   
        \Bigl[ - 22 A X y 
         - 6 A^2
         + 52 X B
         - 4 X^2 (1+y)\Bigr]\,\ln_+\xi(x,y)
\nl
{}&{}&
       - 20   \xi^3(x,y) X\,\ln_+\xi(x,y)
\nl
{}&{}&
\nl
{}&{}&
\nl
J_{21} &=& 
       - \frac{59}{9}  A X^2 B 
       + \frac{182}{75}   A X^3
       + \frac{31}{9}   A^2 X B 
       - \frac{282}{300} A^2 X^2
       + \frac{283}{75}  X^3 B 
       - \frac{1234}{735} X^4 
\nl
{}&{}&
       +     \Bigl[ 
          - 20 B^2 A X
          - \frac{56}{3} B A X^2
          + \frac{592}{15} A X^3
          + \frac{58}{3} B A^2 X
\nl
{}&{}&
          - \frac{124}{5} A^2 X^2
          + 9 B^2 A^2
          + 5 A^3_x X
          - 5 B A^3_x
          + 20 B^2 X^2
          - \frac{4}{5} B X^3
          - \frac{412}{21} X^4 \Bigr]\,L_x
\nl
{}&{}&
       - 4  \ln \xi(x,0)   A^2 B^2 
       -  \ln \xi(x,0) L_x  A^2  B^2 
\nl
{}&{}&
       +    \Bigl[ 
          20 A X B^2 
        + 20 A X^2 B 
        - 40 A X^3
        - 20 A^2 X B 
\nl
{}&{}&
        + 25 A^2 X^2 
        - 5  A^2 B^2 
        - 5  A^3 X 
        + 5  A^3 B 
        - 20 X^2 B^2 
        + 20 X^4 \Bigr]\,\ln \xi(x,1) 
\nl
{}&{}&
       +   \Bigl[  
         4 A X B^2 
       + 4 A X^2 B 
       - 8 A X^3
       - 4 A^2 X B
\nl
{}&{}&
       + 5 A^2 X^2 
       -   A^2 B^2
       -   A^3 X 
       +   A^3 B
       - 4 X^2 B^2
       + 4 X^4\Bigr]\,\ln \xi(x,1) L_x
\nl
{}&{}&
       +   \xi(x,0)   A^2  B^2\,\ln^2 
       +    \Bigl[
        - 4 A X B^2 
        - 4 A X^2 B 
        + 8 A X^3 
        + 4 A^2 X B
\nl
{}&{}&
        - 5 A^2 X^2
        +   A^2 B^2 
        +   A^3 X
        -   A^3 B
        + 4 X^2 B^2
        - 4 X^4 \Bigr]\,\ln^2 \xi(x,1)
\nl
{}&{}&
\nl
{}&{}&
\nl
J_{12} &=&  \xi(x,y)   
        \Bigl[ A   ( 2 + 24 y )
         - 16 X y 
         - 30 B \Bigr]\,L_y  \ln_+\xi(x,y)
       +  30  \xi^2(x,y)\,L_y  \ln_+\xi(x,y)  
\nl
{}&{}&
       +   \xi(x,y)   
        \Bigl[ A   (  - 7 - 64 y )
         + 40 X y
         + 86 B\Bigr]\,L_x
       - 86   \xi^2(x,y)\,L_x
\nl
{}&{}&
       +   \xi(x,y)   
        \Bigl[ A   ( 2 + 24 y )
         - 16 X y
         - 30 B \Bigr]\,\ln \xi(x,y) L_x
       +  30 \ln \xi(x,y) L_x \xi^2(x,y)   
\nl
{}&{}&
       +   \xi(x,y)   
        \Bigl[ A   ( 7 + 64 y )
         - 40 X y
         - 86 B \Bigr]\,\ln \xi(x,y)
       +  86  \xi^2(x,y)\,\ln \xi(x,y)
\nl
{}&{}&
       +   \xi(x,y)   
        \Bigl[ A   (  - 2 - 24 y )
         + 16 X y
         + 30 B \Bigr]\,\ln^2 \xi(x,y)
       - 30  \xi^2(x,y)\,\ln^2 \xi(x,y)
\nl
{}&{}&
       +   \xi(x,y)    
       \Bigl[ - A + 2 X\Bigr]\,\lbpa\frac{\ln\xi(x,y)}{y-1}\rbpa_+
\nl
{}&{}&
       +   \xi(x,y)   
        \Bigl[ A   (  - 1 + 3 y )
         + 2 X (1+y)
         - 6 B\Bigr]\,\ln_+\xi(x,y)
       + 6  \xi^2(x,y)\,\ln_+\xi(x,y)
\nl
{}&{}&
\nl
{}&{}&
\nl
J_{11} &=&
       - \frac{11}{36}  A X 
       - \frac{1}{6}    A^2  
       + \frac{7}{9}    X B
       + \frac{103}{300} X^2
\nl
{}&{}&
       +     \Bigl[
        - \frac{143}{12} A X
        - \frac{7}{2}    A B 
        + 4      A^2
        + \frac{20}{3}   X B 
        + \frac{41}{15}  X^2 \Bigr]\,L_x
\nl
{}&{}&
       +     \Bigl[
         \frac{49}{4} A X
        + \frac{7}{2} A B 
        - 4   A^2  
        - \frac{22}{3} X B 
        - \frac{42}{5} X^2 \Bigr]\,\ln \xi(x,1)
\nl
{}&{}&
       +   L_x   
       \Bigl[ 3 A X
        +       A B
        -       A^2 
        - 2     X B
        - 2     X^2 \Bigr]\,\ln \xi(x,1)
\nl
{}&{}&
       +     \Bigl[
        - 3 A X 
        -   A B 
        +   A^2 
        + 2 X B
        + 2 X^2 \Bigr]\,\ln^2 \xi(x,1)
\nl
{}&{}&
\nl
{}&{}&
\nl
J_{01} &=&  \frac{1}{3}\,\Bigl[ \ln \xi(x,1) - L_x \Bigr]
\qquad
J_{00} = - \frac{7}{72} 
\eqa
\normalsize
\section{Appendix D}
In this Appendix we derive results for the sunset diagram $S_3$ at $p^2 = 0$. 
In this case we can use the notation of ref.~\cite{vanderBij:1984bw}, i.e.
\bqa
{}&{}&
\lbpa M_{11},\dots,M_{1n}\mid M_{21},\dots,M_{2m}\mid M_{31},\dots,M_{3l}\rbpa
=  \nl
{}&{}&
\intmomsii{n}{p}{q}\,\prod_{i=1}^{n}\prod_{j=1}^{m}\prod_{k=1}^{l}\,
\frac{1}{\lbpa p^2 + M^2_{1i}\rbpa\,
\lbpa q^2 + M^2_{3j}\rbpa\,
\lbpa (p-q)^2 + M^2_{2l}\rbpa}.
\eqa
In these notations we have
\bq
S_3\lbpa 0;m_1,m_2,m_3\rbpa = \lbpa m_1\mid m_2\mid m_3\rbpa.
\eq
Furthermore one easily derive
\bq
\lbpa m_1\mid m_2\mid m_3\rbpa = -\,\lbpa m_1\mid m_2\mid m_3\rbpa_{(-2)}\,
\ep^{-2} +
\lbpa m_1\mid m_2\mid m_3\rbpa_{(-1)}\,\ep^{-1} + 
\lbpa m_1\mid m_2\mid m_3\rbpa_0,
\eq
with singular parts given by
\bqa
\lbpa m_1\mid m_2\mid m_3\rbpa_{(-2)} &=& -\,2\,\sum_{i=1}^{3}\,m^2_i  \nl
\lbpa m_1\mid m_2\mid m_3\rbpa_{(-1)} &=& \lbpa m_1\mid m_2\mid m_3\rbpa_{(-2)}
-\,\sum_{i=1}^{3}\,m^2_i\,\Bigl[
1 - 2\,\gamma - 2\,\ln\lbpa \pi m^2_i\rbpa\Bigr],  
\eqa
and a finite part that becomes
\bqa
{}&{}&
\lbpa m_1\mid m_2\mid m_3\rbpa_0 = -\,\lbpa m_1\mid m_2\mid m_3\rbpa_{(-2)} +
\lbpa m_1\mid m_2\mid m_3\rbpa_{(-1)}  \nl
{}&{}&
- \sum_{i=1}^{3}\,m^2_i\Bigl[
-\frac{1}{2} - \frac{\pi^2}{12} - \lpar \gamma + \ln \pi m^2_i \rpar
\lpar \gamma + \ln \pi m^2_i - 1\rpar - F_i\Bigr].
\label{zana}
\eqa
Furthermore, the finite part is given in terms of the function
\bqa
F_i &=& \intfx{x}\,\Bigl[ \li{2}{1-\mu^2_i} - \frac{\mu^2_i}{1-\mu^2_i}\,
\ln\mu^2_i\Bigr],  \nl
\mu^2_i &=& {{a_i x + b_i \lbpa 1 - x\rbpa}\over {x \lbpa 1 - x\rbpa}},
\qquad a_i = \frac{m^2_j}{m^2_i}, \quad b_i = \frac{m^2_k}{m^2_i},
\label{zanb}
\eqa
and 
\bqa
j=3, k=2 \quad &\mbox{for}& \quad i=1, \nl
j=1, k=3 \quad &\mbox{for}& \quad i=2, \nl
j=1, k=2 \quad &\mbox{for}& \quad i=3.
\eqa
Since $p^2 = 0$ we can apply directly the original proposal of
F.~V.~Tkachov~\cite{Tkachov:1997wh}. First we combine the $q_1$-dependent
propagators with a Feynman parameter $x$ and obtain
\bqa
{}&{}&
\intmomi{n}{q_1}\, \frac{1}{\lbpa q^2_1 + m^2_1\rbpa\,
\lbpa\lbpa q_1-q_2\rbpa^2 + m^2_2\rbpa} =  \nl
{}&{}&
i\,\pi^{2-\ep/2}\,\egam{\frac{\ep}{2}}\,\intfx{x}\,\Bigl[
x(1-x)\,q^2_2 + (1-x)\,m^2_1 + x\,m^2_2\Bigr]^{-\ep/2}.
\eqa
We can {\em raise} one power in the integrand and obtain
\bqa
V^{-\ep/2} &=& 4\,{{q^2}\over
{\lbpa q^2 + m^2_2 - m^2_1\rbpa^2 + 4\,q^2 m^2_1}}  \nl
{}&\times& \Bigl\{ 1 - \frac{1}{2-\ep}\,\Bigl[ x - \frac{1}{2}\,
\frac{q^2 - m_+m_-}{q^2}\,\frac{d}{dx}\Bigr]\Bigr\}\,V^{1-\ep/2},
\eqa
where $m_{\pm} = m_1 \pm m_2$ and
\bq
V = x(1-x)\,q^2 + (1-x)\,m^2_1 + x\,m^2_2 = x(1-x)\,\lbpa q^2 + m^2_x\rbpa.
\eq
Next we observe that
\bq
\lbpa q^2 + m^2_2 - m^2_1\rbpa^2 + 4\,q^2 m^2_1 = \lbpa q^2 + m^2_+\rbpa\,
\lbpa q^2 + m^2_-\rbpa,
\eq
and perform the remaining $q$-integration which requires the introduction of 
two additional Feynman parameters, $y$ and $z$.
This procedure will introduce two monomials in $y,z$,
\bq
U_{\pm}(y,z) = \lbpa m^2_{\pm} - m^2_x \rbpa\,y + \lbpa m^2_3 - m^2_{\pm}
\rbpa\,z + m^2_x,
\eq
which appear with power $-1-\ep$, so that we may use the following identity:
\bq
U^{-1-\ep}_{\pm} = \frac{1}{m^2_x}\,\Bigl[ 1 + \frac{y}{\ep}\,
\partial_y + \frac{z}{\ep}\,\partial_z\Bigr]\,U^{-\ep}.
\eq
After that we proceed as usual, namely we integrate by parts and expand in
Laurent series around $\ep = 0$. The final result is
\bqa
S^{\rm fin}_3\lbpa 0;m_1,m_2,m_3\rbpa &=& \frac{1}{m^2_+ - m^2_-}\,\Bigl[
\intfx{x}\intfxy{y}{z}\,K_3  \nl
{}&+& \intfxx{x}{y}\,K_2 + \intfx{x}\,K_1\Bigr] + K_0.
\eqa
We now introduce auxiliary variables
\bqa
W_{\pm}(x,y,z) &=& x \lbpa 1 - x \rbpa\,\Bigl[z m^2_3 + \lbpa y - z \rbpa
m^2_{\pm}\Bigr] + \frac{1}{4}\,\lbpa 1 - y\rbpa\,
\Bigl[\lbpa m_+ + m_-\rbpa^2 - 4\,m_+ m_- x\Bigr],  \nl
\chi_{_{\pm}} &=&  x\,m^2_3 + \lbpa  1 - x\rbpa\,m^2_{\pm},  \quad
A = \lbpa m_+ + m_-\rbpa^2 - 4\,m_+ m_- x,  \quad
X = x\,\lbpa 1 - x\rbpa,  \nl
L_y &=& \ln (1-y), \quad L_x = \ln X,
\eqa
and derive
\bqa
K_{2,3} &=& m^2_+\,K_{2,3+} - m^2_-\,K_{2,3-},
\nl
{}&{}&
\nl
K_{3\pm} &=& \ln W_{\pm}(x,y,z)\,\Bigl\{
        3 (L_x + L_y)\Bigl[ A - 12 W_{\pm}(x,y,z)\Bigr]  
       - \frac{1}{2}  A - 54 W_{\pm}(x,y,z)
\nl
{}&{}&
       - 3 \ln W_{\pm}(x,y,z) \Bigl[ A
       - 12 W_{\pm}(x,y,z)\Bigr] \Bigr\}
       - \frac{3}{2}  A\,\lbpa\frac{\ln W_{\pm}(x,y,z)}{y-1}\rbpa_+ 
\nl
{}&{}&
\nl
K_{2\pm} &=& \frac{1}{2}\,A \ln W_{\pm}(x,1,y) \Bigl\{
 3\,L_y + 1 + 3\,\Bigl[ L_x - \ln W_{\pm}(x,1,y) \Bigr]\Bigr\}
\nl
{}&{}&
\nl
{}&{}&
\nl
K_1 &=&  - m^2_- (  m_+ m_- + \frac{3}{4} m^2_+  + \frac{1}{4} m_-^2 )\,
         \ln \frac{\chi_{_-}}{\chi_{_+}} \ln m^2_1
\nl
{}&{}&
         +  m^2_- ( m_+ m_- - \frac{3}{4} m^2_+ - \frac{1}{4} m_-^2 )\,
         \ln \frac{\chi_{_-}}{\chi_{+}} \ln m^2_2
\nl
{}&{}&
       - \frac{1}{4} m_+ m_- \ln \chi_{_-} \Bigl[ (m^2_+-m^2_-)\, \ln m^2_1
       - \ln m^2_2 \Bigr]
\nl
{}&{}&
       + \frac{1}{4} \Bigl[ 3 m_+ m_- + (m^2_++m^2_-) \Bigr]
         (m^2_+-m^2_-)\,\ln \chi_{_+} \ln \frac{m^2_1}{m^2_2}
\nl
{}&{}&
       - \frac{1}{2} (m^4_+-m^4_-)\,\ln \chi_{_+} ( 1 - 
         \frac{1}{2}\,\ln\chi_{_+} )
       - \frac{1}{4} m^2_-  ( 3 m^2_+ + m_-^2 )  
\nl
{}&{}&
\times ( \ln^2  \chi_{_-} - \ln^2 \chi_{_+} - 2 \ln\frac{\chi_{_-}}{\chi_{_+}})
\nl
{}&{}&
\nl
{}&{}&
\nl
K_0 &=&
       + \frac{1}{2}\,\Bigl[ m^2_3 + \frac{1}{2} (m^2_++m^2_-) \Bigr]\,\zeta(2)
       - \frac{1}{2}\,\Bigl[ m_+ m_- + \frac{1}{2} (m^2_++m^2_-) \Bigr]\,
\ln m^2_1
\nl
{}&{}&
       + \frac{1}{2}\,\Bigl[ m_+ m_- - \frac{1}{2} (m^2_++m^2_-)\Bigr]\,
\ln m^2_2
       + \frac{1}{4}\,\Bigl[ m_+ m_- + \frac{1}{4} (m^2_++m^2_-)\Bigr]\,
\ln^2  m^2_1
\nl
{}&{}&
       - \frac{1}{4}\,\Bigl[ m_+ m_- - \frac{1}{2} (m^2_++m^2_-)\Bigr]\,
\ln^2  m^2_2
       + \frac{13}{4} m^2_3 - \frac{17}{4} (m^2_++m^2_-)
\eqa
\vskip 5pt
The above results have been coded in a FORTRAN program and compared with
those obtained with the analytical result of~\cite{vanderBij:1984bw}.
The comparison is shown in \tabn{zeroc} where we use $\ep = 1$ and also
the $\tHs = 1\,$GeV, where $\tHs$ is the unit of mass.
\begin{table}[hp]\centering
\begin{tabular}{|r|c|c|r|r|}
\hline
 & & & & \\
$m_1\,$[GeV] & $m_2\,$[GeV] & $m_3\,$[GeV] & BT-numerical & 
ref.~\cite{vanderBij:1984bw}  \\
 & & & & \\
\hline
 & & & & \\
3  & 2 & 4 & 147.63632 & 147.63634  \\
 & & & & \\
6  & 2 & 4 & 491.36054 & 491.36060  \\ 
 & & & & \\
9  & 2 & 4 & 1437.2692 & 1437.2693  \\ 
 & & & & \\
90 & 2 & 4 & 585240.60 & 585240.63  \\
 & & & & \\
\hline
\hline
\end{tabular}
\vspace*{3mm}
\caption[]{The sunset topology $S_3$ for $p^2 = 0$. First entry is the 
numerical BT-approach, second entry is the analytical result of 
\eqns{zana}{zanb} (see ref.~\cite{vanderBij:1984bw}). 
The UV-pole is $\ep = 1$ and the unit of mass is $\tHs = 1\,$GeV.}
\label{zeroc}
\end{table}
\normalsize

\clearpage

\begin{figure}[ht]
\hspace{-0.5cm}
\vspace{2.cm}
\[
  \vcenter{\hbox{
  \begin{picture}(350,180)(-15,-15)
\Line(-100.,0.)(100.,0.)
\Line(300.,0.)(400.,0.)
\Line(-30.,200.)(-30.,-200)
\Line(100.,0.)(300.,0.)
\Line(100.,1.)(300.,1.)
\Line(100.,-1.)(300.,-1.)
\Line(100.,0.5)(300.,0.5)
\Line(100.,-0.5)(300.,-0.5)
\DashCArc(0.,0.)(150.,15.,90.){3}
\ArrowArcn(0.,0.)(150.,15.,13.)
\DashCArc(0.,0.)(150.,270.,345.){3}
\ArrowArc(0.,0.)(150.,345.,347.)
\DashCArc(400.,0.)(150.,90.,165.){3}
\ArrowArc(400.,0.)(150.,165.,167.)
\DashCArc(400.,0.)(150.,195.,270.){3}
\ArrowArcn(400.,0.)(150.,197.,195.)
\ArrowLine(-30.,0.)(30.,0.)
\ArrowArcn(65.,0.)(50.,180.,0.)
\ArrowArc(320.,0.)(50.,180.,360.)
\ArrowLine(370.,0.)(400.,0.)
\Text(100.,-3.)[cb]{$\bullet$}
\Text(300.,-3.)[cb]{$\bullet$}
\Text(160.,10.)[cb]{$-$}
\Text(240.,10.)[cb]{$+$}
\Text(160.,-15.)[cb]{$+$}
\Text(240.,-15.)[cb]{$-$}
\Text(395.,15.)[cb]{$1$}
\Text(-40.,15.)[cb]{$0$}
\Text(395.,-3.)[cb]{$\bullet$}
\Text(-30.,-3.)[cb]{$\bullet$}
\SetScale{1.5}
\Text(390.,-25.)[cb]{$\Reb\, x$}
\Text(-50.,190.)[cb]{$\Imb\, x$}
\Text(95.,-25.)[cb]{$x_{\ssL}$}
\Text(300.,-25.)[cb]{$x_{\ssR}$}
\Text(70.,-3.)[cb]{$\bullet$}
\Text(350.,-3.)[cb]{$\bullet$}
\Text(65.,-25.)[cb]{$x^-_-$}
\Text(350.,-25.)[cb]{$x^-_+$}
\SetScale{1.}
  \end{picture}}}
\]
\vspace{6.cm}
\caption[]{The complex $x$-plane with the cut of $\ln\chi$ (\eqn{defchi})
internal to the interval $[x^-_-,x^-_+]$ (\eqn{defxm}). Dashed lines indicate 
the sign of the imaginary part of the logarithm when approaching the real axis 
on the cut. An example of the deformation of the integration path is shown.}
\label{cutplane1}
\end{figure}
\vskip 5pt
\begin{figure}[ht]
\hspace{-1.5cm}
\vspace*{-8mm}
\[
  \vcenter{\hbox{
  \begin{picture}(350,180)(-15,-15)
\Line(120.,0.)(280.,0.)
\Line(-30.,200.)(-30.,-200)
\Line(-100.,0.)(120.,0.)
\Line(-100.,1.)(120.,1.)
\Line(-100.,-1.)(120.,-1.)
\Line(-100.,0.5)(120.,0.5)
\Line(-100.,-0.5)(120.,-0.5)
\Line(280.,0.)(400.,0.)
\Line(280.,1.)(400.,1.)
\Line(280.,0.5)(400.,0.5)
\Line(280.,-1.)(400.,-1.)
\Line(280.,-0.5)(400.,-0.5)
\DashCArc(-40.,0.)(150.,15.,90.){3}
\ArrowArcn(-40.,0.)(150.,15.,13.)
\DashCArc(-40.,0.)(150.,270.,345.){3}
\ArrowArc(-40.,0.)(150.,345.,347.)
\DashCArc(440.,0.)(150.,90.,165.){3}
\ArrowArc(440.,0.)(150.,165.,167.)
\DashCArc(440.,0.)(150.,195.,270.){3}
\ArrowArcn(440.,0.)(150.,197.,195.)
\Text(120.,-3.)[cb]{$\bullet$}
\Text(280.,-3.)[cb]{$\bullet$}
\Text(110.,10.)[cb]{$+$}
\Text(290.,10.)[cb]{$-$}
\Text(110.,-15.)[cb]{$-$}
\Text(290.,-15.)[cb]{$+$}
\Text(395.,15.)[cb]{$1$}
\Text(-40.,15.)[cb]{$0$}
\Text(395.,-3.)[cb]{$\bullet$}
\Text(-30.,-3.)[cb]{$\bullet$}
\ArrowLine(125.,0.)(145.,0.)
\ArrowArcn(172.5,0.)(35.,180.,0)
\ArrowLine(200.,0.)(250.,0.)
\SetScale{1.5}
\Text(390.,-25.)[cb]{$\Reb\, x$}
\Text(-50.,190.)[cb]{$\Imb\, x$}
\Text(125.,-25.)[cb]{$x_{\ssL}$}
\Text(270.,-25.)[cb]{$x_{\ssR}$}
\Text(150.,-3.)[cb]{$\bullet$}
\Text(200.,-3.)[cb]{$\bullet$}
\Text(145.,-25.)[cb]{$x^-_-$}
\Text(200.,-25.)[cb]{$x^-_+$}
\SetScale{1.}
  \end{picture}}}
\]
\clearpage

\vspace{6.cm}
\caption[]{The complex $x$-plane with the cut of $\ln\chi$ (\eqn{defchi})
external to the interval $[x^-_-,x^-_+]$ (\eqn{defxm}). Dashed lines indicate 
the sign of the imaginary part of the logarithm when approaching the real axis 
on the cut. An example of the deformation of the integration path is shown.}
\label{cutplane2}
\end{figure}
 
\bfi
\centerline{
\epsfig{file=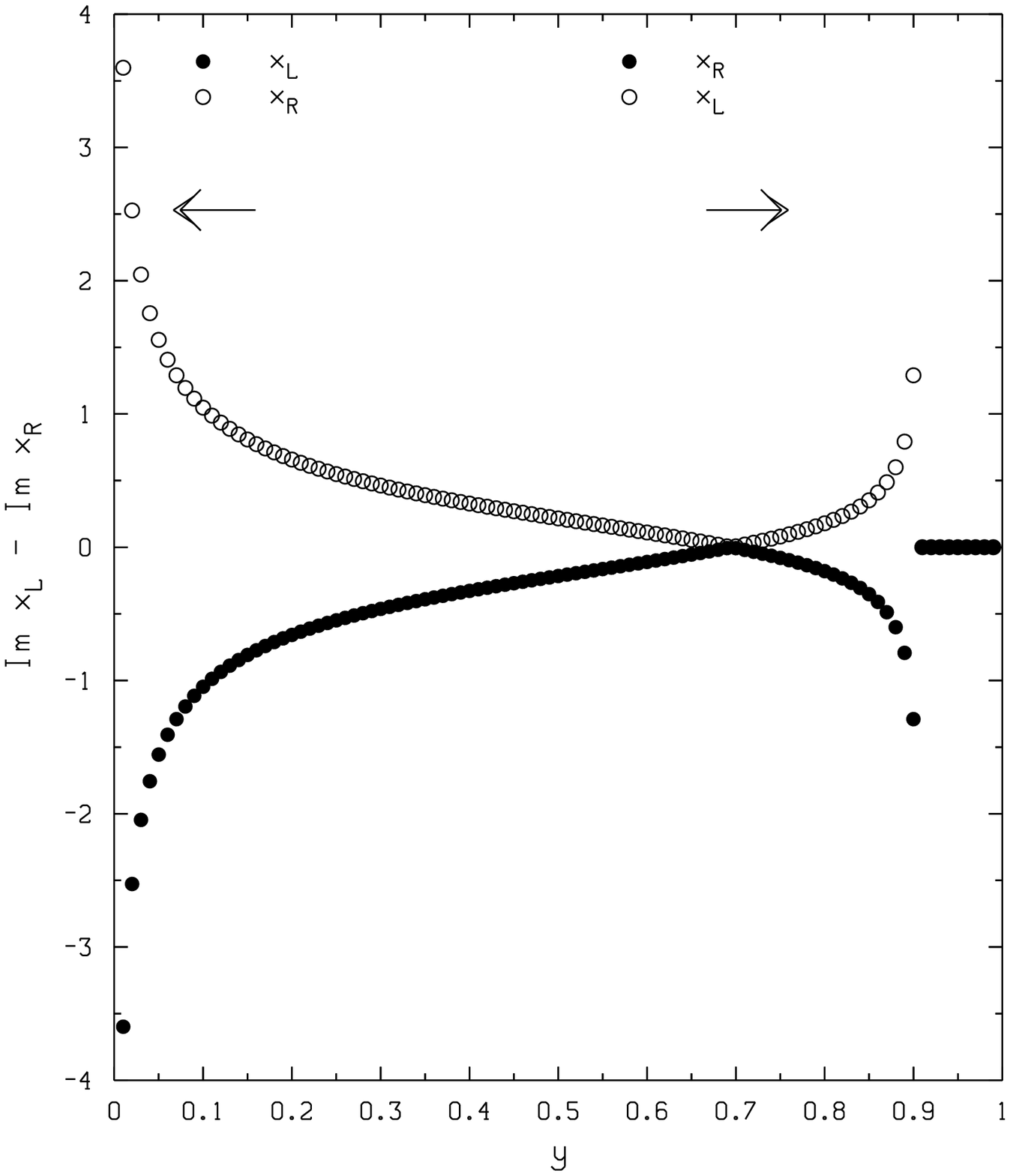,height=12cm,angle=0}
\epsfig{file=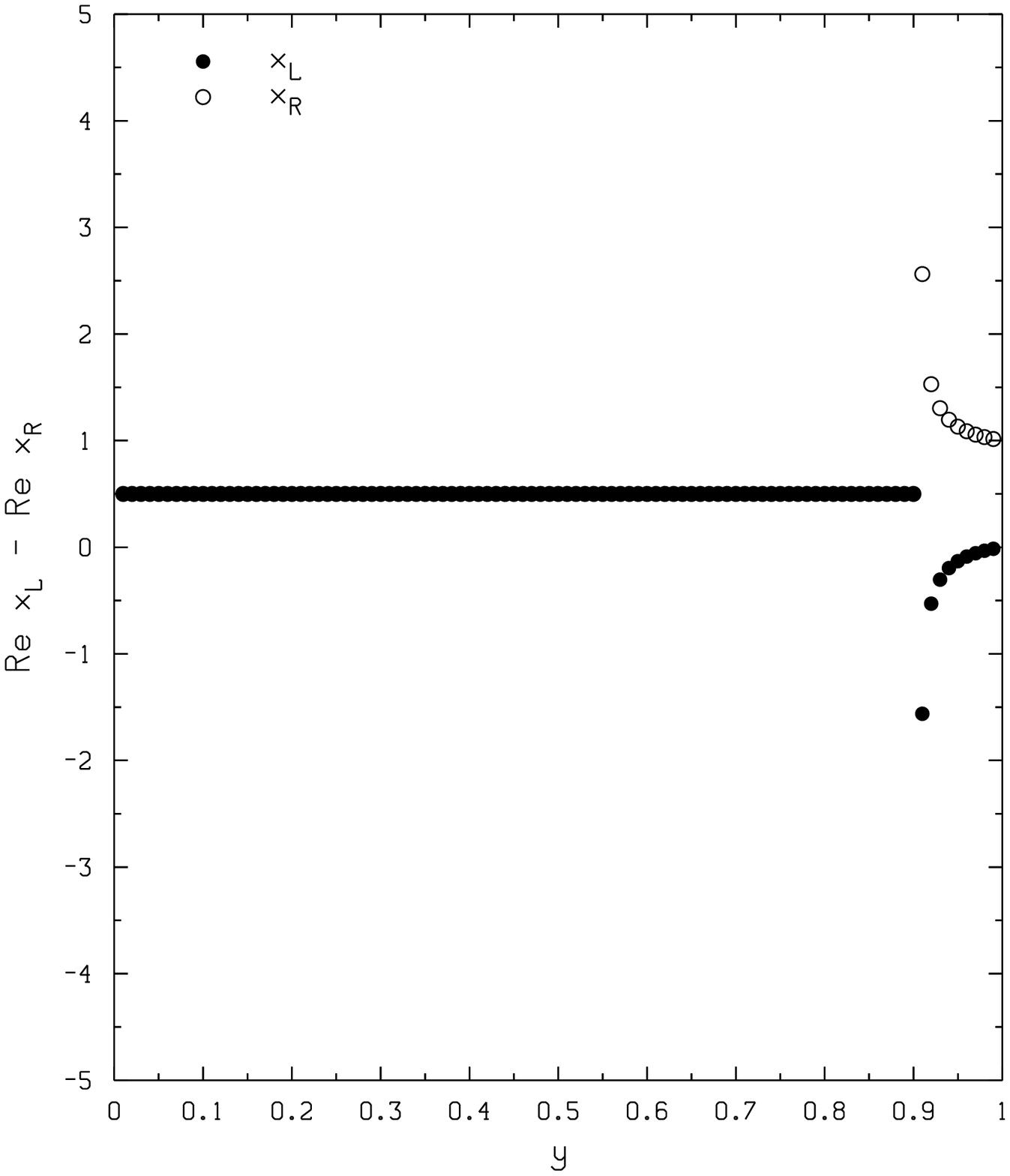,height=12cm,angle=0}}
\caption[]{The behavior of $x_{\ssL}$ and $x_{\ssR}$, the roots of the
quadratic of \eqn{quad}) as a function of $y$ for the sunset diagram at the 
normal threshold $s = (m_1+m_2+m_3)^2$.
The figure shows that the integration contour is pinched by the two
branch points $x_{\ssL,\ssR}$ which are the roots of \eqn{quad}.}
\label{irxlr}
\efi
\bfi
\centerline{
\epsfig{file=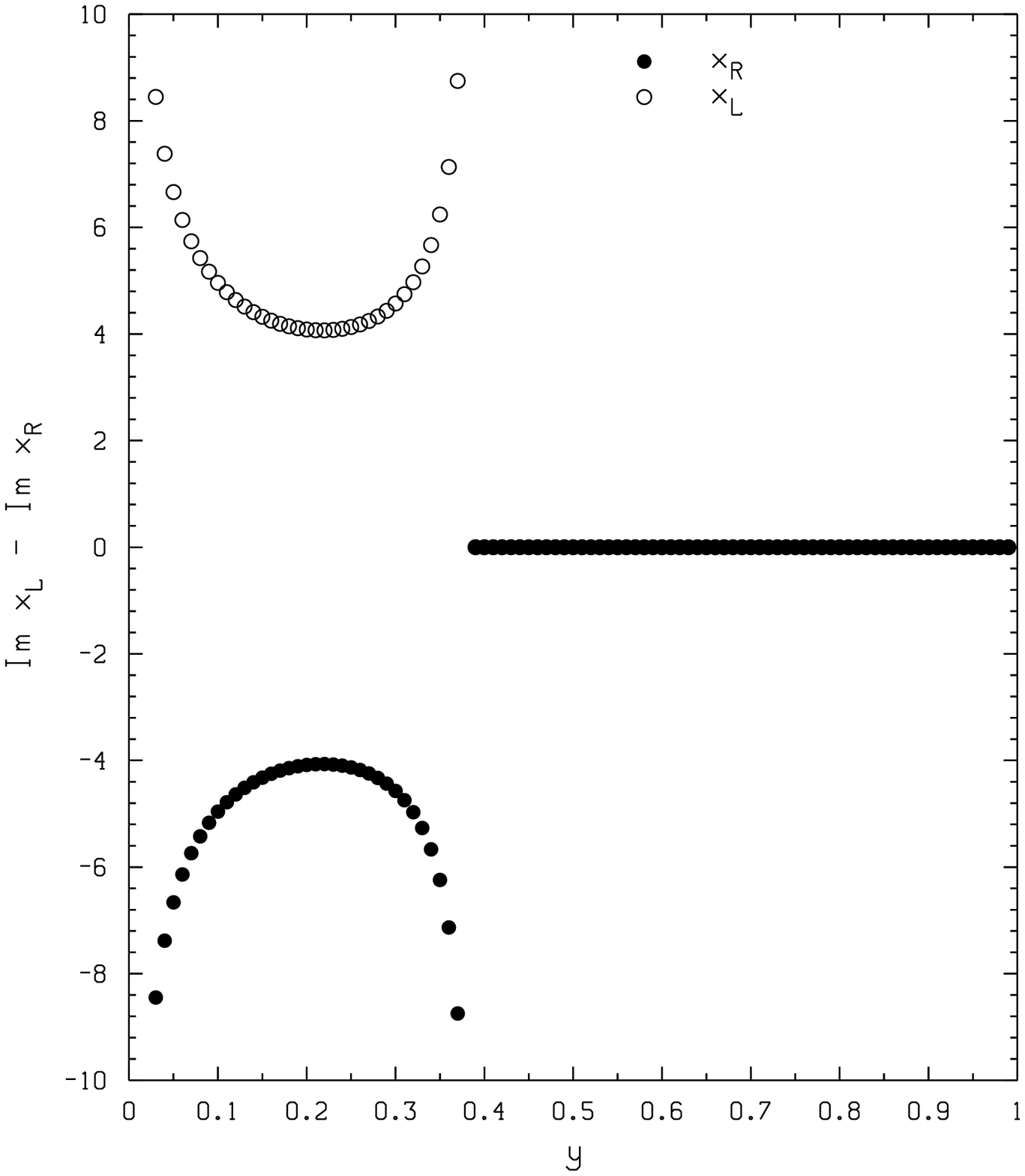,height=12cm,angle=0}
\epsfig{file=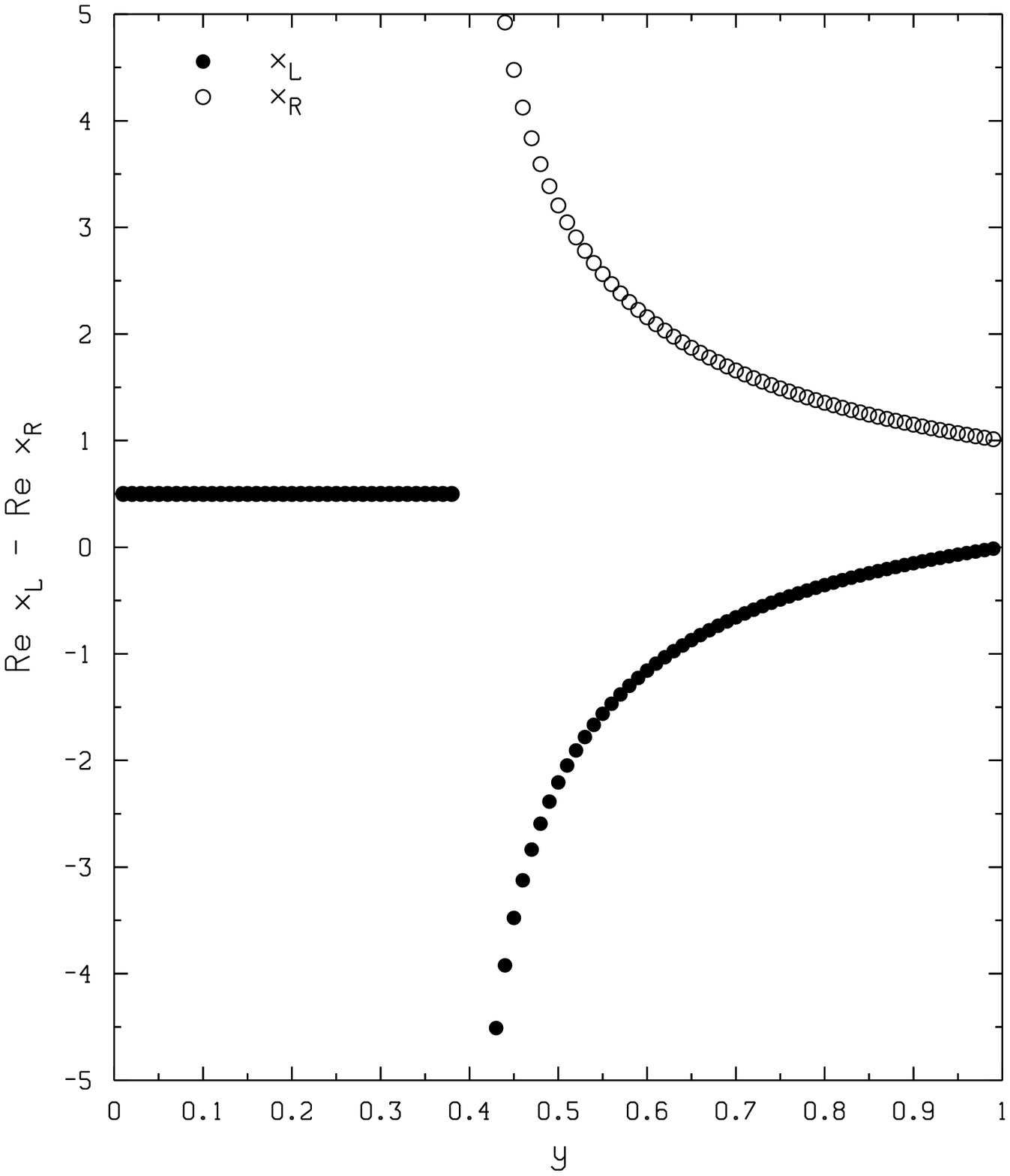,height=12cm,angle=0}}
\caption[]{The behavior of $x_{\ssL}$ and $x_{\ssR}$, the roots of the
quadratic of \eqn{quad}) as a function of $y$ for the sunset diagram at the 
pseudo threshold $s = (m_1+m_2-m_3)^2$.}
\label{irxlrp}
\efi

\end{document}